\documentclass{article}
\usepackage{graphicx}
\usepackage{float}
\usepackage{topcapt}

\usepackage{lipsum}

\usepackage{amsmath}
\usepackage{url}
\usepackage[T2A]{fontenc}
\usepackage[russian,english,british]{babel}
\usepackage{longtable}
\usepackage{fancyvrb}
\usepackage{tocbibind}

\begin{document}

\title{Optimal Trading Strategies as Measures of Market Disequilibrium}
\author{Valerii Salov}
\date{}
\maketitle

\begin{abstract}
For classification of the high frequency trading quantities, waiting times, price increments within and between sessions are referred to as the a-, b-, and c-increments. Statistics of the a-b-c-increments are computed for the Time \& Sales records posted by the Chicago Mercantile Exchange Group for the futures traded on Globex. The Weibull, Kumaraswamy, Riemann and Hurwitz Zeta, parabolic, Zipf-Mandelbrot distributions are tested for the a- and b-increments. A discrete version of the Fisher-Tippett distribution is suggested for approximating the extreme b-increments. Kolmogorov and Uspenskii classification of stochastic, typical, and chaotic random sequences is reviewed with regard to the futures price limits. Non-parametric $L_1$ and log-likelihood tests are applied to check dependencies between the a- and b-increments. The maximum profit strategies and optimal trading elements are suggested as measures of frequency and magnitude of the market offers and disequilibrium. Empirical cumulative distribution functions of optimal profits are reported. A few classical papers are reviewed with more details in order to trace the origin and foundation of modern finance.
\end{abstract}

\tableofcontents

\section{Introduction}

In financial economy theories \textit{equilibrium} plays an important role. William Sharpe defines: \textit{"...a financial economy is in equilibrium when no further trades can be made"} \cite[p. 9]{sharpe2007}. He adds: \textit{"But of course in the real world trading seldom stops, and when it does stop, it is typically because low-cost markets are temporary closed. The implication is that financial markets never really reach a state of equilibrium. ... In actuality, people make trades to move toward an equilibrium target but the target is constantly changing. Despite this completely valid observation, we need to understand the properties of a condition of equilibrium in financial markets, because markets will usually be headed toward such a position."} Solving the task, Sharpe deviates from the mean/variance approach associated with the Modern Portfolio Theory of Harry Markowitz \cite{markowitz1952} and his own Capital Asset Pricing Model \cite{sharpe1964} and applies the state/preference method originated in the works of Kenneth Arrow \cite{arrow1951} and Gerard Debreu \cite{debreu1951}.

In contrast to the task \textit{"to understand the properties of a condition of equilibrium"}, this article concentrates on the "complement", the non-equilibrium market state, considering it as an essential condition of the speculative market existence. If the non-equilibrium state is "people making trades", then it can be expressed in terms of trading. How?

The market equilibrium associates with the perfect distribution of information about the \textit{"equilibrium target"} between the market participants, although Sharpe suggests that the \textit{"target is constantly changing"}. This view on the equilibrium leads to another deep notion - the \textit{efficient market hypothesis}, EMH, developed by Eugene Fama \cite{fama1965}, \cite{fama1970}, \cite{fama1971}, \cite{fama1992}. The market is efficient but speculators continue trading. Why?

Many speculators hardly know about moving \textit{"toward an equilibrium"}. Their striving for making money is so strong that the assortment of means supporting their decisions stretches from science to astrology. The market must have an objective property "explaining" such an aspiration. Ideally, it should be measurable. There must exist something. What?

Twenty years ago the author had to stop the research in analytical and computational chemistry, molecular dynamics and Monte-Carlo simulation of liquid phases and plunged into the world of markets, trading futures, stochastic processes, and models for pricing derivative instruments. He has found markets not less challenging. Coming from a society, where speculation of American jeans could result in a jail term and exchange of substantial amounts of rubles to a foreign currency in a death penalty, the \textit{Rokotov-Faibishenko case}, the author was pleasantly astonished  at the "lower FOREX transaction costs". "Things Have Changed". The English "f-u-t-u-r-e-s" is a frequent word in Russian TV News. This article summarizes the journey and presents the author's answers, prompted by the market, on the questions: How, Why, and What.
 
The speculative markets, \textit{"the front lines of capitalism"} \cite[p. 7]{najarian2001}, are the areas of \textit{"cooperation between human consciousness and technology governed by partly unknown laws of nature"} \cite[p. 34]{salov2011}. Ignoring that electronic markets involve people and programs created by people and programs "bred" by programs \cite{koza1992} created by people is a costly trading experiment. Daniel Kahneman and Amos Tversky \cite{kahneman1979} \cite{tversky1992} show us that a person knowing about mathematical expectations still may act against this knowledge, demonstrating the \textit{"risk aversion"} and \textit{"reflection effect"}. Comparing their research with notes written by Jesse Livermore, an outstanding speculator \cite[pp. 11 - 13]{livermore1940}, the author has made a "small discovery". Livermore describes a \$2 profit per share taken without a risk in a fear to lose it. He also illustrates holding a position already losing \$2 per share in a hope of a price reversion. This is a foresight of the risk aversion and reflection effect. An intuition built during 40 years by large-scale speculation accompanied by making and losing fortunes caught an idea of a remarkable scientific achievement 39 years earlier. Human beings tend to set targets "unexpected mathematically". Can this make equilibrium an exception but non-equilibrium properties a norm? A measure is needed.

George Soros, a large market practitioner and philosopher, designs the \textit{Theory of Reflexivity} \cite{soros2009}. It emphasizes essential interference of the \textit{cognitive function} aiming to bring the knowledge about our world and the  \textit{participating} or \textit{manipulative} or \textit{executive function} changing it. He demonstrates that markets are full of \textit{reflexive situations} and sees this as one of the reasons, why they deviate from equilibrium. This qualitative paradigm could also benefit employing  a measure of the frequency and magnitude of the reflexive situations.

The electronic markets, \textit{"ultra-high-frequency"} coined by Robert Engle \cite{engle2000}, unknown proportion of computer programs making trading decisions and hidden principals embedded into them complicate matters but do not give signs that speculation will cease to exist. It seems that the means distributing information about the "targets" faster also change the latter more frequently.

\section{Inalienable Property of a Speculative Market}

\begin{figure}[h!]
  \centering
  \includegraphics[width=130mm]{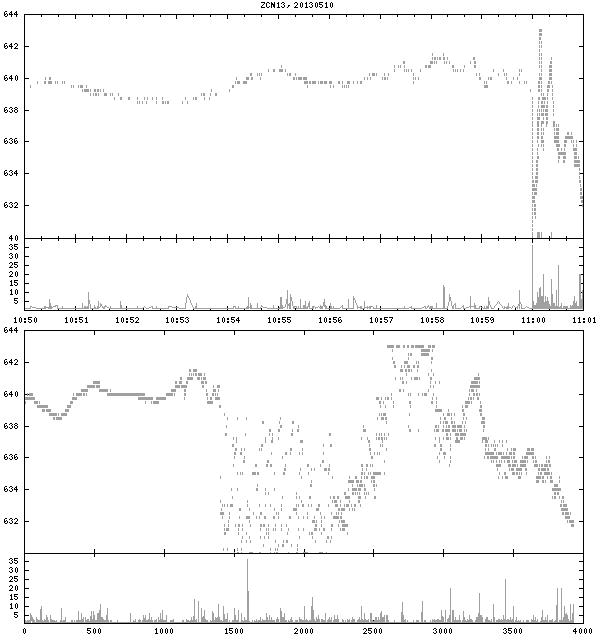}
  \caption[ZCN13 Explosion]
  {Friday May 10, 2013/Globex. "Explosion" of Corn July 2013 contract ZCN13 on news at 11:00:00 AM CT. Price and volume vs. time (top) and transaction index set to 0 at 10:50:00 AM CT (bottom).}
  \label{explosion}
\end{figure}

The market property \textit{to provide frequent opportunities to make large profits} is discussed as an \textit{essential condition of its existence} and the \textit{main law of the speculative market} in the section "Why we speculate" of \cite{salov2011b} and \cite[p. 31]{salov2012}. If the term law will find a use, then it should be recognized as \textit{phenomenological}: it is not deductively derived from other postulates, it is confirmed only by market data, it does not explain why the market posses such properties. A market is people and programs trading something. This activity is reflected by transactions occurred in time, at a certain price and number of contracts or shares. A market can "sleep" but not long. Sometimes it "explodes", Figure \ref{explosion}. Studying millions of transactions, the author did not find exceptions, contracts with horizontal lines of price vs. time, and started thinking towards the terms "regularity" or "law" instead of "hypothesis" in order to denote the phenomenon.

Such a regularity is in opposition to approaches intentionally or unintentionally dismissing these properties. Accounting the large number of market data supporting the regularity, the opposite approaches should be reviewed for correctness.

Even, being formulated as a law, this property does not provide a way to extract potential profits but supplies the \textit{maximum profit strategy framework}, as we shall see, to search for such a way.

The qualitative formulation, \textit{"markets provide frequent opportunities to make large profits"}, is intuitively known to traders. Naming it the law intents to increase the confidence: these properties pass away only with a market. The law becomes quantitative after adding the measure - the maximum profit strategy, MPS. It answers on "how frequently" and "how large". For futures, a MPS can be constructed using the l- (left) and r-(right)algorithms \cite{salov2007}, given chains of prices and transaction costs, margins, and accounting rules. This law is expressed not by an equation but algorithm. The latter circumstance is interesting with regard to the three modern definitions of randomness, each also based on the \textit{theory of algorithms} \cite{kolmogorov1987}.

The market offers profit opportunities in a view of the \textit{optimal trading elements}, OTE. This term is coined in \cite{salov2011b} and \cite{salov2012}.

The law, based on MPS - an objective market property, relates to a human being activity. It has no status of physical laws: socialist revolutions or meteorites can delete the free markets from the face of the Earth and terminate trading, while the Newton's laws will continue working.

\section{A Comment on Equilibrium}

A ping-pong ball hanging from a vertical thread is in equilibrium. The same ball balancing on a top of a vertical needle is in equilibrium too. The first equilibrium is \textit{stable}. After displacement, the sum of gravitation, reaction (caused by tension), and friction forces decreases to zero and returns the ball to the initial state. The second equilibrium is \textit{unstable}. After displacement, the sum of the same three types of forces increases up to a constant taking the ball away of the initial position. The same forces play an opposite role. Studying the two equilibrium states reveals the zero sum of the two vectors: the gravitation and reaction forces. If a reader feels uncomfortable comparing the thread tension with the needle reaction, then the author suggests to replace the first example with a tall wine glass and a pea on its bottom. To "understand the properties of a condition of equilibrium" is important, however, being identical in both cases they do not explains a very different behavior of the ball after displacement. The change and distribution of forces after displacement is crucial.

The word \textit{change} was popular during the fall of 2008 in U.S. election TV News. The changes but not only absolute values are important for anticipation of a behavior involving human beings. In his Nobel Prize lecture \cite[p. 460]{kahneman2002} Daniel Kahneman presents an example, where a person holds the left and right hands in two different buckets with cold and hot water and later places them in a bucket with tepid water. The judgment about the \textit{same} temperature depends on the hand.

Two husbands talk about the third one. "His wife made him a millionaire." "A lucky man!" "No, before he was a billionaire." A trader left with \$100,000 judges differently about the trade after \$110,000 - \$10,000 loss or \$90,000 + \$10,000 gain, although the final state, money, is the same.

These examples demonstrate unstable and stable equilibrium and importance of studying transitions together with the states. A market consisting of the stable equilibrium states only would not guarantee simplicity.

\begin{figure}[h!]
  \centering
  \includegraphics[width=130mm]{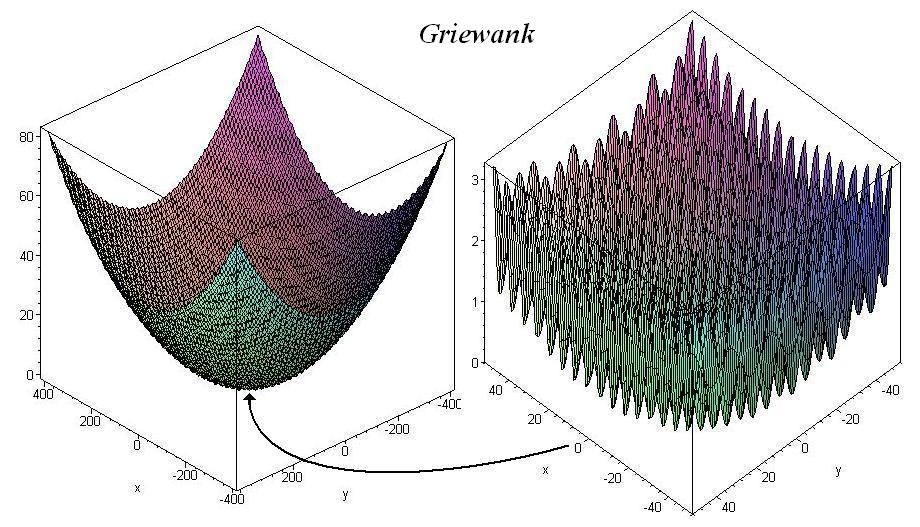}
  \caption[Griewank Function]
   {The Griewank function of two variables. A multidimensional version of this function with a single global minimum and myriads of local ones is a challenge for many solvers.}
  \label{griewank}
\end{figure}

A steepest-descent solver \cite[pp. 21, 22]{nocedal1999} starting from different points stably finds the minimum of the elliptic paraboloid $z = x^2 + y^2$. The mechanism resembles a pea rolling down to the bottom of a tall wine glass. However, the Griewank's surface $z = \frac{x^2 + y^2}{200} + 1 - \cos(x)\cos(\frac{y}{\sqrt{2}})$ \cite{griewank1981} traps the pea in one of the local minimums. The Mars Rover Curiosity's mission would end after landing on the surface on Figure \ref{griewank}. A pea would need \textit{activation energy} to overcome barriers, explore the surface, and reach the lowest point. The local minimums in smooth and convex areas mimic stable equilibrium states. Participants, trading on a market, consisting of such states only, would need enough activation energy in order to reach the most optimal equilibrium state, and analysts would need to find its source and mechanism. Let us notice that a \textit{differential evolution solver} \cite{storn1997} combining directional moves with random selection, a kind of "diffusion tunneling" through the barriers, routinely finds the right answer.

\section{A Comment on Surviving Systems}

Nikita Moiseev, a mathematician widely known due to computer simulation of the "nuclear winter"  - a consequence of a global nuclear war, applies the term \textit{homeostasis} \cite[p. 140]{moiseev1982}. It is a region of critical values of parameters that a system must not exceed in order not to be destroyed. A surviving system aims at the center of the region far from the dangerous boundary. This principal, combining monitoring of the distance from the boundary, a feedback, and adjustment of the parameters affecting the work of the system, leads to autopilots and missiles autonomously navigating to targets. A trading system enriched by money management rules and adopting its behavior to varying market and portfolio conditions would not be an exception.

At first glance, this principal contradicts to heroic life sacrifices on a battle field, when a solder closes an embrasure, dies, but his comrades using the moment win the battle. Surviving is ignored. A \textit{hierarchy} seems save the principal. A higher level system, ideologically cultivating (in a good, patriotic sense) people, follows to the goal of surviving sacrificing its subsystems. The principal on the higher level overrides its violation on the lower one.

In the \textit{Immortal Game} of chess played on June 21, 1851 in London, Adolf Anderssen sequentially sacrificed to Lionel Kieseritsky pawn (by definition in King's Gambit), Bishop, second pawn, Rook, second Rook, Queen, and won on 23d move by Bd6-e7$\times$ (checkmate) \cite[p. 291 - 293, game 227]{tartakower1975}. In chess, sacrifices impress. In life, they are tragedies suddenly supporting the principal.

A speculative market proposes prices, redistributes risk, intensifies cash flows. It does not create a treasure but redistributes it paying a part to the brokers, analysts, and information technologists. For traders, risking their and others' capital, this is not a zero-sum game. In order to win, somebody must lose. Alexander Elder comments \cite[p. 49]{elder1993}: \textit{"Trading means trying to rob other people while they are trying to rob you. It is a hard business."} His provoking definition is \cite[p. 44]{elder1993}: \textit{"Price is what the greater fool is ready to pay."} The market needs frequent transactions offering attractive potential profits. Interest drives speculation more than a fear of losses. However, the latter are the only source of the real profits. The hierarchical principal seems justify partial financial disasters. What is death for a few hedge funds is life for the market, coming out as a steak house with exotic dish names such as \textit{Long-Term Capital Management}, \textit{Tiger Management Corp.}, \textit{Basis Yield Alpha Fund}, \textit{Sowood Capital}, \textit{Lehman Brothers Holdings Inc.}, where the Nobel Prize does not guarantee a seat. The seat can be on a frying-pan. This creates enormous stress.

\section{A Comment on Attractors and Fractals}

\begin{figure}[h!]
  \centering
  \includegraphics[width=130mm]{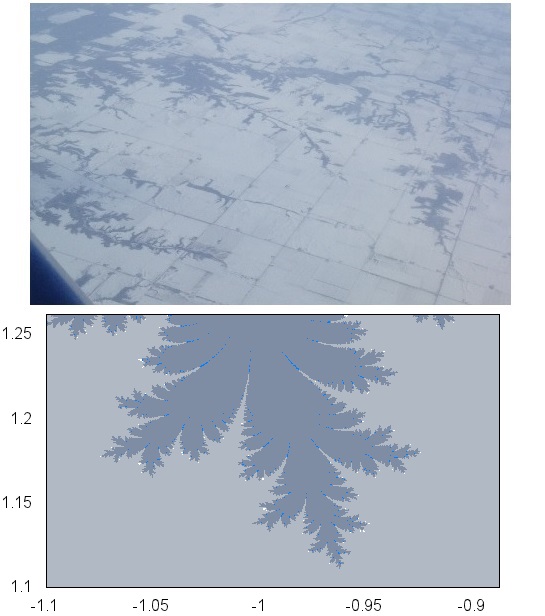}
  \caption[Fractals]
   {Top: photo taken from an illuminator by the author flying from Albuquerque, New Mexico to Chicago, Illinois. Bottom: a filled-in Julia set obtained using iterals of cosine and the C++ program from \cite[pp. 14 - 15]{salov2012c}, and Gnuplot \cite{gnuplot}. More about iterals read in \cite{salov2012b} and \cite{salov2012c}.}
  \label{fractals}
\end{figure}

\begin{figure}[h!]
  \centering
  \includegraphics[width=130mm]{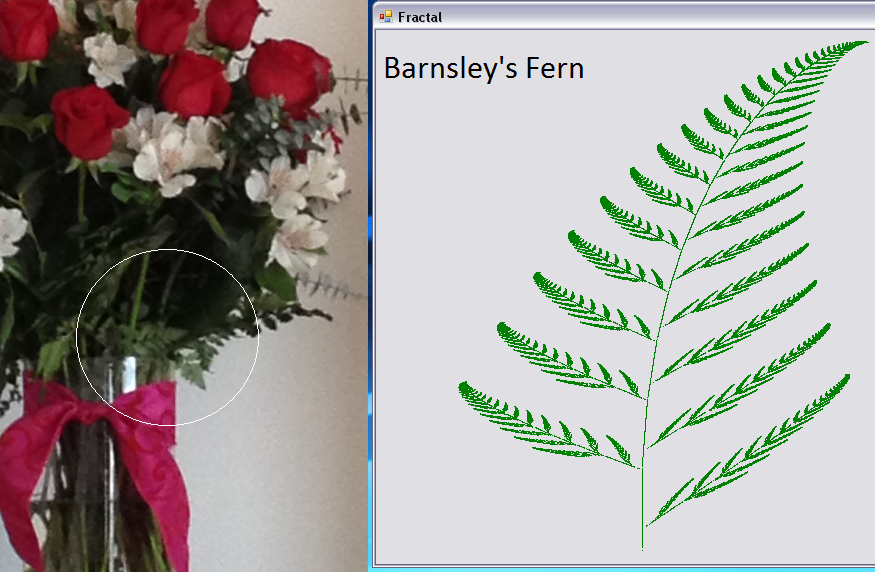}
  \caption[Fractals]
   {Left: photo of a bouquet of roses taken by the author with leather leaf, \textit{rumohra adiantiformis}, inside the circle. Right: the author's Microsoft Visual Basic program building Barnsley's Fern based on John Hutchinson's \textit{iterated fractal systems} \cite{hutchinson1981}.}
  \label{fern}
\end{figure}

If we place the stable and unstable equilibrium of the ping-pong ball on the left side of a complexity axis and hypothetical equilibrium and non-equilibrium market states and transitions on the right side labeled by greater values, then the middle is ready for: the origin of the theory of chaos contained in celestial mechanics of Henri Poincare \cite{poincare1971} \cite[pp. 4 - 7]{arnold2006}, the theory of stability from Alexandr Lyapunov \cite{lyapunov1948}, the theory of rings of operators from John von Neumann \cite{neumann1932} \cite{neumann1949} \cite[pp. 69 - 70]{dyson2010} \cite[pp. 2 - 3]{sinai2010}, the 1940th works of Andrey Kolmogorov on fluid dynamics \cite[pp. 29 - 33]{arnold2004} and his "first seeds in chaos theory" \cite{kolmogorov1954} \cite{kolmogorov1958} \cite[p. 2]{sinai2010}, the entropy of dynamical systems from Yakov Sinai \cite{sinai1958} \cite{sinai2010}, the \textit{Arnold diffusion} \cite[pp. 67 - 70]{arnold2008} \cite[p. 5]{sinai2010} having its roots in the work of Alexandr Andronov, Lev Pontryagin, and Alexandr Vitt in 1930th \cite[pp. 68]{arnold2008}, the contributions of Vladimir Arnold into unstable dynamical systems \cite{arnold1963}, the theorem of Olexander (Alexander) Sharkovsky \cite{sharkovsky1964} \cite{misiurewicz1997}, the term chaos coined by Yorke \cite{li1975} \cite{may1974}, the chain reactions from Nikolay Semenov \cite{semenov1956} (coming to mind after examining Figure \ref{explosion}), the non-equilibrium processes from Ilya Prigogine \cite{prigogine1977} \cite{prigogine1993}, the discovery of a \textit{strange attractor} by Edvard Lorenz \cite{lorenz1963},  the discovery of universality of period-doubling bifurcations by Mitchell Feigenbaum \cite{feigenbaum1978}, the world of fractals presented by Benoit Mandelbrot \cite[p. 402]{mandelbrot1975} \cite{mandelbrot2004} and surrounding us Figures \ref{fractals}, \ref{fern}, the smooth ergodic theory from Yakov Pesin \cite{pesin1977} providing strong links to the \textit{differential dynamical systems} from Stephen Smale and \textit{Anosov systems} from Dmitri Anosov, varieties of billiards \cite{vorobec1992}, \cite{tumanov2006}, the linear search problem \cite{baryshnikov2011}. The author was impressed how Grigory Galperin forces the billiard trajectories to count 50,000,000 of the decimal digits of $\pi$ \cite{galperin2001} and to measure distances in the hyperbolic Lobachevsky's space \cite{galperin2004}. The human being consciousness shifts the markets to the right but inheritance from the middle is likely. Increasing complexity does not seem throw out but accumulates simple rules of the lower levels \cite{goldenfeld1999}, \cite{wolfram2002}. It creates new regularities and laws on a higher level, otherwise, as it is wittily noticed by Nigel Goldenfeld and Leo Kadanoff: \textit{"In order to model a bulldozer, we would need to be careful to model its constituent quarks!"} The author recognizes that "there are a number of glaring omissions in this" list \cite[p. 5]{sinai2010}.

\begin{figure}[h!]
  \centering
  \includegraphics[width=130mm]{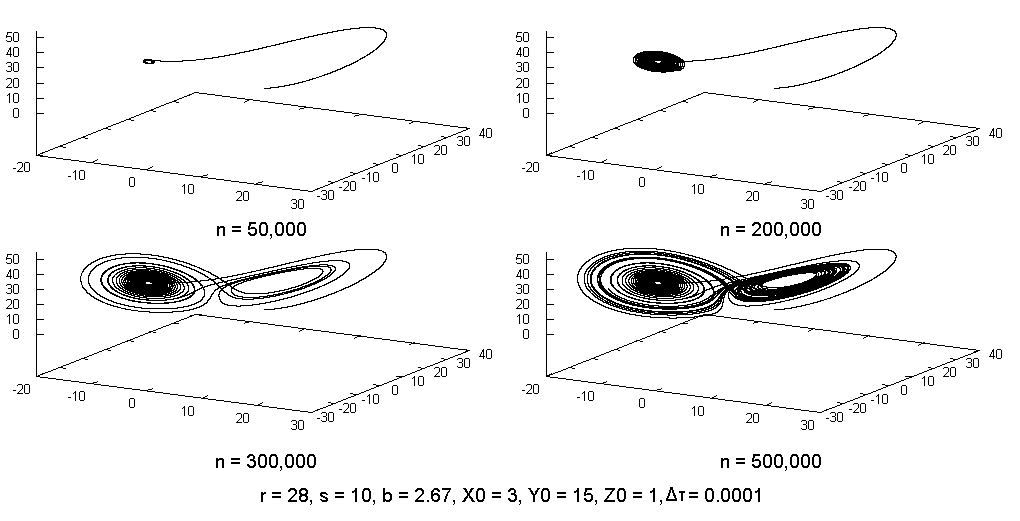}
  \includegraphics[width=130mm]{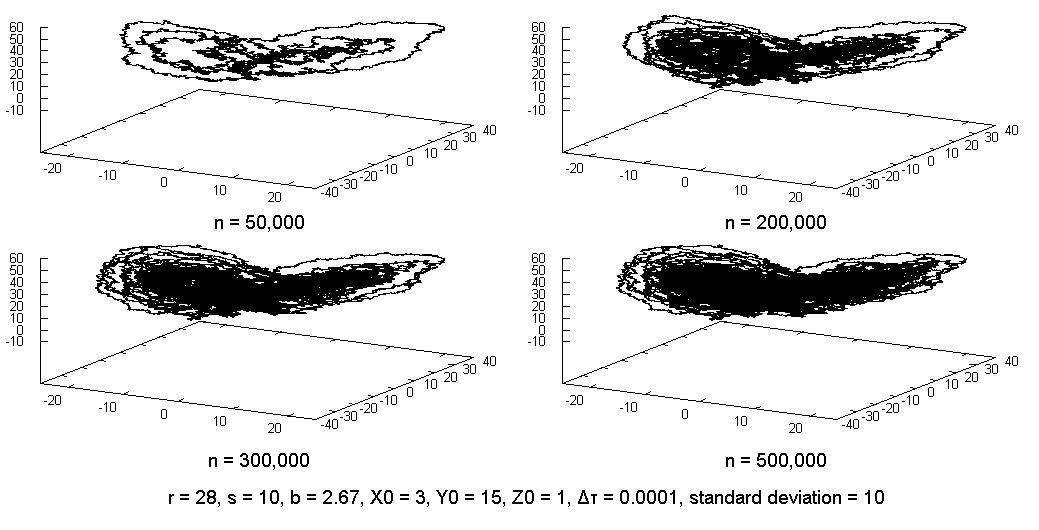}
  \caption[Fractals]
   {Top: four snapshots of a phase trajectory of a Lorenz' system. Evolution reveals two attractors. Bottom: adding a Gaussian noise helps to fill faster the phase space at the areas of the two attractors}
  \label{lorenz}
\end{figure}

Let us check the influence of "diffusion of the phase coordinates" on a phase trajectory in a system with attractors. After modification of the model of Barry Saltzman \cite{saltzman1962}, Lorenz \cite{lorenz1963} has come to the system of three ordinary differential equations named after him today
\begin{displaymath}
\frac{dX}{d\tau} = -\sigma X + \sigma Y, \; \frac{dY}{d\tau} = -XZ + rX - Y, \; \frac{dZ}{d\tau} = XY - bZ,
\end{displaymath}
where $\sigma, r,$ and $b$ are positive constants, and $\tau$ is a dimensionless time. The solution is a phase trajectory $(X(\tau), Y(\tau), Z(\tau))$, where $X, Y,$ and $Z$ define a 3D phase space of a layer of fluid of uniform depth between two surfaces maintained at two different temperatures. Depending on the conditions and \textit{Rayleigh number} the liquid remains steady or gets in motion, \textit{convection}. $X$ is proportional to the intensity of the convection, $Y$ is proportional to the temperature difference between the ascending and descending currents, and $Z$ is proportional to distortion of the vertical temperature profile from linearity. These $X, Y, Z$ should not be mixed with a configuration space of liquid parts. The author has determined the iteral expression \cite[p. 10]{salov2012b} for the Lorenz's \textit{double-approximation procedure} \cite[pp. 133 - 134]{lorenz1963} and written a C++ program outputting phase coordinates triplets to the graphics package Gnuplot \cite{gnuplot}. Figure \ref{lorenz} applies $s$ instead of $\sigma$.

The Lorenz's phase trajectory or orbit has no \textit{fixed points}. It is non-periodic and does not intersect itself. \textit{Its future does not repeat its past}. The bottom picture is obtained after adding to $X, Y, Z$ on each iteration the term $number \times standard \; deviation \times \sqrt{\Delta \tau}$. The \textit{number} is drawn from the standard (mean 0, variance 1) Gaussian random generator reusing the \textit{Box-Muller algorithm} \cite{box1958}. Presence of diffusion speeds up filling of the phase space. Now, during the same 50,000 time steps the system has time to explore both attractors. This behavior, indeed, "correlates" with the qualitative picture predicted by Arnold in \cite[p. 68]{arnold2008}, where he introduces the terms \textit{attractor force}, \textit{attractor pull}, and \textit{tunneling effect}. We should also remember the robust differential evolution solver from a previous section, where random selection is crucial for finding the minimum of the Griewank function. This experiment shows that randomness from a continuous distribution theoretically can place an originally non-periodic chaotic system into a state, which it already had. From such a state and for a while the future trajectory could repeat the past one, especially, if randomness is suppressed after the impact.

\section{The Data}

The Chicago Mercantile Exchange, CME, Group publishes daily, after a market is closed, on the homepage \url{http://www.cmegroup.com/} the Time \& Sales tables for futures listed on CME, the Chicago Board of Trade, CBT, the New York Mercantile Exchange, NYMEX, and some other traded \textit{open outcry} (\textit{pit}) and electronically on Globex. A table is a chain of records containing date, time, price, indicator for pit and date, time, price, size, indicator for electronic sessions. The date is the same for all records belonging to one session, although the latter can last in two calendar dates. The time is accurate to one second. The price quotes follow to the contract specifications and may require conversion to decimal numbers. The size, available only for electronic sessions, is a positive number of traded contracts for transactions and zero for other types of prices. The indicator is a price type such as '-' used for transactions, indicative, open, close, ask, bid, settlement and other. Pit sessions omit size and majority of transactions occurred at the same price in a row. This creates \textit{snake tongue} histograms of the price increments with almost "missed" zeros, the center.

A contract ticker consists of the product name, expiration month and year. The months are F - January, G - February, H - March, J - April, K - May, M - June, N - July, Q - August, U - September, V - October, X - November, and Z - December. For the same commodity, product names may differ for the Globex and pit like ZC and C for corn contracts. Some brokerage companies list such pairs under one ticker. Durations of electronic sessions typically overlap the pit and include nights. This supports people preferring trading to sleeping and not important for trading computer robots. This report is mainly based on Globex transactions, Table \ref{table_futures_properties}. Consult with brokers and review for details contracts specifications available on the CME Group homepage.

\begin{table}[!h]
  \topcaption{Some Properties of Futures Traded on Globex/Open Outcry}
  \begin{tabular}{ccccc}
  Symbol & Commodity & Exchange & Months & Tick $\delta$=\$ \\
  ZB/US & U.S. Treasury Bond & CBT & HMUZ & 0.03125=\$31.25 \\
  ZC/C & Corn & CBT & HKNUZ & 0.25=\$12.50 \\
  ZS/S & Soybean & CBT & FHKNQUX & 0.25=\$12.50 \\
  ZW/W & Wheat & CBT & HKNUZ & 0.25=\$12.50 \\
  6A/AD & AUD/USD & CME & HMUZ & 0.0001=\$10.00 \\
  6B/BP & GBP/USD & CME & HMUZ & 0.0001=\$6.25 \\
  6C/CD & CAD/USD & CME & HMUZ & 0.0001=\$10.00 \\
  6E/EC & EUR/USD & CME & HMUZ & 0.0001=\$12.50 \\
  6J/JY & JPY/USD & CME & HMUZ & 0.0001=\$12.50 \\
  ES & E-mini S\&P 500 & CME & HMUZ & 0.25=\$12.50 \\
  GE/ED & Eurodollar & CME & All & 0.0025=\$6.25 \\
  LE/LC & Live Cattle & CME & GJMQVZ & 0.025=\$10.00 \\
  HE/LH & Lean Hogs & CME & GJKMQVZ & 0.025=\$10.00 \\
  CL & Light Sweet Crude Oil & NYMEX & All & 0.01=\$10.00 \\
  GC & Gold & COMEX & All & 0.1=\$10.00 \\
  HG & Copper & COMEX & All & 0.0005=\$12.50 \\
  NG & Natural Gas & NYMEX & All & 0.001=\$10.00 \\
  SI & Silver & COMEX & FHKNUZ & 0.005=\$25.00 \\
  \end{tabular}
  \label{table_futures_properties}
\end{table}

Due to the high frequency of transactions and limited accuracy of time, records may get one time, where only arriving order defines the chain. The lack of accuracy creates vertical zig-zags in plots of prices vs. time. Plotting the price against the index of a record assists but separates ticks by artificial constant \textit{waiting times}. On Figure 1, one second at 11:00:00 AM CT is responsible for 806 transactions with accumulated volume of 1166 contracts and prices from the range [630.00, 638.75]. The $\delta_{ZC}$, Table 1, gives the dollar equivalent of the range $ \$12.50 \times (638.75 - 630.00) / 0.25 = \$437.50$ before fees and commissions, which could be \$10.66 per contract per \textit{round trip}.

While the words of Sir Maurice Kendall \textit{"the golden rule in publishing work on time-series is to give the original data"} remain imperative, it is difficult to follow them in a paper on high frequency data 60 years later \cite{kendall1953}. DVDs can accompany a book but not an article. Only the 57 sessions of ESM13 collected between March 1 and May 22, 2013 (Wednesday March 27, 2013 is missed) contain 21,364,635 records totaling 835,401,602 bytes in files like ESM13\_20130522.txt:
\begin{verbatim}
...
2013-05-22 08:33:21.000-06 1669.25 6 T
2013-05-22 08:33:22.000-06 1669.25 1 T
2013-05-22 08:33:22.000-06 1669.25 1 T
2013-05-22 08:33:23.000-06 1669 34 T
...
\end{verbatim}
This is 374,818 records per session. An ordinary ESM13 session lasts from 17:00:00 of one day until 15:15:00 of a next day, then 15 minutes pause, trading from 15:30:00 until 16:15:00, and the second pause until 17:00:00. With 23 hours, the mean number of records per second is $374,818/82,800 = 4.5$. Transactions are distributed non-uniformly in time (compare the mean 4.5 with 806), Figure 1.  There is some intraday seasonality of arrivals. The "time" will often mean the "date and time", where reserved milliseconds and time zone are marked as ".000-06" together with the 24 hours "HH:MM:SS" format.

\section{The a-b-c-Process}

The ultra-high-frequency futures transactions is a rich source of time-series. The Globex transactions are \textit{triplets} of time, price, and volume. Such a triplet is called tick. This should not be mixed with the minimal absolute non-zero price fluctuation $\delta$ in Table 1. For classification, the author introduces \textit{a-b-c-increments}, \textit{a-b-c-properties}, and \textit{a-b-c-process}. 

\subsection{Basic formulas}

Each contract starts and expires at some date and time. No futures are traded 24/7. Time-series can be divided into $n$ sessions indexed by $s = 1, \; \dots, \; n$. The $s$ and a \textit{session date} are in one-to-one correspondence. A $s$th session contains $N_s$ transactions. The $N_s$ varies between sessions and contracts. Additional contract and date labels $N_s^{ESM13}$, $N_{20130522}^{ZCN13}$ can be helpful.

A $s$th session may enclose several time ranges indexed by $r = 1, \dots, m_s$. During the contract life, their number, opening and closing times can change. The Globex ZCN13 was trading in one range 17:00:00 (previous date) - 14:00:00 (session date) CT before April 8, 2013 and in two ranges since April 8, 2013: 19:00:00 - 07:45:00 and 08:30:00 - 13:15:00 CT with the 45 minutes pause. The after Fourth of July 2013 holiday session, on July 5, 2013, had only one range 08:30:00 - 13:15:00 CT.  If $m_s = 1$, then the range coincides with the enclosing session. The $N_{s,r}$ is the number of transactions within the $r$th range of the $s$th session and $N$ is the total per contract life number of transactions

\begin{equation}
\label{EqNs}
N_s = \sum_{r=1}^{m_s} N_{s,r}, \; N = \sum_{s=1}^{n} N_{s} = \sum_{s=1}^{n} \sum_{r=1}^{m_s} N_{s,r}.
\end{equation}

The triplets $(t_i, P_i, V_i)$ are indexed by the intra-range $i = 1, \; \dots, \; N_{s,r}$. The $N_{s,r}$ and $N_s$ can be equal to zero. Contracts can be illiquid at the life beginning or expiration. The less liquid \textit{deferred} months and years or expiring contracts coexist with the more liquid \textit{nearby} or getting to become nearby contracts on the same commodity, supporting the Main Law. Futures can be delisted like the Frozen Pork Bellies on Monday July 18, 2011. The latter did not meet hedging needs and balance between hedgers and speculators. Contracts, which cannot frequently propose potential profits, die, confirming the Main Law - essential condition of the market existence. We shall deal mainly with cases $N_{s,r} > 1$.

The \textit{a-} and \textit{b-increments} are the differences between neighboring transaction times and prices within the $(r = 1, \dots, m_s)$-range of the $(s = 1, \dots, n)$-session
\begin{equation}
\label{EqAIncr}
\mathrm{a}\mathrm{-}\mathrm{increment:} \; \Delta t_i^{s,r} = t_i^{s,r} - t_{i-1}^{s,r}, \; i = 2, \dots, N_{s,r},
\end{equation}
\begin{equation}
\label{EqBIncr}
\mathrm{b}\mathrm{-}\mathrm{increment:} \; \Delta P_i^{s,r} = P_i^{s,r} - P_{i-1}^{s,r}, \; i = 2, \dots, N_{s,r}.
\end{equation}
They are undefined for $N_{s,r} < 2$. Summing up the increments, the undefined quantities can be skipped. Replacing them by zeros would give the same sum but affect the sample statistics due to the increasing number of summands. A discretional approach may be needed for each indeterminate case.

The \textit{c-increments} are the differences between the first price of a $s$th session and the last price of a $(s - 1)$th session irrespectively on ranges
\begin{equation}
\label{EqCIncr}
\mathrm{c}\mathrm{-}\mathrm{increment:} \; \Delta P^{s} = P_1^{s} - P_{N_{s-1}}^{s-1}.
\end{equation} 
They are undefined for $n < 2$, $N_{s} = 0$, or $N_{s-1} = 0$. The c-increments over the pauses between regular sessions, weekends, and holidays are the \textit{cr-increments}, \textit{cw-increments} and \textit{ch-increments}. The \textit{ci-increments} (\textit{internal}) are the differences between the first price of a $r$th range and last price of a $(r-1)$th range
\begin{equation}
\label{EqCIIncr}
\mathrm{ci}\mathrm{-}\mathrm{increment:} \; \Delta P^{s,r} = P_1^{s,r} - P_{N_{s,r -1}}^{s,r - 1}.
\end{equation}
They are undefined for $m_s < 2$, $N_{s,r} = 0$, or $N_{s,r-1} = 0$. If $m_s = 1$ for all $s$, then the second index $r$ can be dropped from Equations \ref{EqNs} - \ref{EqBIncr} and some other below. This is true for several futures. Thus, there is a family of c-increments.

The price b-increments and the family of c-increments are defined over different waiting times. The former associate with the a-increments. The latter correspond to the two \textit{a-like} increments plus a known in advance, usually much longer than a-increments, time separating sessions and/or ranges.

The increments can be counted forward affecting initial and final index values
\begin{equation}
\label{EqAIncrForward}
\mathrm{a}\mathrm{-}\mathrm{increment:} \; \Delta t_{i+1}^{s,r} = t_{i+1}^{s,r} - t_{i}^{s,r}, \; i = 1, \dots, N_{s,r} - 1,
\end{equation}
\begin{equation}
\label{EqBIncrForward}
\mathrm{b}\mathrm{-}\mathrm{increment:} \; \Delta P_{i+1}^{s,r} = P_{i+1}^{s,r} - P_{i}^{s,r}, \; i = 1, \dots, N_{s,r} - 1,
\end{equation}
\begin{equation}
\label{EqCIncrForward}
\mathrm{c}\mathrm{-}\mathrm{increment:} \; \Delta P^{s+1} = P_1^{s+1} - P_{N_{s}}^{s}, \; s = 1, \dots, n - 1.
\end{equation}
\begin{equation}
\label{EqCIIncrForward}
\mathrm{ci}\mathrm{-}\mathrm{increment:} \; \Delta P^{s,r+1} = P_1^{s,r+1} - P_{N_{s,r}}^{s,r}, \; r = 1, \dots, m_s - 1.
\end{equation}
Forwarding does not change $n, m_s, N_s, N_{s,r}$, determining whether the quantities are defined. The backward and forward b-increments have and c- have no lower index at $\Delta P$. The c- and ci-increments have one and two upper indexes. 

The time interval corresponding to a c-increment consists of a known interval between the current session opening $T_o^s$ and previous closing $T_c^{s-1}$ times plus two durations: between the current first tick time and $T_o^s$ and between $T_c^{s-1}$ and the previous last tick time, where $s=2, \; \dots, \; n$
\begin{equation}
\label{EqCTimeIncr}
\Delta t^s = (T_o^s - T_c^{s-1}) + (t_1^{s} - T_o^s) +(T_c^{s-1} - t_{N_{s-1}}^{s-1}) = t_1^{s} - t_{N_{s-1}}^{s-1}.
\end{equation}
This can be written as the forward increment, where $s = 1, \; \dots, \; n - 1$
\begin{equation}
\label{EqCTimeIncrForward}
\Delta t^{s+1} = (T_o^{s+1} - T_c^s) + (t_1^{s+1} - T_o^{s+1}) +(T_c^s - t_{N_s}^{s}) = t_1^{s+1} - t_{N_s}^{s}.
\end{equation}
The time intervals associated with the ci-increments for $r=2,\dots,m_s$ are
\begin{equation}
\label{EqCITimeIncr}
\Delta t^{s,r} = (T_o^{s,r} - T_c^{s,r-1}) + (t_1^{s,r} - T_o^{s,r}) +(T_c^{s,r-1} - t_{N_{s,r-1}}^{s,r-1}) = t_1^{s,r} - t_{N_{s,r-1}}^{s,r-1}.
\end{equation}
The corresponding forward time intervals for $r = 1, \dots, m_s - 1$ are
\begin{equation}
\label{EqCITimeIncrForward}
\Delta t^{s,r+1} = (T_o^{s,r+1} - T_c^{s,r}) + (t_1^{s,r+1} - T_o^{s,r+1}) +(T_c^{s,r} - t_{N_{s,r}}^{s,r}) = t_1^{s,r+1} - t_{N_s,r}^{s,r}.
\end{equation}
No names are introduced for the sums in Equations \ref{EqCTimeIncr} - \ref{EqCITimeIncrForward}. The \textit{a1-increments} are $t_1^{s} - T_o^s$ and $t_1^{s,r} - T_o^{s,r}$. If $t_1^{s} = t_1^{s,1}$ and $T_o^s = T_o^{s,1}$, then the a1-increment is included one time to a sample combining a1s from ranges and sessions. The \textit{a2-increments} are $T_c^{s-1} - t_{N_{s-1}}^{s-1}$ and $T_c^{s,r-1} - t_{N_{s,r-1}}^{s,r-1}$.  If $T_c^{s-1} = T_c^{s-1,m_{s-1}}$ and $t_{N_{s-1}}^{s-1} = t_{N_{s-1,m_{s-1}}}^{s-1,m_{s-1}}$, then the a2-increment is included one time to a sample combining a2s from ranges and sessions. The a1- and a2-increments are fractions of $T_o^s - T_c^{s-1}$ or $T_o^{s,r} - T_c^{s,r-1}$. If $N_s > 0, m_s > 1$ but some $N_{s,r} = 0$, then a1- and/or a2-increments get ranges durations. The two \textit{a-like} contributions remain intact. With 24/7 trading, the c-increments will become a history.

\begin{figure}[h!]
  \centering
  \includegraphics[width=130mm]{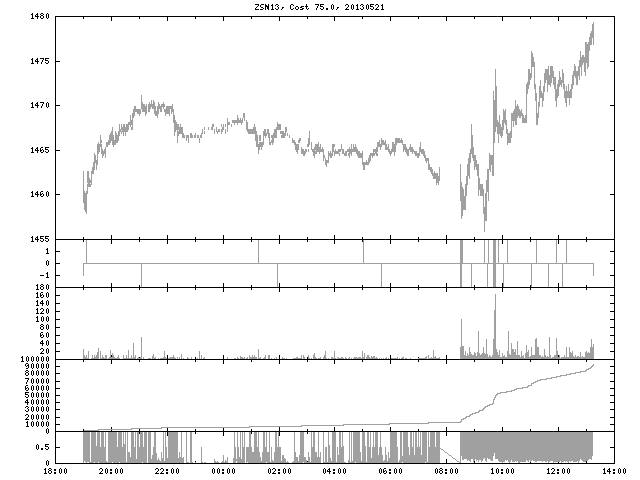}
  \caption[ABCPROC]
   {ZSN13 traded on Tuesday May 21, 2013. From top to bottom: price vs. time, $N_{20130521} = N_{20130521,1} + N_{20130521,2} = 9202 + 58381 = 67,583$ transaction ticks, the a-b-part of the a-b-c-process; MPS with filtering transaction cost \$75; volume ticks; accumulated volume; speed of transactions arrival.}
  \label{abcproc}
\end{figure}

In contrast with classifications of time and price increments based on constant observation intervals, the a-b-c-increments associate with transactions and share the property: they are \textit{indecomposable} - there are no ticks between neighbors. The a-b-c-increments measure the a-b-c-properties. The a-b-c-process evolves as following: a) the \textit{a-property} determines the \textit{time fluctuation} and, thus, the time of a next transaction within a range; b) the \textit{b-property} determines the \textit{price fluctuation} and the price of a next transaction within a range; and c) the \textit{c-property} is responsible for the price change between the \textit{current range/session last} and \textit{next range/session first} prices and concatenates two neighboring ranges/sessions by price.

Any time or price within the $r$th range of the $s$th session is the algebraic sum of the a- or b-increments added to the first time or price
\begin{equation}
\label{EqTimeSession}
t_i^{s,r} = t_1^{s,r} + \sum_{k=2}^{i} \Delta t_k^{s,r}, \; i = 1, \dots, N_{s,r},
\end{equation}
\begin{equation}
\label{EqPriceSession}
P_i^{s,r} = P_1^{s,r} + \sum_{k=2}^{i} \Delta P_k^{s,r}, \; i = 1, \dots, N_{s,r}.
\end{equation}
Conventionally, the sums $\sum$ vanish, when the iterating index is greater than the top value, for instance, if $i = 1$.

\begin{figure}[h!]
  \centering
  \includegraphics[width=130mm]{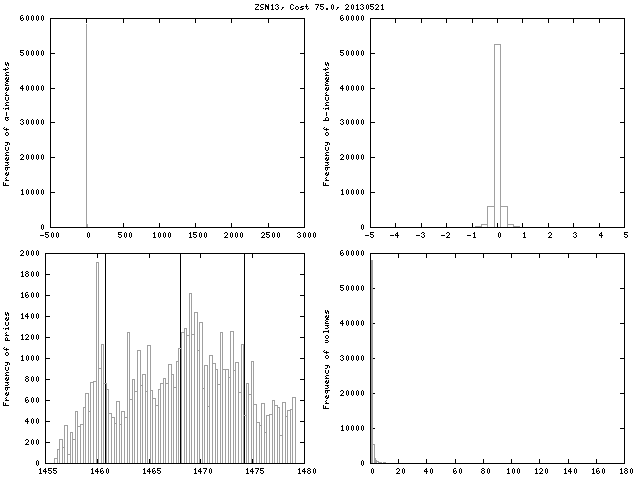}
  \caption[ABCINCR]
   {ZSN13 traded on Tuesday May 21, 2013. Frequency histograms. Top (left to right): a-increments (in seconds), minimal increment is zero; b-increments (almost symmetrical but not Gaussian). Bottom (left to right): prices forming multi-modal empirical distributions (vertical central line is mean and left and right lines mark 70\% of density range, "value area", equally, 15\%, stepped from both ends); volumes, minimal value is one.}
  \label{abcincr}
\end{figure}

The time and price counted from the first contract transaction are
\begin{equation}
\label{EqTimeGlobal}
\begin{split}
t_i^{s,r} = t_1^{1,1} + \sum_{j=1}^{s-1} \left (  \sum_{l=1}^{m_j} \sum_{k=2}^{N_{j,l}} \Delta t_k^{j,l}
 + \sum_{l=2}^{m_j} \Delta t^{j,l} + \Delta t^{j+1} \right ) \\
+ \sum_{l=1}^{r-1} \left ( \sum_{k=2}^{N_{s,l}} \Delta t_k^{s,l} + \Delta t^{s,l+1} \right ) +  \sum_{k=2}^{i} \Delta t_k^{s,r},
\end{split}
\end{equation}
\begin{equation}
\label{EqPriceGlobal}
\begin{split}
P_i^{s,r} = P_1^{1,1} + \sum_{j=1}^{s-1} \left (  \sum_{l=1}^{m_j} \sum_{k=2}^{N_{j,l}} \Delta P_k^{j,l}
 + \sum_{l=2}^{m_j} \Delta P^{j,l} + \Delta P^{j+1} \right ) \\
+ \sum_{l=1}^{r-1} \left ( \sum_{k=2}^{N_{s,l}} \Delta P_k^{s,l} + \Delta P^{s,l+1} \right ) +  \sum_{k=2}^{i} \Delta P_k^{s,r},
\end{split}
\end{equation}
where, $i = 1, \dots, N_{s,r}$, $s = 1, \dots, n$, $r = 1, \dots, m_s$.

Equations \ref{EqNs} - \ref{EqPriceGlobal} are exact, if each range has at least one tick. In Equation \ref{EqPriceGlobal} the first price is added by the a) b-, ci-, and c-increments from all ranges and sessions prior the last requested $s$th session, b) b- and ci-increments from all ranges prior the last requested $r$th range  from the last requested $s$th session, and c) remaining b-increments up to the $i$th one. The time Equation \ref{EqTimeGlobal} is similar. The a-b-c-process, a strategy, liquidity, histograms of a-b-c-increments, price and volume distributions can be depicted, Figures \ref{abcproc}, \ref{abcincr}.

\subsection{Two Shiryaev's tasks}

Describing financial ticks Albert Shiryaev \cite[p. 379]{shiryaev1998} formulates the \textit{two primary tasks} (author's translation from the Russian edition): \textit{(I) "What is the statistics of lengths of [VS: time] intervals between ticks; (II) What is the statistics of changes in prices [VS: between ticks] (in absolute ... or ... relative values)"}. The task (I) relates to the a-property and a-increments. The task (II) relates to the b-property and b-increments. Alfonso Dufour and Robert Engle \cite[p. 2467]{dufour2000} comment on the growing interest to such research: \textit{"The availability of large data sets on transaction data and powerful computational devices has generated a new wave of interest in market microstructure research and has opened new frontiers for the empirical investigation of its hypotheses"}, see the collection of articles \cite{lequeux1999}.

\section{The a-Increments}

The a-increments are nonidentical stones building trading time. Globex transactions are caused by matching book orders. The a-b-c-process depends on the arriving orders, the book state, and the matching algorithms. If an order matches two others of smaller sizes in a queue, then two transactions are triggered with a CPU time between them. CPU instructions take nanoseconds. If the book is waiting an order, then the minimal time is determined by the order transfer. Latencies add. The 806 transactions during one second at 11:00:00 on Figure \ref{explosion} being distributed uniformly would create the a-increment $1/(806+1) = 0.0012$ seconds. A non-uniformity shortens some. The sequential order processing implies non-zero a-increments. However, one second reporting inaccuracy creates an impression of discreteness and zeros.

\subsection{One second inaccuracy}

With the truncation [11:00:00, 11:00:01) $\rightarrow$ 11:00:00, [11:00:01, 11:00:02) $\rightarrow$ 11:00:01, for ticks with one time the a-increment $0.5 \pm 0.5$ seconds is set to zero. For 11:00:00, 11:00:01 it is set to one for $1 \pm 1$ seconds. For 11:00:00, 11:00:02 it is set to two for $2 \pm 1$ seconds. Ironically, 11:00.00.800, 11:00:01.100, 11:00.01.800 reported as 11:00:00, 11:00:01, 11:00:01 mismatch the a-increments 1 and 0 with differences 0.3 and 0.7 seconds. Rounding off time to a second has problems too: $11:00:00.499 \rightarrow 11:00:00$, $11:00:00.500 \rightarrow 11:00:01$, the a-increment 1 stays for 0.001 second. This "incorrectly reshapes" empirical distributions of a-increments at high liquidity and complicates their approximation by theoretical continuous distributions with zero probability density at zero. True zero a-increments require simultaneous transactions. Eventually, it can be implemented similar to parallel and multi-thread computer applications.

\subsection{Irregular waiting times}

Benoit Mandelbrot and Howard Taylor \cite[p. 1057]{mandelbrot1967} write: \textit{"... the number of transactions in any time period is random ..."}. Thus, durations between ticks are irregular. These authors, and Peter Clark \cite{clark1970} \cite{clark1973}, are among the first researchers, who have emphasized the importance of this fact to finance. The \textit{random time} comes out as a \textit{subordinator} of the \textit{random price}. The theory of \textit{subordinated processes} is developed by Salomon Bochner \cite{bochner1955}. High frequency trading presses on the tradition to collect prices at regular times. Charles Goodhart and Maureen O'Hara comment \cite[pp. 80 - 81 and p. 74]{goodhart1997}:

\textit{Traditional studies of financial market behavior have relied on price observations drawn at fixed time intervals. This sampling pattern was perhaps dictated by the general view that, whatever drove security prices and returns, it probably did not vary significantly over short time intervals. Several developments in finance have changed this perception. ... A fundamental property of high frequency data is that observations can occur at varying time intervals.}

Transactions occur at irregular times independently on "observations". The observations at regular times would miss many ticks. Table \ref{a-increments} summarizes sample statistics of a-increments. Samples from time ranges within a session are treated separately and marked by 1 and 2 in the date column. They are also combined in one sample. The dates mark such aggregates and one range sessions.

\subsection{The main regularities found}

\begin{figure}[h!]
  \centering
  \includegraphics[width=130mm]{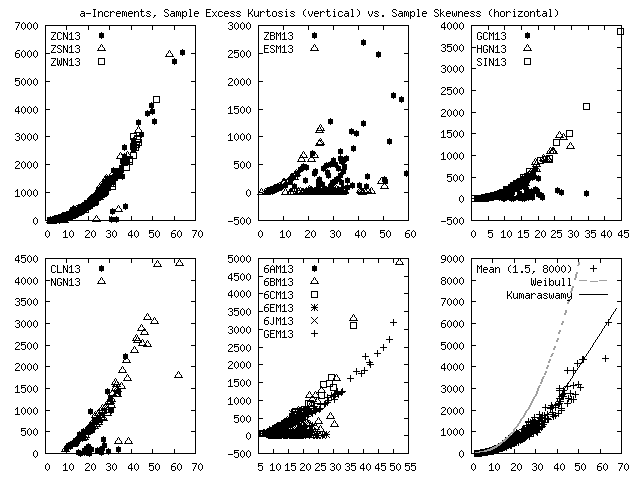}
  \caption[a-increments statistics]
   {Dependences of the sample excess kurtosis vs. the sample skewness for liquid futures, March - July, 2013.}
  \label{a_incr_skewness_ekurtosis}
\end{figure}

The author notices non-linear dependences between the sample excess kurtosis and skewness of the a-increments, Figure \ref{a_incr_skewness_ekurtosis}. The grains are the winners. It is interesting that points from ranges and aggregates belong to one curve. The four outliers for ZCN13 on April 29, 30 and May 13, 29, 2013 correspond to the mean $< 1$. For liquid contracts, the time inaccuracy badly affects the statistics. The two outliers for ZSN13 on May 21 and 23, 2013 confirm it too. The mean a-increments of the ZBM13 and ESM13 are less than a second in many sessions, Table \ref{a-increments}. Conclusions about time differences would be suspicious for them. Other "clusters" sympathize with grains. The right bottom plot on Figure \ref{a_incr_skewness_ekurtosis} combines 991 entries with the 1.5 < mean < 8000. It resembles a known dependence between the Weibull's kurtosis and skewness \cite{rousu1973}. The points from lines "ALL" deviate and are excluded.

\begin{figure}[h!]
  \centering
  \includegraphics[width=130mm]{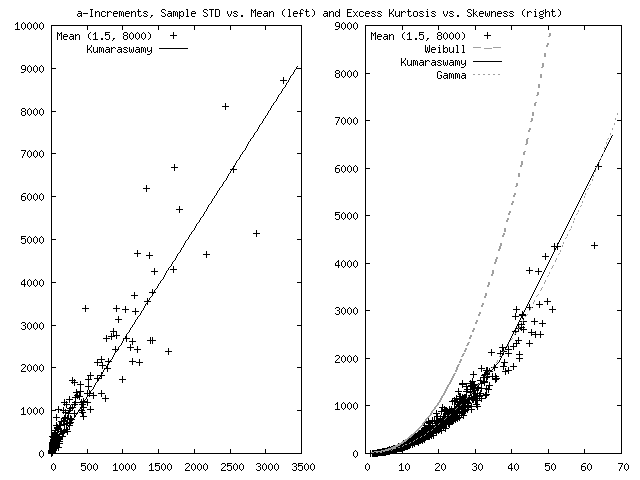}
  \caption[a-increments statistics]
   {The linear correlation between the standard deviation and mean (left) and parabolic dependence of the excess kurtosis vs. skewness (right) of a-increments combined for sixteen futures traded in March - July, 2013.}
  \label{a_mean_std}
\end{figure}

For the same 991 entires the standard deviations and means correlate with the coefficient of correlation 0.957, slope 2.65, and intercept 54.2, Figure \ref{a_mean_std}.

\begin{figure}[h!]
  \centering
  \includegraphics[width=130mm]{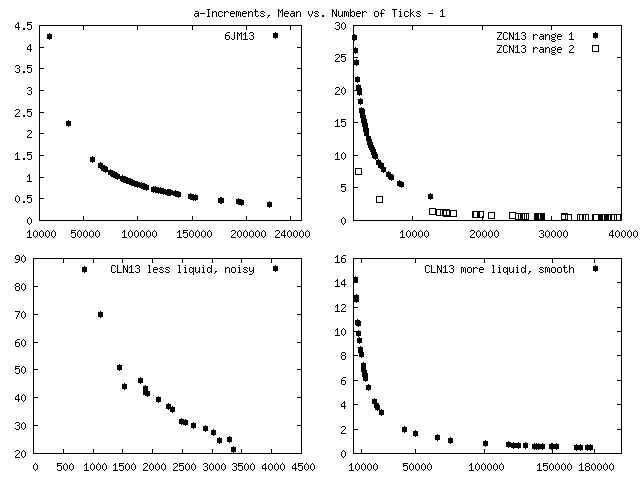}
  \caption[a-increments statistics]
   {Mean a-increment is a predictable function of the number of ticks.}
  \label{a_m_nsr}
\end{figure}

The sum of all a-increments in a range is close to its duration promoting a hyperbolic curve: mean vs. $N_{s,r}$. Short pre-holiday and last trading day ranges and sessions create outliers. Noise is larger for less liquid sessions, where the a1- and a2-increments become comparable with the range duration, Figure \ref{a_m_nsr}.

\begin{figure}[h!]
  \centering
  \includegraphics[width=130mm]{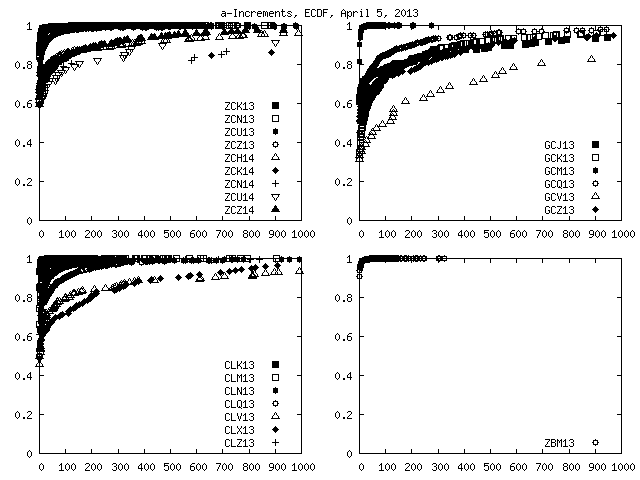}
  \caption[a-increments statistics]
   {Typical ECDF of a-increments, April 5, 2013}
  \label{a_incr_ecdf}
\end{figure}

A histogram of a-increments, Figure \ref{abcincr}, can be converted into the \textit{empirical cumulative distribution function}, ECDF, Figure \ref{a_incr_ecdf}. A \textit{cumulative distribution function}, CDF, is a full characteristic of a probability distribution. Kolmogorov \cite{kolmogorov1941} defines: \textit{"Let $x_1, x_2, \ldots, x_n$ be mutually independent random variables following the same distribution law $P\{x_i \le \xi\} = F(\xi).$ ... Put $F_n(\xi) = \frac{N(\xi)}{n}$ where $N(\xi)$ denotes the number of those $x$'s who's observed values do not exceed $\xi$".} $F(\xi)$ is CDF. $F_n(\xi)$ is ECDF. This definition implies counting repetitions. The one second inaccuracy is conductive to them. Thus, for the sorted chain $1, 2, 2, 5, 5, 5, 7, 8, 9, 9$ the author builds ECDF as (value, probability): $(1, 0.1), (2, 0.3), (5, 0.6), (7, 0.7), (8, 0.8), (9, 1)$. Kolmogorov's definition makes computing ECDF, the pairs, straight forward. But is the probability to get a value $\le 7$ equal to 0.6 or 0.7? Boris Gnedenko \cite[p. 201]{gnedenko1988} clarifies: for a sample sorted in the ascending order $x_1^* \le x_2^* \le \ldots \le x_n^*$ the ECDF is
\begin{equation}
 \label{EqECDF}
 F_n(x) = \left\{
 \begin{array}{ll}
 0 & \textrm{for} \; x \le x_1^*, \\
 \frac{k}{n} & \textrm{for} \; x_k^* < x \le x_{k+1}^*, \\
 1 & \textrm{for} \; x > x_n^*.
 \end{array}\right.
\end{equation}
This does not assume repetitions but being applied to our example leads to the inequalities: $(x \le 1, 0), (1 < x \le 2, 0.1), (2 < x \le 5, 0.3), (5 < x \le 7, 0.6), (7 < x \le 8, 0.7), (8 < x \le 9, 0.8), (9 < x, 1)$. Thus, the ECDF used is left continuous. It is a \textit{consistent estimator uniformly converging over $x$} to a CDF with $n \rightarrow \infty$ (the \textit{Glivenko-Cantelli main theorem of statistics} \cite[pp. 201 - 207]{gnedenko1988}). Equation \ref{EqECDF} adjusted for repetitions is applied for computing points on Figure \ref{a_incr_ecdf}.

The means in A1-ALL and A2-ALL in Table \ref{a-increments} are often greater than in A-ALL, while individual a2s and especially a1s are frequently zeros in ranges and sessions: transactions arrive during the first and last second. Each studied session has one or two ranges giving one or two a1s and a2s found big for illiquid contracts. In contrast, the numbers of a-increments $N_{s,r} - 1$ are usually greater than two. A single illiquid "outlier" affects stronger a1 and a2 than a-increments statistics. Let us consider two sessions with one range $T$, $N_1 \gg N_2 > 1$, and uniform distributions of ticks: $a_1 = a1_1 = a2_1 = \frac{T}{N_1+1}$ and $a_2 = a1_2 = a2_2 = \frac{T}{N_2+1}$. The mean a-increment is $\frac{(a_1(N_1-1)+a_2(N_2-1))}{N_1 + N_2 - 2} \approx \frac{2T}{N_1}$. The mean a1 and a2 are $\frac{a1_1 + a1_2}{2} = \frac{a2_1 + a2_2}{2} \approx \frac{2T}{N_2}$ but $\frac{2T}{N_2} \gg \frac{2T}{N_1}$. The 157 ZCN13's a1s consist of 153 zeros, 13, 916, 60, and 30 seconds. In the range opened on July 7, 2013 at 19:00:00 CT, the first transaction had arrived at 19:15:16 creating a1 = 916.  The mean 6.4904 is greater than the mean a-increment 2.1373.

\section{A Comment on the Weibull Distribution}

It is suggested \cite[pp. 2475 - 2476]{dufour2000} that the Weibull distribution is a \textit{"plausible assumption"} for describing durations between transactions: \textit{"The Weibull distribution is to be preferred if the data show overdispersion with extreme values (very short and long durations) more likely than the exponential would predict."} The exponential distribution $CDF(x) = 1 - e^{-\lambda x}, x \ge 0$ with the constant skewness 2 and excess kurtosis 6 is unsound for the author due to Figure \ref{a_incr_skewness_ekurtosis}.

Waloddi Weibull proposed in 1939 a statistical distribution \cite{weibull1939} named after him since 1951 \cite{weibull1951}. Nancy Mann suggests \cite{mann1968} that it is similar to \textit{the Fisher-Tippett Type III distribution} \cite{fisher1928}. The \textit{Weibull's three parameters} CDF is
\begin{equation}
\label{EqWeibullCDF}
CDF(x) = F(x) = 1 - \exp\left[- \left(\frac{x - x_u}{x_o} \right)^m\right],
\end{equation}
where, $x_u$ is the location or threshold, $x_o$ is the scale, and $m$ is the shape or modulus. Differentiating by $x$ gives the probability density function, PDF,
\begin{equation}
\label{EqWeibullPDF}
PDF(x) = \frac{dF(x)}{dx} = \frac{m}{x_o}\left(\frac{x - x_u}{x_o}\right)^{m - 1}\exp\left[- \left(\frac{x - x_u}{x_o} \right)^m\right].
\end{equation}
CDF and PDF are set to zero for $x < x_u$. Setting $x_u = 0$ gives the Rosin-Rammler equation \cite{rosin1927} \cite[discussion 1952]{weibull1951}. The distribution mean $\alpha_1$ can be derived using the substitutions $z = (\frac{x - x_u}{x_o})^m$, $x = x_u + x_o z^{\frac{1}{m}}$, $dx = \frac{x_o}{m}z^{\frac{1-m}{m}}dz$ and the properties of the Euler integral of the second kind, the Gamma function, for real $x > 0$ \cite{davis1959} \cite{korn1968} $\Gamma(x) = \int_0^\infty e^{-t}t^{x-1}dt$, $\Gamma(x+1) = x\Gamma(x)$, $\Gamma(1) = 1$
\begin{equation}
\label{EqWeibullMean}
\alpha_1 = \int_{x_u}^\infty x PDF(x) dx = \int_0^\infty (x_u + x_o z^{\frac{1}{m}})e^{-z}dz = x_u + x_o \Gamma\left(1 + \frac{1}{m}\right).
\end{equation}
With the $z$, Equations \ref{EqWeibullPDF}, \ref{EqWeibullMean}, and \textit{Newton's binomial} $(a + b)^k = \sum_{j=0}^k C_k^j a^{k-j} b^j$, where $C_k^j=\frac{k!}{j!(k-j)!}$ \cite{korn1968}, we get the $k$th \textit{central moment} (about the mean)
\begin{displaymath}
\mu_k = \int_{x_u}^{\infty} (x - \alpha_1)^k PDF(x) dx = \int_0^{\infty}\left[x_u + x_o z^{\frac{1}{m}} - \alpha_1\right]^k e^{-z}dz =
\end{displaymath}
\begin{equation}
\label{EqWeibullKMoment}
= x_o^k \sum_{j=0}^k C_k^j (-1)^j \Gamma^j\left(1+\frac{1}{m}\right) \Gamma\left(1 + \frac{k-j}{m}\right) = x_o^k W(m, k)
\end{equation}
These moments do not depend on $x_u$. The variance, $k = 2$, is equal to
\begin{equation}
\label{EqWeibullVariance}
variance = \mu_2= x_o^2\left[\Gamma\left(1+\frac{2}{m}\right) - \Gamma^2\left(1+\frac{1}{m}\right)\right]
\end{equation}
The ratio $\mu_q/\mu_p^{\frac{q}{p}} = W(m, q)/W^{\frac{q}{p}}(m, p)$ does not depend on $x_o$. Thus,
\begin{equation}
\label{EqWeibullSkewness}
skewness = \frac{\mu_3}{\mu_2^{\frac{3}{2}}} = \frac{\Gamma\left(1 + \frac{3}{m}\right)-3\Gamma\left(1 + \frac{2}{m}\right)\Gamma\left(1 + \frac{1}{m}\right) + 2\Gamma^3\left(1 + \frac{1}{m}\right)}{\left[\Gamma\left(1+\frac{2}{m}\right) - \Gamma^2\left(1+\frac{1}{m}\right)\right]^{\frac{3}{2}}},
\end{equation}
\begin{equation}
\label{EqWeibullEKurtosis}
\begin{split}
excess \; kurtosis = kurtosis - 3 = \frac{\mu_4}{\mu_2^2} - 3 = -3 + \\
+ \frac{\Gamma\left(1 + \frac{4}{m}\right) - 4\Gamma\left(1 + \frac{3}{m}\right)\Gamma\left(1 + \frac{1}{m}\right)                                       + 6\Gamma\left(1 + \frac{2}{m}\right)\Gamma^2\left(1 + \frac{1}{m}\right) - 3\Gamma^4\left(1 + \frac{1}{m}\right)                                   }{\left[\Gamma\left(1+\frac{2}{m}\right) - \Gamma^2\left(1+\frac{1}{m}\right)\right]^2}
\end{split}
\end{equation}
depend on $m$ only and \textit{m-parametrically} each on other. The Lanczos's approximations \cite{lanczos1964} are sufficient for the Gamma function computation. Equations \ref{EqWeibullSkewness} and \ref{EqWeibullEKurtosis} are applied to plot the Weibull curve on Figure \ref{a_incr_skewness_ekurtosis}. It resembles the data but experiences a systematic shift up increasing with the skewness.

\section{A Comment on the Kumaraswamy Distributions}

Conducting hydrologic research, Ponnambolam Kumaraswamy has invented three probability distributions \cite{kumaraswamy1976} - \cite{kumaraswamy1980}. One \cite{kumaraswamy1980} for $z_{min} \le z \le z_{max}$ is
\begin{equation}
\label{EqKumaraswamyCDF}
CDF(z) = F(z) = F_0 + (1 - F_0)\left (1 - \left (1 - \left (\frac{z - z_{min}}{z_{max} - z_{min}} \right )^a \right )^b \right ),
\end{equation}
where $F(z_{min}) = F_0$, $F(z_{max}) = 1$. $F_0$ is cumulative probability of $z_{min}$. Differentiating $F(z)$ with respect to $z$ gives the PDF
\begin{equation}
\label{EqKumaraswamyPDF}
\begin{split}
PDF(z) = \frac{ab(1 - F_0)}{z_{max} - z_{min}} \left (\frac{z - z_{min}}{z_{max} - z_{min}} \right )^{a-1}      \left (1 - \left (\frac{z - z_{min}}{z_{max} - z_{min}} \right )^a \right )^{b-1}
\end{split}
\end{equation}
This differs from the original $f(z)$ \cite[p. 81, Equation 3]{kumaraswamy1980} by the factor $\frac{1}{z_{max} - z_{min}}$. In fact, Kumaraswamy differentiates $F(z)$ with respect to $x=\frac{z-z_{min}}{z_{max}-z_{min}}$ instead of the promised $z$ and $z = z_{min}+(z_{max} - z_{min})x, dz = (z_{max} - z_{min})dx$. The $F_0$ should be taken care because $\int_{z_{min}}^{z_{max}}PDF(z)dz = 1 - F_0$ but not one. The beginning $k$th moment can be expressed as
\begin{displaymath}
\alpha_k = \int_{-\infty}^{z_{min}}z^kPDF(z)dz + \int_{z_{min}}^{z_{max}}z^kPDF(z)dz + \int_{z_{max}}^{\infty}z^kPDF(z)dz,
\end{displaymath}
where the last integral is zero but the first is $z_{min}^kF_0$ accounting the probability mass (not density) at $z_{min}$. With $Q=z_{max}-z_{min}$ and the Kumaraswamy's $x$,
\begin{displaymath}
\int_{z_{min}}^{z_{max}}z^kPDF(z)dz= \int_0^1(z_{min}+Qx)^kab(1-F_0)x^{a-1}(1-x^a)^{b-1}dx=
\end{displaymath}
\begin{displaymath}
=ab(1-F_0) \sum_{j=0}^k C_k^jz_{min}^{k-j}Q^j \int_0^1 x^{a-1+j}(1-x^a)^{b-1}dx.
\end{displaymath}
The Newton's binomial is applied. If $y=x^a, x=y^{\frac{1}{a}}, dx=\frac{1}{ax^{a-1}}dy$, then
\begin{displaymath}
\int_{z_{min}}^{z_{max}}z^kPDF(z)dz=b(1-F_0) \sum_{j=0}^k C_k^jz_{min}^{k-j}Q^j \int_0^1 y^{\frac{j}{a}}(1-y)^{b-1}dy.
\end{displaymath}
The right integral is the Euler integral of the first kind, the Beta-function, \cite{korn1968}
\begin{displaymath}
\textrm{B}(p, q) = \frac{\Gamma(p)\Gamma(q)}{\Gamma(p + q)} = \int_0^1 t^{p-1}(1-t)^{q-1}dt, \; \textrm{Re} \; p > 0, \textrm{Re} \; q > 0.
\end{displaymath}
Finally,
\begin{equation}
\label{EqKumaraswamyKBeginning}
\alpha_k=z_{min}^kF_0+b(1-F_0) \sum_{j=0}^k C_k^jz_{min}^{k-j}Q^j  \textrm{B}(1+\frac{j}{a},b).
\end{equation}
Since $\Gamma(1)=1, \; \Gamma(z+1)=z\Gamma(z)$ \cite{korn1968}, the $\textrm{B}(1,b)=\frac{1}{b}$ and the mean, $k=1$, is
\begin{equation}
\label{EsKumaraswamyMean}
\alpha_1 = z_{min} + b(1 - F_0)(z_{max}-z_{min})\textrm{B}(1+\frac{1}{a},b).
\end{equation}
With $R = z_{min}-\alpha_1$ and the mass $F_0$ at $z_{min}$, the $k$th central moment is
\begin{displaymath}
\mu_k = R^kF_0 + ab(1-F_0) \int_0^1(Qx + R)^kx^{a-1}(1-x^a)^{b-1}dx.
\end{displaymath}
\begin{displaymath}
= R^kF_0 + b(1-F_0) \sum_{j=0}^k C_k^j Q^{k-j}R^j \int_0^1y^{\frac{k-j}{a}}(1-y)^{b-1}dy.
\end{displaymath}
Finally,
\begin{equation}
\label{EqKumaraswamyKMoment}
\mu_k= R^kF_0+ b(1-F_0) \sum_{j=0}^k C_k^j Q^{k-j}R^j \textrm{B}(1 + \frac{k - j}{a}, b).
\end{equation}
\begin{equation}
\label{EqKumaraswamyVariance}
\mu_2= R^2F_0 + b(1-F_0) \left (Q^2\textrm{B}(1+\frac{2}{a},b) + 2QR\textrm{B}(1+\frac{1}{a},b) + \frac{R^2}{b}\right ).
\end{equation}
\begin{equation}
\label{EqKumaraswamyMu3}
\mu_3= R^3F_0 + b(1-F_0)(Q^3\textrm{B}(1+\frac{3}{a},b) + 3Q^2R\textrm{B}(1+\frac{2}{a},b) + 3QR^2\textrm{B}(1+\frac{1}{a},b) + \frac{R^3}{b}).
\end{equation}
\begin{equation}
\label{EqKumaraswamyMu4}
\begin{split}
\mu_4=R^4F_0 + b(1-F_0)(Q^4\textrm{B}(1+\frac{4}{a},b) + 4Q^3R\textrm{B}(1+ \frac{3}{a},b) \\
+ 6Q^2R^2\textrm{B}(1+\frac{2}{a},b) + 4QR^3\textrm{B}(1+\frac{1}{a},b) + \frac{R^4}{b}).
\end{split}
\end{equation}
The skewness is $\frac{\mu_3}{\mu_2^{\frac{3}{2}}}$. The excess kurtosis is $\frac{\mu_4}{\mu_2^2} - 3$. The $\mu_2=\alpha_2-\alpha_1^2$ holds.

On histograms of a-increments the highest bar is often at zero, Figure \ref{abcincr}. A PDF can approximate it. For $a = 1$, Equation \ref{EqKumaraswamyPDF} gives $\frac{b(1-F_0)}{z_{max} - z_{min}}$ at $z = z_{min}$, where Equation \ref{EqKumaraswamyCDF} returns $F_0$ suitable for ECDFs on Figure \ref{a_incr_ecdf}. However, zero times between transactions is likely a consequence of the one second inaccuracy.

Alternative is to apply theoretically natural $z_{min} = 0, F_0=0$ and fit the heights of the bars by the integrals of the PDF, Equation \ref{EqKumaraswamyPDF}, on some intervals. We can use (a-increment, [integration interval]): $(0, [0, 0.5]), (1, [0.5, 1.5]), \dots$ or $(0, [0, 1]), (1, [1, 2]), \dots$ and the PDF
\begin{equation}
\label{EqKumaraswamyPDF2}
PDF(z) = \frac{ab}{z_{max}} \left (\frac{z}{z_{max}} \right )^{a-1} \left (1 - \left (\frac{z}{z_{max}} \right )^a \right )^{b-1},
\end{equation}
where $z_{max}$ can be greater than $T_c^{s,r} - T_o^{s,r}$. If a simulated a-increment being added to the current time exceeds the range/session closing time, then no transaction is completed and trading terminates until a new range/session. Using $z_{max} = T_c^{s,r}-t_{current}^{s,r}$ as the Kumaraswamy's bound would never stop trading because any a-increment will fit within the remaining time interval. This would lead to a-increments decreasing closer to the end but not observed.

The $z_{min}=0, F_0=0$ leave three degrees of freedom $a, b, z_{max}$. The parametric curve of the excess kurtosis vs. skewness using $b=3.42, a \in [0.041, 1.5]$, and arbitrary $z_{max} > 0$ fits the data, Figure \ref{a_incr_skewness_ekurtosis}. The related equations are
\begin{equation}
\label{EqKumaraswamyMean0}
\alpha_1 = b\textrm{B}(1+\frac{1}{a},b)z_{max}, \; \mu_2 = b\left ( \textrm{B}(1+\frac{2}{a},b) - b\textrm{B}(1+\frac{1}{a},b)^2\right )z_{max}^2
\end{equation}
\begin{equation}
\sqrt{\mu_2} = \frac{\sqrt{\textrm{B}(1+\frac{2}{a},b) - b\textrm{B}(1+\frac{1}{a},b)^2}}{\sqrt{b} \textrm{B}(1+\frac{1}{a},b)}\alpha_1,
\end{equation}
\begin{equation}
\mu_3=b\left ( \textrm{B}(1+\frac{3}{a},b)-3b\textrm{B}(1+\frac{1}{a},b)\textrm{B}(1+\frac{2}{a},b)+2b^2\textrm{B}(1+\frac{1}{a},b)^3 \right) z_{max}^3,
\end{equation}
\begin{displaymath}
\mu_4=b( \textrm{B}(1+\frac{4}{a},b)-4b\textrm{B}(1+\frac{1}{a},b)\textrm{B}(1+\frac{3}{a},b)+6b^2\textrm{B}(1+\frac{1}{a},b)^2\textrm{B}(1+\frac{2}{a},b) -
\end{displaymath}
\begin{equation}
3b^3\textrm{B}(1+\frac{1}{a},b)^4) z_{max}^4.
\end{equation}
\begin{figure}[h!]
  \centering
  \includegraphics[width=130mm]{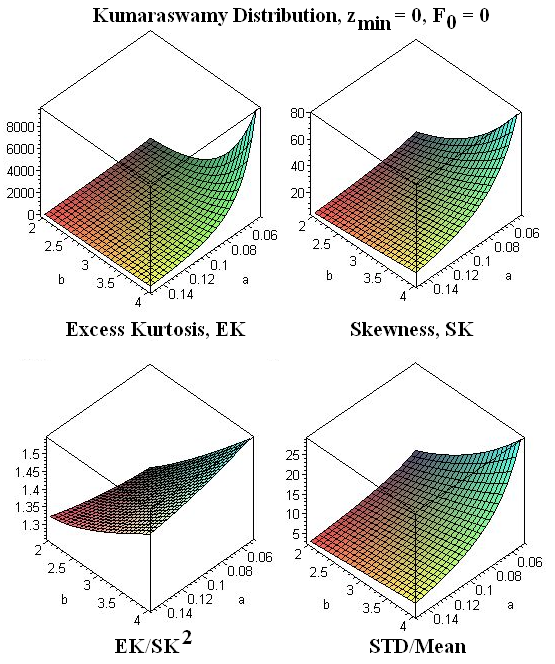}
  \caption[a-increments statistics]
   {Dependencies of the quantities of Kumaraswamy distribution on the parameters $a$ and $b$. Plots are done using Maple 10 from Maplesoft.}
  \label{kumaraswamy}
\end{figure}
Now, the skewness, excess kurtosis, and $\frac{\mu_k}{\mu_2^{\frac{k}{2}}}$ do not depend on $z_{max}$. We can choose $a$ and $b$ to fit the sample skewness and excess kurtosis and $z_{max}$ to adjust the mean and variance depending linearly and quadratically on $z_{max}$. This is useful, since the standard deviation and mean linearly correlate. The theoretical excess kurtosis and skewness within $[0, 8000]$ and $[0, 80]$ support the experimental dependence $\frac{\mu_4}{\mu_2^{2}} - 3 \approx 1.5 \frac{\mu_3^2}{\mu_2^2}$, keeping the ratio within [1.30, 1.55], Figure \ref{kumaraswamy}. The theoretical ratio of the standard deviation to the mean [2, 25] is less supportive for the experimental slope 2.65, Figure \ref{a_mean_std}. Fitting the excess kurtosis and skewness  varying $a$ and $b$ and then reusing $a$ and $b$ to fit the mean and standard deviation changing $z_{max}$ works for many entries in Table \ref{a-increments}. Table \ref{experiment_kumaraswamy} is an illustration, where rows with date are self-explanatory and rows "Theory" contain the ticker in column $a$, the sum of relative errors taken by absolute value for skewness and excess kurtosis in column $b$, and the  sum of relative errors taken by absolute value for mean and standard deviation in column $z_{max}$. These two sums of relative errors have been minimized in two steps using the Microsoft Excel Solver and the GAMMALN for expansion of the Beta function. With these cost functions the three parameters Kumaraswamy distribution describes better the skewness and excess kurtosis than mean and standard deviation. Reproduction of the four sample moments with relative errors from 0.3\% to 37\% is satisfactory for many purposes.
\begin{table}[!h]
  \topcaption{a-Increments. Sample and Computed Kumaraswamy Moments.}
  \begin{tabular}{cccccccc}
  Date & $\alpha_1$ & $\sqrt{\mu_2}$ & $\frac{\mu_3}{\mu_2^{\frac{3}{2}}}$ & $\frac{\mu_4}{\mu_2^2} - 3$ & $a$ & $b$ & $z_{max}$ \\
  2013-03-01 & 7.3041 & 45.495& 13.4 & 234.2 & 0.08021 & 2.565 & 1642.2\\
  Theory & 6.4605 & 45.495& 13.4 & 234.2 & ZCN13 & 0.35\%  & 12\%\\
  2013-03-04 & 5.6324 & 58.506& 41.1 & 2559 & 0.06680 & 3.807 & 10317.7\\
  Theory & 3.5658 & 58.506 & 41.4 & 2559 & ZCN13 & 0.66\%  & 37\%\\
  2013-04-05 & 3.4886 & 29.191 & 19.4 & 493.6 & 0.1179 & 4.016 & 2105.7\\
  Theory & 3.4886 & 26.754 & 18.3 & 493.6 & ZCN13 & 5.6\%  & 8.3\%\\
  2013-06-17 & 15.213 & 56.331 & 9.66 & 152.6 & 0.06680 & 3.807 & 10317.7\\
  Theory & 12.091 & 56.331 & 10.3 & 152.6 & ZCN13 & 6.6\%  & 21\%\\
  \end{tabular}
  \label{experiment_kumaraswamy}
\end{table}

\begin{figure}[h!]
  \centering
  \includegraphics[width=130mm]{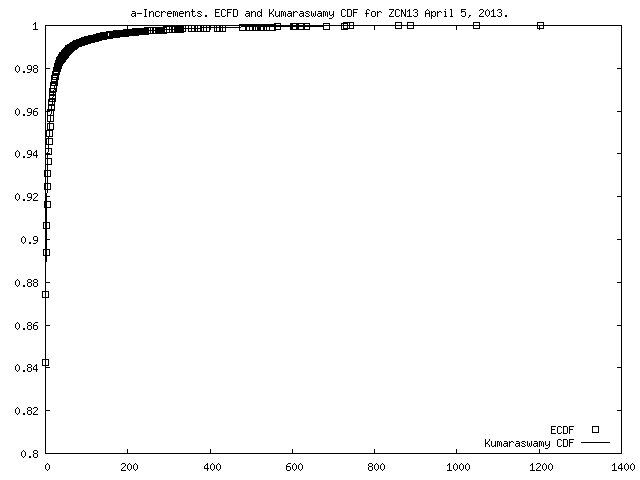}
  \caption[a-increments statistics]
   {Kumaraswamy CDF with $a=0.1179, \; b=4.016, \; z_{max}=2105.7, \; z_{min} = 0, \; F_0=0$ against the ECDF of the July 2013 Corn futures traded on April 5, 2013.}
  \label{a_ecdf}
\end{figure}

A general method for estimating distribution parameters based on fitting an ECDF by a CDF is suggested by Weibull \cite{weibull1969}. The method of moments applied here can provide an initial guess for a solver. Figure \ref{a_ecdf} plots the CDF, Equation \ref{EqKumaraswamyCDF}, together with ECDF, Equation \ref{EqECDF}.
The results of the Pearson's goodness of fit test \cite{korn1968} are in Table \ref{a_incr_pearson}.
\begin{table}[!h]
  \centering
  \topcaption{a-Increments. The $\chi^2$-Test. ZCN13, April 5, 2013, $N=21670$.}
  \begin{tabular}{crrrrrr}
  Class $k$ & $t_k^{min}$ & $t_k^{max}$ & $p_k$ & $N_k$ & $Np_k$ & $\frac{(N_k-Np_k)^2}{Np_k}$ \\
1 & 0 & 16 & 0.9639 & 20890 & 20886.6 & 0.0005438\\
2 & 16 & 135 & 0.03041 & 667 & 658.99 & 0.09727\\
3 & 135 & 260 & 0.003513 & 60 & 76.126 & 3.416\\
4 & 260 & 390 & 0.001199 & 22 & 25.976 & 0.6087\\
5 & 390 & 610 & 0.0006975 & 20 & 15.116 & 1.578\\
6 & 610 & 732 & 0.0001485 & 6 & 3.2185 & 2.404\\
7 & 732 & 1203 & 0.0001659 & 5 & 3.595 & 0.5492\\
& & Sum & $\approx 1$ & 21670 & 21669.7 & 8.653\\
  \end{tabular}
  \label{a_incr_pearson}
\end{table}
Each class has at least five observations $N_k$. With one second inaccuracy, it was difficult to follow to the Mann-Wald technique choosing classes \cite{mann1942}, \cite{williams1950} and requiring to divide one second interval of the highest probability. With three parameters the number of degrees of freedom is $f = 7 - 1 - 3 = 3$. The probabilities $p_k$ are computed by Equation \ref{EqKumaraswamyCDF} using $a$, $b$, $z_{max}$ from Table \ref{experiment_kumaraswamy} for 2013-04-05 and $z_{min} = 0, F_0 = 0$. The tabulated $\chi^2$ values for the levels 0.05, 0.02, 0.01, and 0.001 and $f=3$ are 7.815, 9.837, 11.341, and 16.268 \cite{korn1968}. The observed value is 8.653. The Kumaraswamy distribution hypothesis cannot be rejected. It is perspective for fitting the sample statistics of a-increments.

\section{A Comment on the Gamma Distribution}

The Gamma distribution \cite{korn1968}, \cite[p. 930]{abramowitz1972} has
\begin{displaymath}
PDF(x) = \frac{\beta^{\alpha}}{\Gamma(\alpha)} x^{\alpha-1} e^{-\beta x}, \; x > 0, \; \alpha > 0, \; \beta > 0,
\end{displaymath}
\begin{displaymath}
\alpha_1 = \frac{\alpha}{\beta}, \; \mu_2 = \frac{\alpha}{\beta^2}, \; \frac{\mu_3}{\mu_2^{\frac{3}{2}}}=\frac{2}{\sqrt{\alpha}}, \; \frac{\mu_4}{\mu_2^2} - 3=\frac{6}{\alpha}.
\end{displaymath}
This implies that $\frac{\mu_4}{\mu_2^2} - 3= 1.5 \frac{\mu_3}{\mu_2^{\frac{3}{2}}}$ and $\sqrt{\mu_2} = \frac{\alpha_1}{\sqrt{\alpha}}$, Figure \ref{a_mean_std}. We get
\begin{displaymath}
\alpha = \frac{\alpha_1^2}{\mu_2} = \frac{4}{\textrm{skewness}^2} = \frac{6}{\textrm{excess kurtosis}}.
\end{displaymath}
For moments in Table \ref{experiment_kumaraswamy} the $\alpha$ is: 2013-03-01 $0.02578 \approx 0.02228 \approx 0.02562$; 2013-03-04 $0.009268 \approx 0.002368 \approx 0.002345$; 2013-04-05 $0.01428 \approx 0.01063 \approx 0.01216$; 2013-0-17 $0.07293 \approx 0.04287 \approx 0.03932$. These values indicate insufficient flexibility of the Gamma distribution.

\section{The b-Increments}

The b-increments are nonidentical stones building price, Table \ref{b-increments}.

\subsection{Discreteness of prices and their increments}

Futures prices are conventionally discrete. A ZBM13 price can be 140.00000 and 140.03125 but not between them. The dollar equivalent of 0.03125 is \$31.25 - a lunch. Tom Baldwin, a trader, was known for trading 6,000 bond futures per day in 1980th \cite[p. 321]{bernstein1987}. A 2,000 contracts position fluctuates \$62,500 with each $\delta$-tick. The large positions and high leverage do not allow ignoring price discreteness
\begin{equation}
\label{EqWholePrice}
P_i = n_i \delta; \; \textrm{b-increment}_i = P_i - P_{i-1} = (n_i - n_{i-1}) \delta; \; n_i, n_{i-1} \in \textrm{N}_0
\end{equation}
The ratios of prices or b-increments are \textit{rational numbers} common in accounting. To recollect it, try to withdraw \textit{exactly} $\pi$ dollars from a bank. The asset return $\ln(\frac{P_i}{P_{i-1}})$ is the difference of the logarithms of integers. A theory is wrong, if it suggests that Probability$\{140 < P_{bond}\ < 140.03125\} > 0$. \textit{Discrete} or \textit{lattice distributions} are suitable for b-increments.

The words \textit{"God made the integers; all the rest is the work of man"} are ascribed to Leopold Kronecker. Modern finance is carried away by using continuous price models and distributions starting from the Gaussian one. In contrast, the author is more impressed by the following Kolmogorov's thought: \textit{"It is very likely, that with development of modern computational technology it will be understood that in many cases it is wise to study real phenomena avoiding intermediate stage of their stylization in the spirit of mathematical conceptions of infinite and continuous, moving directly to discrete models"} \cite{kolmogorov1983}.

\subsection{Almost zero mean b-increments}

Even, when the difference between the last and first price in a range is substantial, the number of ticks $N_{s,r}$ is so big, that the mean
\begin{equation}
\label{EqMeanBIncrement}
\textrm{mean b-increment} = \overline{\Delta P}^{s,r} = \frac{P_{N_{s,r}^{s,r}}^{s,r} - P_1^{s,r}}{N_{s,r}-1} = \frac{\sum_{i=2}^{N_{s,r}}\Delta P_i^{s,r}}{N_{s,r}-1}
\end{equation}
is close to zero. The imbalance between negative and positive b-increments in a range or session exists on the background of their large total number including zeros. This keeps skewness small too. However, the distribution is not a Gaussian bell not only because it is essentially discrete but due to the non-zero sample excess kurtosis, Table \ref{b-increments}.

\subsection{Futures limit prices}

\begin{figure}[h!]
  \centering
  \includegraphics[width=130mm]{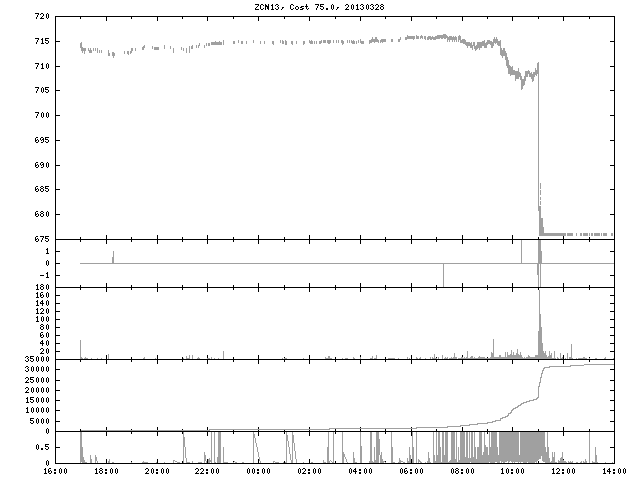}
  \caption[ZCN13 Trading at the Limit Down]
   {Trading ZCN13 at the limit down on Thursday March 28, 2013.}
  \label{ZCN13_20130328_price_limit_down}
\end{figure}

\begin{figure}[h!]
  \centering
  \includegraphics[width=130mm]{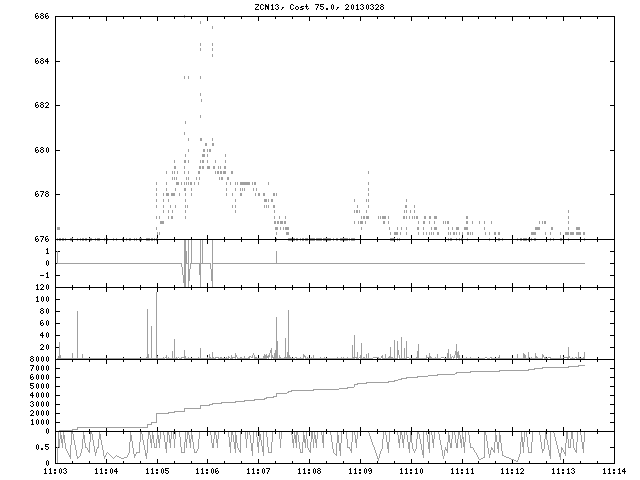}
  \caption[ZCN13 Trading at the Limit Down. Zooming.]
   {Trading ZCN13 at the limit down on Thursday March 28, 2013. Zooming.}
  \label{ZCN13_20130328_price_limit_down_zooming}
\end{figure}

\begin{figure}[h!]
  \centering
  \includegraphics[width=130mm]{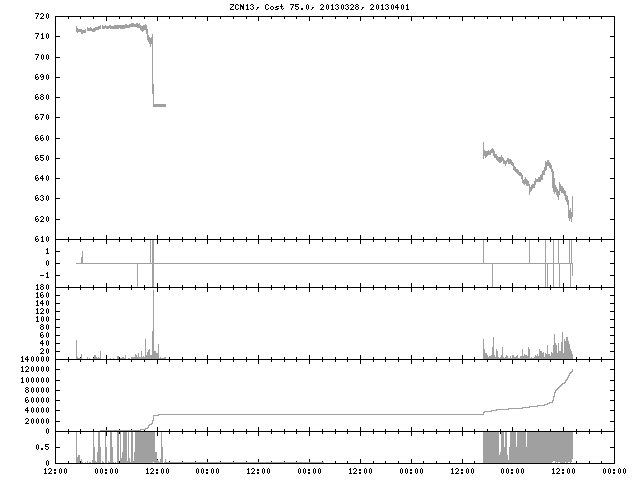}
  \caption[ZCN13 Trading at the Limit Down and in a Next Session.]
   {Trading ZCN13 at the limit down on Thursday March 28, 2013 and the next Monday April 1, 2013 session after long weekend.}
  \label{ZCN13_2_20130401_price_two}
\end{figure}

If a grain futures contract is traded prior its expiration month, then there is a limit price condition. Currently, the price of corn during the first limit session cannot move more than 40 points from the previous settlement price. On March 28, 2013 the bearish market news at 11:00:00 CT brought the price down to the limit 676 at 11:03:02 CT, Figure \ref{ZCN13_20130328_price_limit_down}.
Trading does not stop. With the panic liquidation of long positions, a limit buy 676 order placed right after 11:00:00 would be filled. The price still may go up. That happened in the session prior Good Friday. Zooming to [11:03:02, 11:13:24] shows 1,835 transactions with the total volume 3,776 contracts at the price greater than 676, Figure \ref{ZCN13_20130328_price_limit_down_zooming}. With the commission and fees \$10.66 per trade per contract, a limit sell order for a few contracts placed two - three points above the limit would be very likely filled within the next 10 minutes resulting in a profit \$89.36 - \$139.36 per contract. The prices 686 indicate, even, bigger but less probable potential profit \$489.36, Figure \ref{ZCN13_20130328_price_limit_down_zooming}. The loss could not exceed \$10.66. If the position could not be closed, then after the long weekend it would be a shocking > \$1,000 loss after the open price gapped down, Figure \ref{ZCN13_2_20130401_price_two}. The price distribution on March 28, 2013 has a cut off tail known in advance. This can affect the b-increments.


\subsection{The chicken and egg question}

Sample moments are symmetric functions independent on the order of data. Holding the first price and changing the order of b-increments affects prices and their statistics but not b-increments. Reordering prices changes b-increments and their statistics but not prices. The relationship between the sample means $\bar{P}^{s,r}$ and $\overline{\Delta P}^{s,r}$ includes the prices order. From Equations \ref{EqPriceSession} and \ref{EqMeanBIncrement} \cite{salov2011},
\begin{equation}
\label{EqMeanPMeanBIncr}
\bar{P}^{s,r} = \frac{\sum_{i=1}^{N_{s,r}}P_i^{s,r}}{N_{s,r}} = P_1^{s,r} + (N_{s,r} - 1) \overline{\Delta P}^{s,r} - \frac{\sum_{i=2}^{N_{s,r}}i \Delta P_i}{N_{s,r}}.
\end{equation}
The product $i \Delta P_i$ embeds dependence on the prices order, rising the chicken and egg question: which variables are fundamental prices or increments? The modern financial stochastic differential equations, SDE, taking the baton from Louis Bachelier's Brownian motion \cite{bachelier1900} and Paul Samuelson's geometric or economic Brownian motion \cite{samuelson1965} focus on absolute and relative price changes. Stochastic integration of increments creates prices leaving the latter a secondary role. This role is bigger in the geometric Brownian motion, where price denominators ensure the bigger risk and gain at higher prices. The Ornstein-Uhlenbeck process \cite{uhlenbeck1930} extended to the mean reversion SDE \cite{hull1997} applied for simulation of interest and exchange rates has an embedded level coming out as an attractor. In contrast, often criticized in scientific literature technical analysis \cite{murphy1999} puts significant accent to prices, their patterns, and trends. The author thinks that markets have many modes replacing each other in time, where prices or increments get varying accents.

\section{A Comment on Science and Pseudoscience}

Specialists on no-arbitrage pricing derivatives based on SDE and apologists of the technical analysis are two irreconcilable camps. The former exploit sophisticated mathematics. The latter draw lines and recognize patterns requiring more imagination than knowledge of geometry.

\subsection{Sir Isaak Newton - a trader}

Not every scientific worker trades and not every trader is a scientist. Sir Isaak Newton was such a "combination". His scientific authority is indisputable. His trading the South Sea Company stock since creation of the company in 1711 until the bubble in 1720 is an example, when an outstanding mind \textit{"can calculate the movements of the stars but not the madness of men"} - the words attributed to Newton. This article is about the attractive market mean to withdraw specialists from professions - high frequency of big potential profits. The market keeps them in professions using losses. What kind of society would it be, if everybody would only speculate.

\subsection{A common element of successful patterns}

The source of income of a SDE, C++, software specialist in finance is profits made by traders. The traders watch the a-b-c-process. It, external, and internal events affect their minds and trading robots. They make decisions influencing on the process. The snake bites its tail. Its goal is survival. This is achieved by showing to the majority past frequent potential profits and hiding upcoming losses under the profit attire. If a market is \textit{efficient} in something, then it is in this ability to fool. Larry Williams, a known trader and educator, writes: \textit{"The best patterns I have found have a common element tying them together: patterns that present extreme market emotions reliably set up trades for price swings in the opposite direction"} \cite[p. 95]{williams1999}.

\subsection{On technical analysis}

Technical analysts appeal to Newton's mechanics seeking a basis for price trends: \textit{"... a trend in motion is more likely to continue than to reverse. This corollary is ... an adaptation of Newton's first law of motion"} \cite[p. 4]{murphy1999}. They criticize the theory of randomly walked prices. Ironically, the molecular-kinetic theory of heat served to Albert Einstein as an explanation of the Brownian particle displacement proportional to the square root of the time \cite[Equation 11]{einstein1905}  employs statistics on a top of consideration of molecules as classical Newton's particles. The system establishes the Laplace determinism. Quantum mechanics dismisses the latter paradigm axiomatically applying \textit{uncertainty} on the micro level and making the classical mechanics a macro limit. This lack of causality seemed never satisfy Einstein.

Trend lines were drawn prior Bachelier. His ingenious mathematical model of Brownian motion coming five years prior the Einstein's paper creates a parallel. Quantum mechanics has consumed the Newton's one replacing the fully deterministic picture with uncertainty. Bachelier introduces uncertainty into price changes and presses determinism of trend lines. In both cases (physics and markets), applications continue using each paradigm depending on a situation. This parallel ends, if we recollect that the market, involving human beings, extends beyond the physical Brownian motion. Einstein derives his model for a physical phenomenon. His last equation for determination of the Avogadro number suggests experimental conditions for verification. Bachelier's attempt to apply the same model to a phenomenon involving human consciousness and "the madness of men" requires serious confirmations. Its inadequacy leading to negative prices patched by Samuelson switching to a lognormal distribution and further attempts to eliminate inadequacy (\textit{volatility smile, volatility term, discrete dividend adjustments}) of the advanced analog - the Black-Scholes-Merton option pricing formulas \cite{hull1997} - look important but minor details within the entire picture of complexity. Effects discovered by Kahneman and Tversky, masterly measuring human being behavior, is only the beginning of understanding of their influence on the a-b-c-process.

Since the market is \textit{"a result of cooperation of modern technology and human being consciousness governed by partly unknown laws of nature"} \cite{salov2012b}, there is a danger to fall into pseudoscience. Indeed, some pay attention to moon phases. If they trade, should moon phases influence on prices? There are daily charts showing the days of full and new moon coinciding with significant local minimums and maximums of silver, corn, soybean, and wheat \cite[pp. 94 - 96]{williams1979}. Are these events independent? Should we dismiss the dependence, if some minimums and maximums are one - three days apart from the full and new moon? Was it only in 1971 - 1973? The Karl Popper's \textit{falsifiability} - criterion of demarcation between science and pseudoscience implies that in this case we can formulate a hypothesis and prove or disprove it. In general, it is not easy to demarcate them \cite{shermer2011} and a \textit{halo of scientist} can play a terrible role. Serguei Kara-Murza, a philosopher of science, reminds about the Stanley Milgram's experiments on obedience to authority figures conducted in the 1960th \cite{karamurza1990}.

Sometimes technical analysis presents patterns without algorithmic definitions. Often it does not supply enough evidences of claims. Its representatives are seem busy with trading and have no time for a rigorous research: \textit{"technical analysis is a broad class of prediction rules with unknown statistical properties, developed by practitioners without reference to any formalism."} \cite{neftci1991}. Neftci is a constructive critic. His research is promising for technical analysis and cautiously gives a hope: \textit{"... if the processes under consideration were nonlinear, then the rules of technical analysis might capture some information ignored by Wiener-Kolmogorov prediction theory"}. Barton Malkiel is more skeptical: \textit{"In ... simulated stock charts derived from student coin-tossings, there were head-and-shoulder formations, triple tops and bottoms, and other more esoteric chart patterns"} \cite[p. 131]{malkiel2007}. One formation was found very bullish by a chartist. However, Malkiel does not insist that market is a perfect random walk.

A technical analyst needs to turn his or her face to modern pattern recognition and machine learning techniques based on genetic programming \cite{koza1992}, neural networks, support and relevance vector machines, probabilistic principal component analysis, Bayesian optimization (as a way to overcome overfitting) \cite{bishop2006} with a solid theoretical foundation laid in works of Vladimir Vapnik and Alexey Chervonenkis \cite{vapnik1974}. The task here is to formalize algorithmic definitions of patterns and signals and automate their recognition with the purpose of objective statistical analysis.

\subsection{Computer generated random walk vs. a-b-c-process}

\begin{figure}[h!]
  \centering
  \includegraphics[width=130mm]{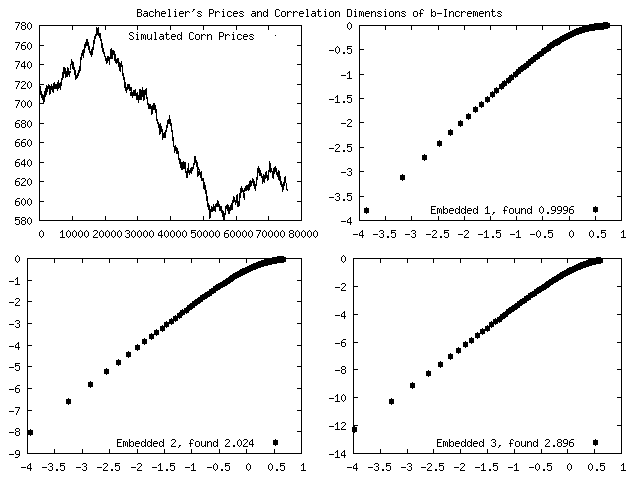}
  \caption[Correlation dimensions of simulated b-increments]
   {Bachelier's model: prices, and embedded, and found correlation dimensions of b-increments.}
  \label{ci_sim}
\end{figure}

\begin{figure}[h!]
  \centering
  \includegraphics[width=130mm]{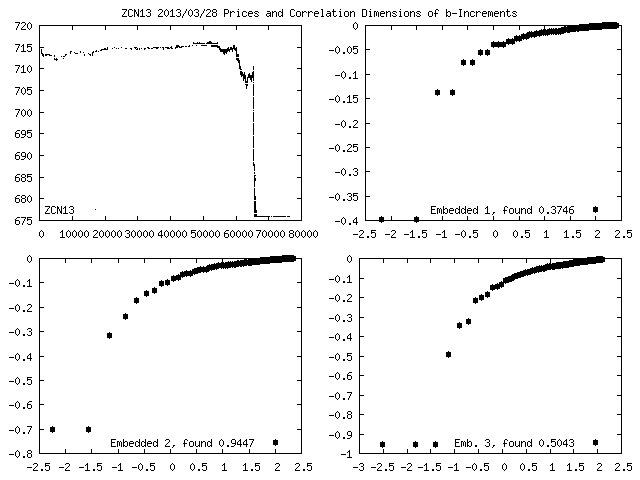}
  \caption[Correlation dimensions of real b-increments]
   {ZCN13 20130328: prices, and embedded, and found correlation dimensions of b-increments.}
  \label{ci_b_ZCN13_20130328}
\end{figure}

Visible similarity of a random walk with prices can be misleading. Techniques magnifying hidden differences of the two time series are valuable. A fair coin tossing, well studied Bernoulli trials, can be simulated with a uniform pseudo random numbers generator \cite{lecuyer1988}. The normal generator \cite{box1958} can be applied for simulation of the Bachelier's normal and Samuelson's lognormal time series. With the normal generator the variance must be set proportional to the time step of a Brownian motion, if time steps vary. For a constant step, it can be scaled to a single suitable value. The finite difference equations of the Bachelier's model applied for generation of data on Figure \ref{ci_sim} are
\begin{displaymath}
P_1^{ZCN13, 20130328} = 714, \; P_i = P_{i-1} + \Delta P_i, \; i = 2, \dots, N_{20130328}^{ZCN13} = 19611,
\end{displaymath}
\begin{displaymath}
\Delta P_i = -0.001937775 + 0.52755 \times NormalGenerator(\alpha_1 = 0, \; \mu_2 = 1),
\end{displaymath}
\begin{displaymath}
t_1 = 0, \; t_i = \textrm{int}(t_1 + 3.8548 i); \textrm{int truncates a number to a lower integer}.
\end{displaymath}
The chains of simulated b-increments and prices depend on a \textit{seed} of the generator. It was 21325476. Presenting the seed makes sense, if a generating algorithm is given. The 19611 points $(t_i, P_i)$ are plotted, Figure \ref{ci_sim} left top. While the price looks realistic for a chartist, the model cannot reproduce many important properties of the real a-b-c-process, Figure \ref{ci_b_ZCN13_20130328} left top: discreteness of prices and their increments, limit prices, distributions of b-increments changing in time within sessions, non-Gaussian properties of b-increments discussed later, volatility clusters. It ignores varying distributions of a-increments and implies \textit{independent} normal b-increments distributed identically.

\subsection{Computing the correlation integral}

The author wants to attract attention to the \textit{correlation integral} and \textit{dimension} computed for both time-series. For time series $x_1, ..., x_n$ the data is subdivided in chunks of size $m$: $(x_1, \dots, x_m), (x_{m+1}, \dots, x_{2m}), \dots$, where small number of extra points not matching $n$ is neglected. The $m$ is \textit{embedded dimension}. A norm, the distance between points, is computed for $\frac{N(N-1)}{2}$ uniques pairs. Then, the fraction  of pairs, where the distance is shorter than some $r$, is computed
\begin{displaymath}
C(N, m, r) = \frac{2}{N(N-1)}\sum_{i=2}^N\sum_{j=1}^i Indicator(|\vec{x_i},\vec{x_j}| < r).
\end{displaymath}
The indicator returns one, if the condition is true and zero otherwise. The $\lim_{N \rightarrow \infty} C(m, r)$ is referred to as the correlation integral. For small $r$ dependence $C(m, r)$ vs. $r$ is often a power law and the correlation dimension $\nu$ is
\begin{displaymath}
\nu = \lim_{r \rightarrow 0} \frac{\ln C(m, r)}{\ln r}.
\end{displaymath}
We are indebted for these notions and formulas to Peter Grassberger, Itamar Procaccia \cite{grassberger1983}, \cite{grassberger1983b}, Floris Takens. Evaluation of $\frac{N(N-1)}{2}$ distances is suitable for $N \approx 20000$ and $m \ge 1$. A ES session may get a half of million of ticks. This creates unmanageable 125 billions of unique pairs. James Theiler has invented the \textit{box-assisted algorithm} \cite{theiler1987} with the complexity reaching $O(N\log_2N)$. The author's C++ programs $gptd1$ and $gptd$ implement both possibilities. Chains of pseudo random uniform and normal numbers are evaluated up to 1,000,000 points ($gptd$). The slope well coincides with the embedded dimension. Changing variance shifts the curve in bi-logarithmic coordinates but does not change the slope. Real b-increments produce a lower correlation dimension than embedded one. The discreteness of b-increments creates horizontal plateaus of separated points complicating evaluation of $\nu$. On charts, the found dimensions are given for the steepest slopes. Figures \ref{ci_sim} and \ref{ci_b_ZCN13_20130328} emphasize additional property distinguishing pseudo random and real price increments.

\section{A Comment on the Limit Theorems}

Often, price changes are claimed to be sums of a number of \textit{hidden} random factors. Consequently, if the number tends to infinity, they may obey a Gaussian or another \textit{stable law} \cite{levy1934} \cite{khintchine1936} \cite{khintchine1937} \cite[p. 76, p. 86]{gnedenko1949}, \cite{mandelbrot1963}, \cite{fama1965}. The stable laws are a subclass of \textit{infinitely divisible probability distributions} important for the \textit{limit theorems} \cite{finetti} \cite[pp. 73 - 100]{gnedenko1949} \cite{mainardi2008}. An example of infinitely divisible not stable laws is distributions given by the \textit{incomplete gamma function} \cite[p. 13]{gnedenko1949}.

\subsection{"Non-Gaussian atoms"}

The a-b-c-classification implies that differences of two transaction prices are sums of the b- and c-increments, Equation \ref{EqPriceGlobal}. Times between transactions are sums of the a-increments and durations of the c-increments, Equation \ref{EqTimeGlobal}. For example, the difference of the last and first ZBM13 transaction prices on May 30, 2013 $P_{105351} - P_1 = 141.68750 - 141.71875 = -0.03125 = -\delta$ is the sum of the $N=105350$ b-increments. Their sample has mean $-2.97 \times 10^{-7} = -9.50 \times 10^{-6}\delta$, standard deviation $6.72 \times 10^{-3} = 0.215 \delta$, skewness $0.194$, and kurtosis $26.3$. Table \ref{nongaussian} presents the empirical probability mass distribution.

\begin{table}[!h]
  \topcaption{b-Increments of ZBM13, May 30, 2013}
  \begin{tabular}{cccccc}
  In $\delta$ & In StdDev & $m$ & $m/N$ & Gaussian, $p$ & $\frac{(m - Np)^2}{Np}$\\
  -7 & -32.6 & 2 & $1.90 \times 10^{-5}$ & $4.4 \times 10^{-201}$ & $8.6 \times 10^{195}$\\
  -6 & -27.9 & 2 & $1.90 \times 10^{-5}$ & $1.2 \times 10^{-144}$ & $3.2 \times 10^{139}$ \\
  -5 & -23.3 & 2 & $1.90 \times 10^{-5}$ & $1.4 \times 10^{-97}$ & $2.7 \times 10^{92}$ \\
  -4 & -18.6 & 1 & $9.49 \times 10^{-6}$ & $7.0 \times 10^{-60}$ & $1.4 \times 10^{54}$ \\
  -3 & -14.0 & 14 & $1.33 \times 10^{-4}$ & $1.5 \times 10^{-31}$  & $1.2 \times 10^{28}$ \\
  -2 & -9.30 & 59 & $5.60 \times 10^{-4}$ & $1.5 \times 10^{-12}$ & $2.2 \times 10^{10}$ \\
  -1 & -4.65 & 1808 & 0.0172 & 0.010 & 540 \\
   0 & 0 & 101598 & 0.964 & 0.98 & 26.2 \\
  1 & 4.65 & 1770 & 0.0168 & 0.010 & 487 \\
  2 & 9.30 & 65 & $6.17 \times 10^{-4}$ & $1.5 \times 10^{-12}$ & $2.7 \times 10^{10}$  \\
  3 & 14.0 & 18 & $1.71 \times 10^{-4}$ & $1.5 \times 10^{-31}$  & $2.1 \times 10^{28}$  \\
  4 & 18.6 & 7 & $6.64 \times 10^{-5}$ & $7.0 \times 10^{-60}$  & $6.6 \times 10^{55}$  \\
  5 & 23.3 & 1 & $9.49 \times 10^{-6}$ & $1.4 \times 10^{-97}$  & $6.8 \times 10^{91}$  \\
  6 & 27.9 & 2 & $1.90 \times 10^{-5}$ & $1.2 \times 10^{-144}$  & $3.2 \times 10^{139}$  \\
  7 & 32.6 & 0 & 0 & $4.4 \times 10^{-201}$ & $4.6 \times 10^{196}$   \\
  8 & 37.2 & 1 & $9.49 \times 10^{-6}$ & $6.6 \times 10^{-267}$ & $1.4 \times 10^{261}$ \\
 & $\sum$ & 105350 & 1 & 1 & $1.4 \times 10^{261}$ \\
  \end{tabular}
  \label{nongaussian}
\end{table}
The hypothetic Gaussian $(\mu = -9.5 \times 10^{-6}\delta, \; \sigma = 0.215 \delta)$ probabilities are for the unit intervals, like $[-7.5, -6.5]$, with the centers from the first column. The distribution is quite symmetrical. The kurtosis and astronomical summands in the last column evaluated for the $\chi^2$ test demonstrate absurdity of a Gaussian hypothesis for these b-increments, see also \cite[p. 35]{salov2011}.

\begin{figure}[h!]
  \centering
  \includegraphics[width=130mm]{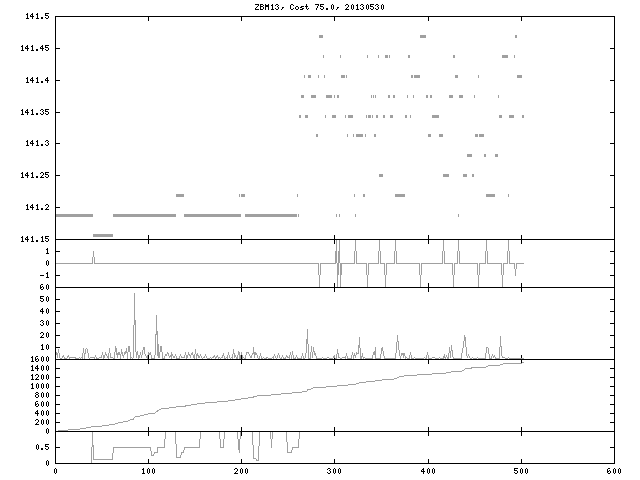}
  \caption[noniid]
   {ZBM13 trading on Thursday May 30, 2013. Price, MPS with filtering cost \$75, Volume, Accumulated Volume, Arrival Speed  vs. transaction index (non-proportional time). Two consecutive intervals 07:28:39 - 07:30:00 and one second starting at 07:30:01 contain 262 and 241 ticks but look very differently.}
  \label{noniid}
\end{figure}

The b-increments constituting intra-session price changes are not hidden. They are not mathematical abstractions but empirical indivisible "non-Gaussian atoms" limiting the range of \textit{self-similarity} \cite{mandelbrot2004} and infinite divisibility.

\subsection{Wisdom from the first source}

A finite variance of i.i.d. variables guarantees approaching a Gaussian sum for non-Gaussian components. A varying number of summands $N_j - 1$ compromises comparison of the sums. A violation of the i.i.d. property can be even a more serious obstacle for finding a limit distribution. The classics' opinion is \cite[p. 13]{gnedenko1949} (author's translation from the Russian Edition): \textit{"If the assumption about identity of the distribution laws of random variables in one series is refused, then the task to find possible limit law $V(z)$ ... becomes contentless: the limit law can be absolutely arbitrary. ... The requirement $m_n \to \infty$ has an illusive meaning: it does not prevent, for example, that a single summand $\xi_{nk}$ could play a dominating role."} Figure \ref{noniid} exhibits two different behaviors answering on \textit{"I.I.D., or not I.I.D, that is the question."} The sample statistics of b-increments, left|right, are size $N_1 = 261$ | $N_2 = 240$, mean 0$\delta$|0$\delta$, standard deviation 0.196$\delta$|1.75$\delta$, skewness 0|-0.0984, kurtosis 26.3|8.34, ticks 262|241, elapsed seconds 82|1. In the "chain reaction" the standard deviation is nine times greater, kurtosis is three times less, duration getting almost the same number of transactions is 82 times shorter.

\begin{table}[!h]
  \topcaption{b-Increments of ZBM13, May 30, 2013, 07:28:39 - 07:30:00 and 07:30:01}
  \begin{tabular}{ccccccc}
  In $\delta$ & $m_1$ & $m_2$ & Gaussian, $p_1$ & Gaussian, $p_2$ & $\frac{(m_1 - N_1p_1)^2}{N_1p_1}$ & $\frac{(m_2 - N_2p_2)^2}{N_2p_2}$\\
  -7 & 0 & 2 & $1.82 \times 10^{-241}$ & $9.28 \times 10^{-5}$ & $4.75 \times 10^{-239}$ & 176 \\
  -6 & 0 & 2 & $1.46 \times 10^{-173}$ & $7.35 \times 10^{-4}$ & $3.81 \times 10^{-171}$ & 18.9 \\
  -5 & 0 & 2 & $5.96 \times 10^{-117}$ & $4.23 \times 10^{-3}$ & $1.56 \times 10^{-114}$ & 0.955 \\
  -4 & 0 & 1 & $1.27 \times 10^{-71}$ & $1.77 \times 10^{-2}$ & $3.31 \times 10^{-69}$ & 2.48 \\
  -3 & 0 & 8 & $1.46 \times 10^{-37}$ & $5.38 \times 10^{-2}$  & $3.81 \times 10^{-35}$ & 1.87 \\
  -2 & 0 & 12 & $9.81 \times 10^{-15}$ & 0.119 & $2.56 \times 10^{-12}$ & 9.60 \\
  -1 & 5 & 17 & 0.00537 & 0.192 & 9.24 & 18.4 \\
   0 & 251 & 154 & 0.989 & 0.225 & 0.197 & 185 \\
  1 & 5 & 13 & 0.00537 & 0.192 & 9.24 & 23.7 \\
  2 & 0 & 11 & $9.81 \times 10^{-15}$ & 0.119 & $2.56 \times 10^{-12}$ & 10.8 \\
  3 & 0 & 10 & $1.46 \times 10^{-37}$ & $5.38 \times 10^{-2}$  & $3.81 \times 10^{-35}$ & 0.657  \\
  4 & 0 & 5 & $1.27 \times 10^{-71}$ & $1.77 \times 10^{-2}$  & $3.31 \times 10^{-69}$ & 0.133  \\
  5 & 0 & 0 & $5.96 \times 10^{-117}$ & $4.23 \times 10^{-3}$  & $1.56 \times 10^{-114}$ & 1.02 \\
  6 & 0 & 2 & $1.46 \times 10^{-173}$ & $7.35 \times 10^{-4}$  & $3.81 \times 10^{-171}$ & 18.9  \\
  7 & 0 & 0 & $1.82 \times 10^{-241}$ & $9.28 \times 10^{-5}$ & $4.75 \times 10^{-239}$ & 0.0223 \\
  8 & 0 & 1 & $1.16 \times 10^{-320}$ & $8.51 \times 10^{-6}$ & $3.03 \times 10^{-318}$ & 488 \\
  $\sum$ & 261 & 240 & 1 & 1 & 18.7 & 956
  \end{tabular}
  \label{chain}
\end{table}
The value $18.7 < \chi^2_{1-0.01}(f = 16 - 3 = 13) = 27.688 < 956$ does not allow rejecting the Gaussian hypothesis for the left side and does not allow accepting it for the right side distribution. It is interesting that the sample kurtosis 26.3 for the left side deviates more from the Gaussian 3 than 8.34 for the right side.

Conclusions about the limit theorems cannot be made mechanically. It is needed to study the rate of convergence and pay attention to variation of distributions of b-increments within a range/session, dominating price fluctuations, presence of up and/or down limit prices, and possible price dependencies.

\section{A Comment on Discrete Distributions}

The multinomial distribution assumes fixed $K > 2$ events with the probabilities $p_1, \dots, p_K$, where $\sum_{i=1}^K p_i=1$. It generalizes the binomial distribution $p_1, p_2 = 1 - p_1$. For a $s$th session, $K^s = \textrm{b-increment}_{max}^s - \textrm{b-increment}_{min}^s + 1$, where b-increments are expressed in $\delta$, and empirical frequencies approximate $p_i^s$. Since b-increments can be negative, zero, or positive, $K^s = K^{-,s} + K^{0,s} + K^{+,s}$, where $K^{0,s} = 1$, if $K^{-,s} > 0$ and $K^{+,s} > 0$. Prior opening a session with the limit $\Delta P_{lim}^s > 0$ and previous settlement price $P_{settle}^{s-1} > \Delta P_{lim}^s$,
\begin{displaymath}
P_{lim \; up}^s = P_{settle}^{s-1} + \Delta P_{lim}^s, \; P_{lim \; down}^s = P_{settle}^{s-1} - \Delta P_{lim}^s, \; K_{max}^s=\frac{2\Delta P_{lim}^s}{\delta} + 1,
\end{displaymath}
\begin{equation}
\label{EqMultinomialK}
 K_{max}^{-,s}=K_{max}^{+,s}=\frac{\Delta P_{lim}^s}{\delta}, \; K_{max}^s= K_{max}^{-,s} + 1 + K_{max}^{+,s}.
\end{equation}
Usually, $K^s \ll K_{max}^s, \; K^{-,s} \ll K_{max}^{-,s}, \; K^{+,s} \ll K_{max}^{+,s}$. After the opening,
\begin{displaymath}
K_{min,i}^{-,s,r}=\frac{P_i^{s,r} - P_{lim \; down}}{\delta}, \; K_{max,i}^{+,s,r}=\frac{P_{lim \; up} - P_i^{s,r}}{\delta},
\end{displaymath}
\begin{equation}
\label{EqMultinomialKi}
K_{max}^s=K_{max,i}^{-,s,r} + \; K_{max,i}^{+,s,r} + 1.
\end{equation}
The limit $\Delta P_{lim}^s$ implies the theoretical limits: $\textrm{b-increment}_{max,i}^{s,r} = K_{max,i}^{+,s,r}$ and $\textrm{b-increment}_{min}^{s,r} = -K_{min,i}^{-,s,r}$. These are equal to $\frac{2\Delta P_{lim}^s}{\delta}$ or $-\frac{2\Delta P_{lim}^s}{\delta}$ for the current down or up limit price. The author has not seen prices gapping from the down to up limit or vice versa within a session. The open price gapping to the limit occasionally takes place. The Pork Bellies had been famous for making several limit sessions in a row.

For futures without the limit, theoretically the next $\textrm{b-increment}_{min,i}^{s,r} = -\frac{P_i^{s,r}}{\delta}$ and $\textrm{b-increment}_{max,i}^{s,r}$ is unlimited. The minimum and maximum b-increments together with the numbers of occurrences in ranges and sessions are in Table \ref{b-increments}. By absolute values, all are less than the theoretical limits and $K^s$ varies from session to session. While the distributions are almost symmetrical, $K^{-,s}$ is rarely exactly equal to $K^{+,s}$. The extreme value occurring just a few times with $N_s$ counted in thousands insignificantly influences on the skewness. They are more critical for the risk of triggering a \textit{stop loss order}, because related frequencies  are not negligible like in a Gaussian distribution. A multinomial distribution is not interesting unless the $K^s$, $K^{-,s}$, $K^{+,s}$ are allowed to be random variables governed by another distribution type and combined with a binary random variable selecting the negative or positive sign of the increment. An alternative to this is finding one distribution responsible for the absolute values of b-increments including their extreme values and combining it with a variable selecting the sign.

\subsection{Zipf-Mandelbrot, Riemann and Hurwitz Zeta distributions}

The author has reviewed the Zipf-Mandelbrot $Q>0 \wedge S > 0$ \cite[pp. 198 - 218]{mandelbrot1997}, where the Zipf case is $Q=0$,
\begin{displaymath}
PDF_{ZM}(k) = \frac{(k + Q)^{-S}}{\sum_{i=1}^N(i + Q)^{-S}}, \; \textrm{the rank } k = 1, \; \dots, \; N,
\end{displaymath}
Riemann zeta \cite[p. 35]{khintchine1938}, \cite[p. 82]{gnedenko1949}, \cite[p. 821, Eq. 8]{lin2001}, \cite[related results]{biane2001}
\begin{displaymath}
PDF_R(k) = \frac{k^{-S}}{\sum_{i=1}^{\infty}i^{-S}}=\frac{k^{-S}}{\zeta(S)}, \; k \in \textrm{N} \wedge S \in R \wedge S > 1,
\end{displaymath}
Hurwitz zeta \cite[related results]{duncan1987}, \cite{hu2006}
\begin{displaymath}
PDF_H(k) = \frac{(k+Q)^{-S}}{\sum_{i=0}^{\infty}(i+Q)^{-S}}=\frac{(k+Q)^{-S}}{\zeta(S,Q)}, \; k \in \textrm{N}_0 \wedge S, Q \in R \wedge S > 1 \wedge Q > 0,
\end{displaymath}
and power law \cite[p. 29]{arnold2004}, \cite[p. 30]{mandelbrot1997} distributions. It is interesting how simple words can trigger a research. The words of Chung Kai Lai \cite[p. 259]{chung2001}: \textit{So far as known, this famous relationship between two "big names" has produced no important issue} - initiated \cite{lin2001} and \cite{hu2006}. The author should recognize that his research on the maximum profit strategy has been triggered by the words of Robert Pardo \cite[p. 125]{pardo1992}: \textit{The measurement of the potential profit that a market offers is not a widely understood idea}, \cite[Preface]{salov2007}.

The Zipf-Mandelbrot law assumes the maximum rank $N$. While it can be set big, this is inconvenient because the upper bound of absolute b-increments can be unknown. The Riemann zeta distribution is less flexible than the Hurwitz zeta distribution. The latter is crafted so that it can start from rank zero, and there is plenty of zero b-increments. All equations imply
\begin{displaymath}
\ln PDF_*(k) = -S \ln(k+Q) - \ln C^*, \; \textrm{where * is ZM, R, H, or P - power law}.
\end{displaymath}
This is an equation of a straight line with the points $(x=\ln(k+Q) \approx \ln(k), y=\ln PDF_*(k))$, if $Q \rightarrow 0$ or $k \gg Q$. The latter cannot be guaranteed for ZM, where $k \le N$. The H looks the most flexible from this group. Such lines express power laws $y=Cx^a$. Vladimir Arnold recollects (the author's traslation from Russian): \textit{From the stories of eye-witnesses, I know that Kolmogorov's similarity laws in the theory of turbulences have been obtained by him not from consideration of dimensions (used for their explanation today) but due to covering the floors of the summer house in Komarovka by paper sheets with thousands of experimental data"} and earlier \textit{"... my results are not proved (VS: mathematically) but correct, and this is more important"} \cite[p. 29]{arnold2004}. Komarovka, a small Russian village outside Moscow, was in that time a Mecca for mathematicians from all over the world.
\begin{figure}[h!]
  \centering
  \includegraphics[width=130mm]{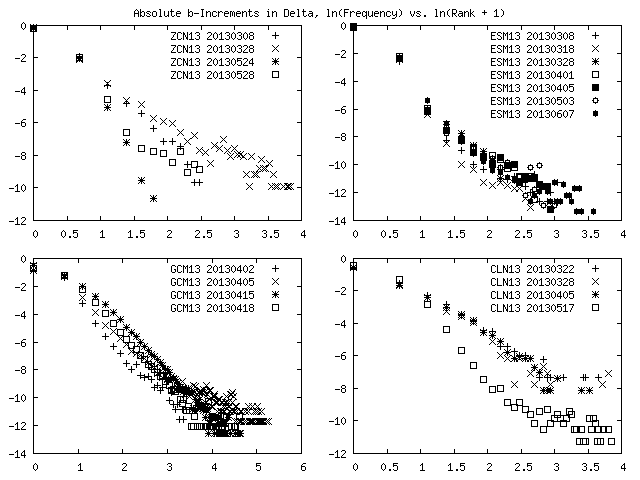}
  \caption[noniid]
   {Bi-logarithmic dependencies of the frequencies of absolute b-increments expressed in $\delta$ for corn, E-mini, gold, and crude oil futures traded in March - June 2013.}
  \label{abs_wb_distr}
\end{figure}

The points on Figure \ref{abs_wb_distr} accumulate large numbers of b-increments. For ES-mini it achieves 614991 on June 7, 2013. The lines for ZCN13 and CLN13 split between sessions. The points for ESM13 and GCM13 are closer to one approximating line. The "straight" lines have a tendency to bent at larger ranks meaning that frequencies are greater than predicted. The approximating lines underestimate risk but are better than the Gaussian distribution. An eye suggests that a parabola is more suitable than a straight line.
\begin{figure}[h!]
  \centering
  \includegraphics[width=130mm]{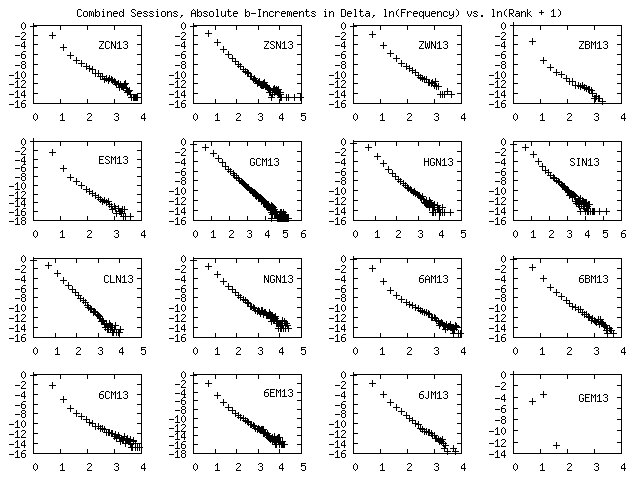}
  \caption[noniid]
   {Bi-logarithmic dependencies of the frequencies of absolute b-increments expressed in $\delta$ for contracts traded in March - July 2013. The increments are combined from all sessions.}
  \label{abs_wb_all_distr_4_4}
\end{figure}

Combining b-increments from sessions prior computing EPDF creates smoother plots, Figure \ref{abs_wb_all_distr_4_4}. The number of ticks used for the plot of ESM13 is equal to 27,438,059. This is greater than the number in Table \ref{b-increments} because the last extra session on June 21, 2013 is included. On these plots the initial artificial single zero increments are added. Their number is negligible and equal to the number of sessions. Also, ci-increments between ranges within a session are treated as b-increments. This cannot create a principal difference. The GEM13 significantly differs from other plots. The NGN13 b-increments expressed in $\delta$ are divided by 10 comparing with Table \ref{b-increments}. For this contract the minimal change in reported quotes is found equal to 0.01, while the official $\delta = 0.001$. Parabolas can better than straight lines approximate these data. While the plots indicate reasonable dependencies, the author could not find suitable $Q$ and $S$ for b-increment frequencies combined from all sessions so that the result would satisfy the Pearson goodness of fit criterion. For b-increments obtained from a single session this is possible. Particularly, the one second ZBM13 May 20, 2013 data from Table  \ref{chain} are well fitted by the Hurwitz Zeta distribution Table \ref{hurwitzzetaZBM13}. The symmetric negative and positive $\delta$ are combined in one class. The number of degrees of freedom $f = 9 - 1 - 2 = 6$. The corresponding $\chi^2(6, 0.02) = 15.033$, $\chi^2(6, 0.01) = 16.812$, $\chi^2(6, 0.001) = 22.457$ \cite{korn1968} are greater than the experimental 13.215.

\begin{table}[!h]
  \topcaption{b-Increments, ZBM13, May 30, 2013, Hurwitz Zeta Distribution with $S =2.385873201, Q =$1.510384234.}
  \begin{tabular}{cccccc}
  $|\delta|$ & $m$ & $\frac{m}{\sum m}$ & $p = PDF_{HZ}(|\delta|)$ & $Np=p\sum m$ & $\frac{(m - Np)^2}{Np}$\\
0 & 154 & 0.641666667 & 0.584630058 & 140.3112139 & 1.335480326 \\
1 & 30 & 0.125 & 0.173952889 & 41.74869328 & 3.306254231 \\
2 & 23 & 0.095833333 & 0.078165303 & 18.7596727 & 0.958458917 \\
3 & 18 & 0.075 & 0.042982386 & 10.31577261 & 5.723987218 \\
4 & 6 & 0.025 & 0.026656006 & 6.397441496 & 0.024691081 \\
5 & 2 & 0.008333333 & 0.017906011 & 4.297442571 & 1.228228715 \\
6 & 4 & 0.016666667 & 0.012733336 & 3.056000653 & 0.291601628 \\
7 & 2 & 0.008333333 & 0.009449749 & 2.267939762 & 0.031655037 \\
8 & 1 & 0.004166667 & 0.007249441 & 1.739865873 & 0.314622821 \\
$\sum$ & 240 & 1 & 0.953725178 & 228.8940428 & 13.21497997 \\
  \end{tabular}
  \label{hurwitzzetaZBM13}
\end{table}

In our case the Riemann and Hurwitz zeta functions should be evaluated for real arguments $S > 1$ and $(S > 1, Q > 0)$. This task is simpler than evaluation of the Riemann Zeta function of a complex argument $\sigma + t\sqrt{-1}; \; \sigma, t \in R$. For $\sigma = \frac{1}{2}$ this has become a competition. Finding 1,500,000,001 values of $t$ \cite{lune1986}, where $\zeta(\frac{1}{2} + t \sqrt{-1}) = 0$, cannot prove the Riemann Hypothesis \cite{bombieri2000} but contributes into computer science. The author has written the C++ functions RiemannZeta and HurwitzZeta and exported them from an XLL, a form of dynamically linked library DLL applied as Add-In for Microsoft Excel \cite{microsoft1997}. This helps to use the Microsoft Solver and Goal Seek in order to optimize the parameters $S$ and $Q$ under the constraints $S > 1.001$ and $Q > 0.001$. The cost function for Table \ref{hurwitzzetaZBM13} is the experimental $\chi^2$ with the nine classes.

\subsection{Euler-Maclaurin formula for Hurwitz Zeta}

The chosen computational methods are based on the \textit{Euler-Maclaurin summation} \cite[pp. 114 - 117]{edwards2001}. The \textit{Bernoulli numbers} are taken from \cite[p. 810]{abramowitz1972}. The derivations are lengthy and the author presents the final formula only for the Hurwitz zeta function, which he could not find in literature. However, the idea is the same as in \cite{edwards2001} for the Riemann zeta. Since convergence is slow for the direct sum, the Euler-Maclaurin summation is applied to the difference
\begin{displaymath}
\zeta(S,Q) - \sum_{i=0}^{N-1} (i + Q)^{-S} = \sum_{i=N}^{\infty}(i + Q)^{-S},
\end{displaymath}
\begin{displaymath}
\zeta(S,Q) = \sum_{i=0}^{N-1} (i + Q)^{-S} + \frac{(N+Q)^{1-S}}{S-1}+\frac{1}{2(N+Q)^{-S}} +
\end{displaymath}
\begin{displaymath}
+ \sum_{k=1}^M \frac{B_{2k}(N+Q)^{1-S-2k} \prod_{j=0}^{2k-2}(S + j)}{(2k)!}+ E(S, Q, N, M),
\end{displaymath}
where $B_{2k}$ are the Bernoulli numbers and $E(S,Q,N,M)$ is the error term. Using the estimates of Harold Edwards \cite{edwards2001} and Linas Vep\v{s}tas \cite{vepstas2008}, the $N=20$ and $M=13$ are selected to ensure 16 decimal digits of accuracy supported by the C++ built-in type \textit{double} for the values of $S$ hinted by Figure \ref{abs_wb_all_distr_4_4}. The Hurwitz Zeta distribution is perspective for describing distributions of b-increments without an attempt to combine multiple sessions or in smaller ranges. Generalization is likely prevented by the fact that the distribution changes in time, even, within a range/session.

\section{A Comment on Parabolic Fractals}

Arnold mentions that in many cases power laws remain experimental facts \cite[pp. 36 - 41]{arnold2004} and searching for asymptotic behavior and logarithmic corrections can provide theoretical explanations. He applies the Kolmogorov's technique to smoothed mean minimal periods of remainders obtained after division of powers of two by odd numbers and finds an interesting bi-logarithmic dependence \cite[p. 39, Figure 1]{arnold2004}. His seven examples from botanics, literature, medicine, volcano activity, genetics, number of scientific publications, graph theory related to compact arrangement of elements in space important for computer science include the Olof Arrhenius Law: the number of species in a district is proportional to a power of its area. The author adds the first name to distinguish the son and his father - Svante Arrhenius (Nobel Prize in Chemistry 1903 for \textit{"... electrolytic theory of dissociation"}), who's equation for the temperature dependence of the constant of the chemical reaction rate is also known under the name "Arrhenius Law". It can be added to the Arnold's list due to good straight lines on plots of logarithm of the constant vs. the reciprocal temperature in \textit{Kelvin degrees}.

\subsection{Plotting Olof Arrhenius's data}

\begin{figure}[h!]
  \centering
  \includegraphics[width=130mm]{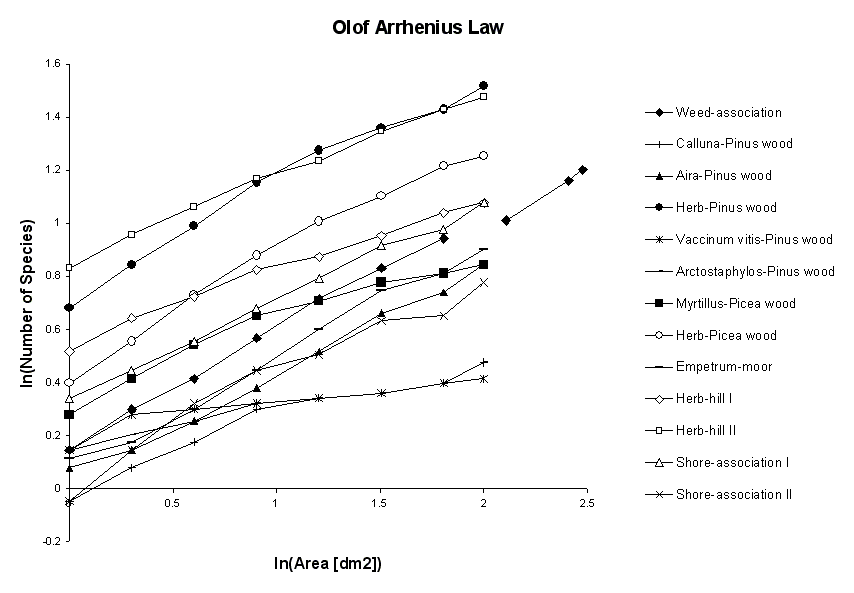}
  \caption[noniid]
   {The data from \cite{arrhenius1921} entered and plotted in the Microsoft Excel.}
  \label{arrhenius}
\end{figure}

The author has reviewed the article \cite{arrhenius1921} and entered 106 pairs (area in decimeter$^2$, number of species) \cite[p. 96, Table]{arrhenius1921} into the Microsoft Excel. For the first time we \textit{see} the Olof's results on Figure \ref{arrhenius}. My eye sees: 1) pieces of parabolas would be better for Calluna-Pinus wood, Herb-Pinus wood, Myrtillus-Picea wood, Herb-Picea wod, Herb-hill II, and Shore-association II; 2) Vaccinum vitis-Pinus wood has a big outlier. In the original Table, the 13 associations are accompanied by 10 - 30 percents deviations between experimental and computed data for greater areas: upper two - three from eight - ten observations. Arrhenius explains: \textit{"It is easily seen that the values calculated and observed agree very well. Generally there is an increase in the deviation corresponding to increasing area. This depends on the fact that the values of the smaller areas are the average of a greater number of observations than those of the larger"}.
\subsection{Parabola's shortcomings}

Deviations from a straight line in log-log coordinates on frequencies vs. ranks plots got a collective name \textit{parabolic fractal}. The so-called \textit{King Effect} relates to the highest frequency rank outlier. A common example is the town-size relationship, where in France Paris deviates from the curve. Several phenomena are claimed to follow to the parabolic fractal: galactic intensities, the distribution of town-sizes, spoken languages, species, and hydrocarbons accumulations by petroleum system \url{http://www.hubbertpeak.com/laherrere/fractal.htm}. They extend the Arnold's list. The cited reference \cite{deheuvels1995} proves that \textit{"the set of points where exceptional oscillations of empirical and related processes occur infinitely often is a random fractal"} and suggests how to evaluate its \textit{Hausdorff dimension}. Figures \ref{abs_wb_distr}, \ref{abs_wb_all_distr_4_4}, and \ref{arrhenius} demonstrate that parts of parabolas can be a better choice than a straight line. However, a parabola has a drawback: in many cases it cannot extrapolate far outside of the observed interval without violation of a natural monotonicity. More data is needed to confirm or reject the parabolic fractal effects in economics. The market ticks are valuable to clarify it for financial time-series.

\section{Extreme b-Increments}

\begin{figure}[h!]
  \centering
  \includegraphics[width=130mm]{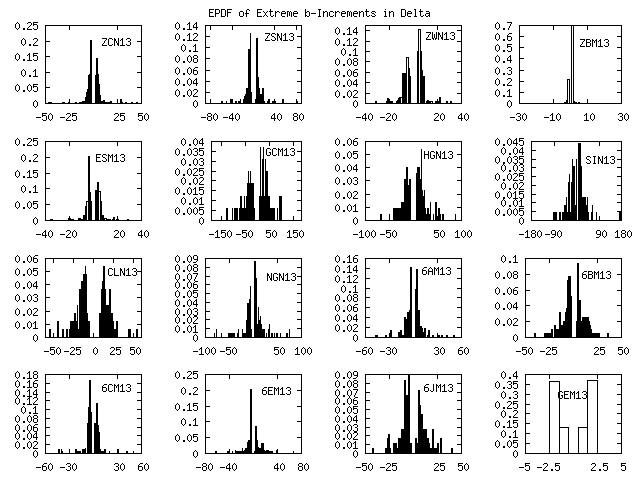}
  \caption[extreme_wb_freq_4_4]
   {Frequencies of extreme b-increments expressed in $\delta$ for contracts traded in March - July 2013.}
  \label{extreme_wb_freq_4_4}
\end{figure}

\begin{figure}[h!]
  \centering
  \includegraphics[width=130mm]{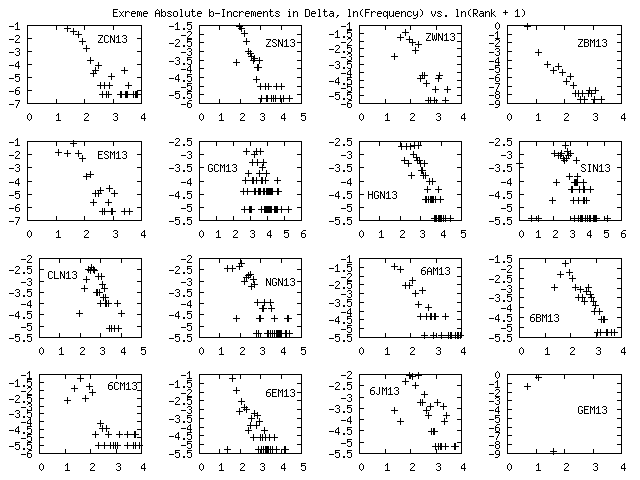}
  \caption[extreme_abs_wb_all_distr_4_4]
   {Bi-logarithmic plots of absolute extreme b-increments frequencies vs. ranks expressed in $\delta$ for contracts traded in March - July 2013.}
  \label{extreme_abs_wb_all_distr_4_4}
\end{figure}

In Table \ref{b-increments}, columns Min, $n_{min}$, Max, and $n_{max}$ present extreme b-increments and numbers of their occurrences within ranges and sessions. For each contract, the values are combined in a sample, where $\textrm{b-increment}_{min}^s$ and $\textrm{b-increment}_{max}^s$ are taken $n_{min}^s$ and $n_{max}^s$ times. The values are extracted from the session rows with dates. EPDFs are evaluated, Figure \ref{extreme_wb_freq_4_4}. Again, the ranks for NGN13 are divided by 10 prior plotting. The same extreme b-increments taken by absolute value are plotted in bi-logarithmic coordinates, Figure \ref{extreme_abs_wb_all_distr_4_4}.

While the chances to get $0\delta$ and $\pm1\delta$ b-increments are the highest, Figures \ref{abs_wb_distr}, \ref{abs_wb_all_distr_4_4}, a chance to get them as extremes in a session is negligible, Figure \ref{extreme_wb_freq_4_4}. Even, if we draw straight lines above the clouds of points on Figure \ref{extreme_abs_wb_all_distr_4_4}, it would be wrong to extrapolate them to the left. The ZSN13, ZWN13, GCM13, SIN13, CLN13, NGN13, 6BM13, 6CM13, 6EM13, and 6JM13 on Figure \ref{extreme_abs_wb_all_distr_4_4} confirm that frequencies of absolute extreme b-increments have a maximum. Others confirm it indirectly: there were no sessions of ZCN13, ZBM13, ESM13, HGN13, and 6AM13 in March - July, 2013, with the extreme $0\delta$ and $\pm1\delta$ b-increments.

\subsection{Fr\'{e}chet, Fisher, Tippett, von Mises, Gnedenko, Gumbel, Haan}

\begin{figure}[h!]
  \centering
  \includegraphics[width=130mm]{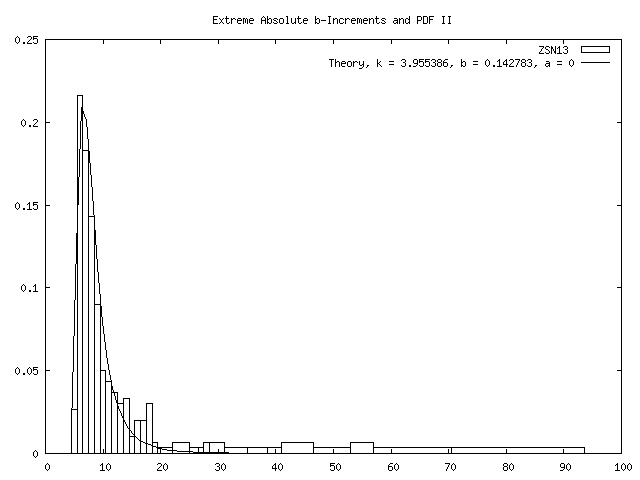}
  \caption[ftg]
   {Plots of absolute extreme b-increments frequencies vs. ranks expressed in $|\delta|$ and the approximating scaled and shifted Fr\'{e}chet-Fisher-Tippett-Gnedenko-Type-2-Gumbel PDF, $PDF^\textrm{II}$, for ZSN13, March - July 2013.}
  \label{ftg}
\end{figure}

The modern theory of extreme values is influenced by \cite{fisher1928}, \cite{mises1936}, \cite{gnedenko1943}, \cite{balkema1974}. Fisher and Tippett have presented three extreme limit distributions based on the functional relation which they must satisfy. Mises has proved a sufficient condition for the weak convergence of the largest order statistics to each of the three types. Gnedenko has given a rigorous proof of the necessary and sufficient conditions for the weak convergence of the extreme order statistics. Haan has improved exposition of the Gnedenko's results. Differentiation of the CDFs of Gnedenko's, G, \cite[p. 423]{gnedenko1943} with respect to $x$ gives the PDFs of Fisher and Tippett, FT, \cite[pp. 211 - 212]{fisher1928}
\begin{displaymath}
\textrm{I.} \; PDF^{FT}(x) = e^{-x - e^{-x}} = \frac{d\Lambda^G(x)}{dx}, \; -\infty < x < \infty,
\end{displaymath}
\begin{displaymath}
\textrm{II.} \; PDF^{FT}(x) = \frac{k}{x^{k+1}}e^{-x^{-k}} = \frac{d\Phi_{\alpha}^G(x)}{dx}, \; x > 0, \; k = \alpha > 0,
\end{displaymath}
\begin{displaymath}
\textrm{III.} \; PDF^{FT}(x) = k(-x)^{k-1}e^{-(-x)^{k}} = \frac{d\Psi_{\alpha}^G(x)}{dx}, \; x < 0, \; k = \alpha \le 0.
\end{displaymath}
For a sample of combined absolute extreme b-increments the II might be useful. Maurice Fr\'{e}chet wrote about II in 1927 \cite{frechet1927}. It is used under his name too. By definition \cite[p. 45]{gnedenko1949}, the distribution functions $F_1(x)$ and $F_2(x)$ are of one type, if for $b > 0$ and $a$, $F_2(x)=F_1(bx+a)$. It is easy to see that $F_1(x)=\Phi_{\alpha=k}^G(x)=e^{-x^{-k}}, \; x > 0$ and $F_2(x)=e^{-(bx+a)^{-k}}, \; x > -\frac{a}{b}$ are valid CDFs, belonging to one type for $k > 0, \; b > 0$.  While changing the scale and origin of the coordinate system does not create a new type, we get a better fitting instrument
\begin{equation}
PDF^\textrm{ II}(x) = \frac{kb}{(bx+a)^{k+1}}e^{-(bx+a)^{-k}}, \; x > -\frac{a}{b}, \; k > 0, \; b > 0, \; a \ge 0.
\label{EqPDFII}
\end{equation}
Minimization of $\sum[PDF^\textrm{II}(|\delta\textrm{-size}|)-EPDF^{ZSN13}(|\delta\textrm{-size}|)]^2$ gives the solution $(k=3.955386, b = 0.142783, a=0)$, Figure \ref{ftg}. With the Microsoft Solver's constraint $a \ge 0$, the optimal $a=0$ is stably obtained for guesses $a > 0$. If $a=0$, then the PDF is the \textit{Type-2 Gumbel distribution} marking the contribution of Emil Gumbel \cite{gumbel1958}. The optimal $b \ne 1$ excludes the Fr\'{e}tchet's case. It is inconvenient to use a continuous PDF like Equation \ref{EqPDFII} for minimization of the Pearson's $\chi^2$ goodness of fit quantity: the fractional boundaries of classes would be far-fetched. Extreme and ordinary b-increments are discrete.

\subsection{We need a discrete distribution}

While the plots, Figure \ref{ftg}, look matching, there are concerns: 1) the theoretical density is continuous, 2) the frequency of large $|\delta|$ is underestimated. Indeed, the point $|\delta|=82$ is obtained with the frequency 0.003322 but the theoretical density is 0.00000286, 1162 times less. An infinite sequence of positive numbers, which mimic unimodal $PDF^\textrm{II}$, with the converging series is needed. Ideally, it should fit in the interval of integers $[0, 100]$. The author has checked the sequence $(PDF^{\textrm{II}}(n)), \; n \in \textrm{N}$, reusing its reciprocal series as the normalizing multiplier ensuring a valid discrete \textit{probability mass function}, PMF,
\begin{equation}
PMF^{\textrm{II}}(n) = \frac{\frac{kb}{(bn+a)^{k+1}}e^{-(bn+a)^{-k}}}{\sum_{i=1}^{\infty} \frac{kb}{(bi+a)^{k+1}}e^{-(bi+a)^{-k}}  }, \; n \in \textrm{N}, \; k > 0, \; b > 0, \; a \ge 0,
\label{EqExtremePMF}
\end{equation}
where $PMF^{\textrm{II}}(0)=0$. The denominator converges.  The \textit{Maclaurin-Cauchy integral test of convergence of series} \cite[p. 281, item 373]{fihtengoltz1970} proves it: the $PDF^{\textrm{II}}(x)$ is positive and monotonously decreasing for $x > n_0 > 0$, the integral $\int_{-\frac{a}{b}}^{\infty}PDF^{\textrm{II}}(x)dx = 1$, and, thus, converges for any lower bound $x \ge -\frac{a}{b}$.

Comparing the denominator in Equation \ref{EqExtremePMF} with the \textit{general Dirichlet's series} \cite[p. 1]{hardy1915} $f(s)=\sum_1^{\infty}a_ne^{-\lambda_ns}$, where $(\lambda_n)$ is a sequence of real increasing numbers whose limit is infinity, and $s = \sigma + t\sqrt{-1}$ is a complex variable whose real and imaginary parts are $\sigma$ and $t$, we notice that setting $\sigma = 1, \; t = 0, \; a_n = \frac{kb}{(bn+a)^{k+1}}, \; \lambda_n = (bn + 1)^{-k}$ yields $Denominator = f(1)$. However, our $(\lambda_n)$ is a sequence of real positive decreasing numbers whose limit is zero for $k > 0, \; b > 0, \; a \ge 0$ and $\lim_{n \rightarrow \infty}e^{-\lambda_n} = 1$. In the Gumbel's case $a = 0$, the $a_n=\frac{k}{b^{k}}\frac{1}{n^{k+1}} = \frac{\nu-1}{b^{\nu-1}}\frac{1}{n^{\nu}}, \; \nu = k + 1 > 1$ and $\sum_{n=n_0}^{\infty}a_n e^{-\lambda_n} \approx \frac{\nu-1}{b^{\nu-1}}\sum_{n=n_0}^{\infty}\frac{1}{n^\nu} = \frac{\nu-1}{b^{\nu-1}}\zeta_{n_0}(\nu)$ for large $n_0 \in \textrm{N}$, where $\zeta_{n_0}(\nu)$ denotes the remaining part of the Riemann's zeta function with the real argument greater than one - the domain of convergence. The generic case $a>0$ also ensures convergence because each summand remaining positive decreases. Being an alternative convergence proof, this consideration hints that it might be difficult to find an expression for the denominator and a numerical method is required. The Euler-Maclaurin summation is a candidate.

\subsection{The Euler-Maclaurin formula for the proposed distribution}

\begin{figure}[h!]
  \centering
  \includegraphics[width=130mm]{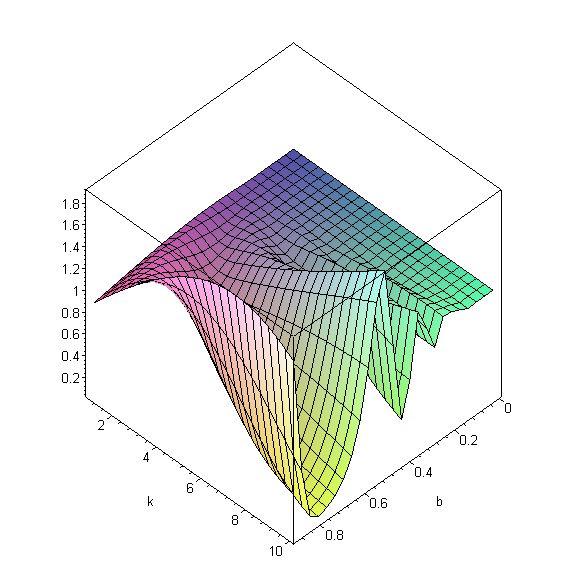}
  \caption[noniid]
   {Dependence of the extreme $PMF^{\textrm{II}}$ denominator Equation \ref{EqExtremePMF} expressed by Equation \ref{EqPMFDenominator} vs. $k$ and $b$, where $a = 0$ and $M=200$. Plot is done using Maple 10 from Maplesoft.}
  \label{PMFIIDenomiator}
\end{figure}

Similar to the Riemann and Hurwitz zeta, we directly evaluate the sum of terms from one to $M - 1$ and the remaining sum \cite[p. 106]{edwards2001}, adapted to Equation \ref{EqExtremePMF}, as the three summands
\begin{displaymath}
\sum_{n=M}^{\infty} PDF^{\textrm{II}}(n) \approx \int_M^{\infty} PDF^{\textrm{II}}(x)dx + \frac{PDF^{\textrm{II}}(M)}{2} + \sum_{j=1}^{m}\frac{B_{2j}}{(2j)!}PDF^{\textrm{II}(2j-1)}(x)\big |_M^{\infty},
\end{displaymath}
where top $(2j-1)$ is the derivative order and $B_{2j}$ are Bernoulli numbers. The error of this approximation is equal to
\begin{displaymath}
R_{2m}=\frac{1}{(2m+1)!}\int_M^{\infty}\bar{B}_{2m+1}(x)PDF^{\textrm{II}(2m+1)}(x)dx,
\end{displaymath}
where $\bar{B}_{2m+1}(x)=B_{2m+1}(x-\lfloor x \rfloor)$ is the Bernoulli polynomial of degree $2m+1$. The $\bar{B}_{2m+1}(x)$ alternates in sign. If $PDF^{\textrm{II}(2m+1)}(x)$ is monotonic on $[M, \infty)$, then evaluation of the error $|R_{2m}|$ requires to compute only the first omitted term: $|R_{2m}|$ does not exceed twice the absolute value of that term. The first summand and the sum of the first two summands are equal to
\begin{displaymath}
\int_M^{\infty}\frac{kbe^{-(bx+a)^{-k}}}{(bx+a)^{k+1}}dx = \int_{-(bM+a)^{-k}}^0 e^ydy=1-e^{-(bM+a)^{-k}},
\end{displaymath}
\begin{displaymath}\int_M^{\infty} PDF^{\textrm{II}}(x)dx + \frac{PDF^{\textrm{II}}(M)}{2} = 1 - e^{-(bM+a)^{-k}}\left(1 - \frac{kb}{2(bM+a)^{k+1}}\right).
\end{displaymath}
For the third term and $|R_{2m}|$ we need the derivatives up to $(2m+1)$. Let
\begin{displaymath}
f(x)=V(x)S(x)=f^{(0)}=V^{(0)}S^{(0)}, \; V^{(0)} = kbe^{-(bx+a)^{-k}}, \; S^{(0)} = (bx+a)^{-k-1}.
\end{displaymath}
The Leibniz's formula \cite[pp. 236 - 238]{fihtengoltz1962} gives $f^{(n)}=\sum_{i=0}^n C_i^n V^{(n-i)}S^{(i)}$. With $S^{(0)}$ defined above and $S^{(1)}=b(-k-1)(bx+a)^{-1}S^{(0)}$, we guess that
\begin{displaymath}
 S^{(n)}=b^n(bx+a)^{-n}S^{(0)}\prod_{i=1}^n(-k-i).
\end{displaymath}
For $n=0, \; 1$ it is valid. Let it be valid for $n > 1$. Then, for $n + 1$ the formula gives $S^{(n+1)}=b^{n+1}(bx+a)^{-n-1}S^{(0)}\prod_{i=1}^{n+1}(-k-i)$. However, differentiation of $f^{(n)}$ gives the same: $b^n\prod_{i=0}^n(-k-i)[-nb(bx+a)^{-n-1}S^{(0)} + (bx+a)^{-n} b(-k-1)(bx+a)^{-1}S^{(0)}]=b^{n+1}(bx+a)^{-n-1}S^{(0)}\prod_{i=1}^{n}(-k-i)(-k-(n + 1))$. This completes the mathematical induction proof for $S^{(n)}$. Getting a formula for $V^{(n)}$ is problematic: $V^{(1)}=kbf^{(0)}= kb V^{(0)}S^{(0)}$ and the Leibniz's formula recursively arises in the branches of $V^{(n-i)}$ on each step. The first three derivatives of $PDF^{\textrm{II}}(x)$ of the odd orders 1, 3, 5 are
\begin{equation}
PDF^{\textrm{II}(1)}(x)=PDF^{\textrm{II}}(x)\frac{b}{bx+a}\left(\frac{k}{(bx+a)^k} - k - 1\right),
\label{EqPDFIIDerivative1}
\end{equation}
\begin{equation}
\begin{split}
PDF^{\textrm{II}(3)}(x)=PDF^{\textrm{II}}(x)\left(\frac{b}{bx+a}\right)^3\left(-\frac{k^3}{(bx+a)^{3k}} +\right. \\
\left. +\frac{6(k^3+k^2)}{(bx+a)^{2k}}-\frac{7k^3+18k^2+11k}{(bx+a)^k}+k^3+6k^2+11k+6\right),
\label{EqPDFIIDerivative3}
\end{split}
\end{equation}
\begin{equation}
\begin{split}
PDF^{\textrm{II}(5)}(x)=PDF^{\textrm{II}}(x)\left(\frac{b}{bx+a}\right)^5
\left(\frac{k^5}{(bx+a)^{5k}}-\frac{15(k^5+k^4)}{(bx+a)^{4k}} + \right. \\
\left. + \frac{65k^5+150k^4+85k^3}{(bx+a)^{3k}}
-\frac{90k^5+375k^4+510k^3+225k^2}{(bx+a)^{2k}} + \right. \\
\left. + \frac{31k^5+225k^4+595k^3+675k^2+274k}{(bx+a)^k} + \right. \\
\left. -k^5-15k^4-85k^3-225k^2-274k-120\right).
\label{EqPDFIIDerivative5}
\end{split}
\end{equation}
Discussion of the regularities found in Equations \ref{EqPDFIIDerivative1} - \ref{EqPDFIIDerivative5} is omitted because they were insufficient to build a generic formula. The derivatives approach zero at $x \rightarrow \infty$. The three relevant Bernoulli numbers are $B_2 = \frac{1}{6}, \; B_4 = -\frac{1}{30}, \; B_6 = \frac{1}{42}$. With $M=200, \; k = 3.955386, \; b = 0.142783, \; a= 0$, the first three terms in the third summand are equal to $\frac{B_2}{2}PDF^{\textrm{II}(1)}(M)=-0.7130872262 \times 10^{-10}$, $\frac{B_4}{24}PDF^{\textrm{II}(3)}(M)=0.1230723154 \times 10^{-14}$, $\frac{B_6}{720}PDF^{\textrm{II}(5)}(M)=-0.5219062910 \times 10^{-19}$. Using only the first and third derivative under these conditions gives the error $\approx 10^{-19}$. For the values of the denominator in Equation \ref{EqExtremePMF} close to one this is better than the accuracy of modern eight bytes C++ built-in type double \cite[pp. 74 - 76, pp. 628 - 629]{stroustrup2000} keeping 16 decimal digits of mantis. The final formula approximating the denominator of the discrete $PMF^{\textrm{II}}(n)$ is
\begin{equation}
\begin{split}
\sum_{i=1}^{\infty}PDF^{\textrm{II}}(x) \approx \sum_{n=1}^{M-1} PDF^{\textrm{II}}(n)  + 1 - e^{-(bM+a)^{-k}} \left(1 - \frac{kb}{2(bM+a)^{k+1}}\right) + \\
+\frac{PDF^{\textrm{II}(1)}(M)}{12} - \frac{PDF^{\textrm{II}(3)}(M)}{720} + \frac{PDF^{\textrm{II}(5)}(M)}{30240}.
\label{EqPMFDenominator}
\end{split}
\end{equation}
The denominator expressed by Equation \ref{EqPMFDenominator} is not always close to one, Figure \ref{PMFIIDenomiator}. Evaluation of the enumerator in Equation \ref{EqExtremePMF} is straightforward.
With the $PMF^{\textrm{II}}(n)$, selection of the Pearson's $\chi^2$ classes' boundaries is natural, Table \ref{extremePMFIIZSN13chi2}.
\begin{table}[!h]
  \topcaption{Fitting Extreme b-Increments of ZSN13 with $PMF^{\textrm{II}}(|\delta|), \; k =2.50205129050786, \; b =0.145521989804209$; $a=0$ is fixed; $f = 10 - 1 - 2 = 7$.}
  \begin{tabular}{ccccc}
  $|\delta|$ & $\sum m_{|\delta|}$ & $p = \sum PMF^{\textrm{II}}(|\delta|)$ & $Np=p\sum m_{|\delta|}$ & $\frac{(\sum m - Np)^2}{Np}$\\
5, 6, 7 & 128 & 0.396078909 & 119.2197517 & 0.646644187 \\
8 & 43 & 0.107926455 & 32.48586293 & 3.402928794 \\
9 & 27 & 0.085068712 & 25.60568221 & 0.075925417 \\
10 & 15 & 0.066185687 & 19.92189185 & 1.215999943 \\
11 & 13 & 0.051504694 & 15.50291288 & 0.404090053 \\
12 & 11 & 0.040335941 & 12.14111814 & 0.107251292 \\
13 & 9 & 0.031877993 & 9.595276012 & 0.036929999 \\
14, 15 & 13 & 0.045979023 & 13.83968581 & 0.050945684 \\
16 - 29 & 32 & 0.096558428 & 29.06408697 & 0.296571688 \\
30 - 82 & 10 & 0.023772516 & 7.155527325 & 1.13073774 \\
Sum & 301 & 0.945288358 & 284.5317959 & 7.368024797 \\
  \end{tabular}
  \label{extremePMFIIZSN13chi2}
\end{table}
The $\chi^2(7, 0.05)=14.067$ is greater than the optimal 7.368 with the 10 classes. With the $\chi^2$-optimal $k, b, a=0$ we get $PMF^{\textrm{II}}(82) = 6.159 \times 10^{-5}$. This is only 54 times less than the experimental 0.003322. Since the denominator under these conditions is equal to 1.0000039587885819, the same density values can be obtained with the continuous $PDF^{\textrm{II}}(x)$. Using the latter is less convenient due to discreteness of $|\delta|$. It is likely that limitations of the extreme values theory, $PMF^{\textrm{II}}(n)$, and $PDF^{\textrm{II}}(x)$ are caused by violation of the theoretical assumptions such as  I.I.D. variables forming samples.

\section{A Second Comment on Discrete Distributions}

Conventional and computational discreteness requires discrete and lattice probability distributions. The Kolmogorov's foresight implies that the demand will grow. In the previous section, an existing continuous distribution $PDF^{\textrm{II}}(x)$ is converted into a discrete one. The transformation steps can be generalized: 1) evaluate existing $PDF(x)$ at integer arguments $n$; 2) apply the reciprocal sum of "all" $PDF(n)$ as a factor ensuring that $\sum_{n = M}^N \frac{PDF(n)}{\sum_{i=M}^N PDF(i)} = 1$, where $M$ and $N$ can be $-\infty$ and $\infty$; 3) establish convergence of the denominator, if the limits are $\pm \infty$. The latter step can be simple: PDFs are often integrable, positive, and monotonic functions on infinite subintervals. This supports the Euler-Cauchy integral test of convergence of series. The role of series summation algorithms increases also due to the moments $\alpha_m=\sum_{n=1}^{\infty}n^mPMF(n)$. The continuous parent and discrete child distributions relate each to other via the $PDF(x)$. A different relationship between the continuous normal and discrete binomial distributions is reminded by Stephen Stigler \cite{stigler2008}: \textit{"When we think of the normal approximation to the binomial, we usually think in terms of large samples. Pearson discovered that there is a sense in which the two distributions agree exactly for even the smallest number of trials. ... the normal density is characterized by the differential equation $\frac{f'(x)}{f(x)}=-\frac{x-\mu}{\sigma^2}$. Pearson discovered that $p(k)$, the probability function for the symmetric binomial distribution ($n$ independent trials, $p = 0.5$ each trial), satisfies the analogous difference equation exactly $2\frac{p(k+1)-p(k)}{p(k+1)+p(k)} = \frac{(k+\frac{1}{2}) - \frac{n}{2}}{(n+1)\frac{1}{2}\frac{1}{2}}$ for all $n, k$".}

\section{Last Minus First Price as the Sum of b-Increments}

If b-increments are random variables, then the difference between the last and first prices in a range is the sum of the random variables. The number of summands is also random due to the discussed properties of a-increments. Information about the increments $P_{N_{s,r}}^{s,r}-P_1^{s,r}$ expressed in $\delta$ can be evaluated from Table \ref{b-increments}. In order to compute this increment for a range multiply two values from the columns Size and Mean. The column Size contains the number of b-increments equal to $N_{s,r}-1$. The column Mean contains values rounded to five meaningful digits. For instance, for ZCN13 traded on March 1, 2013 in one range we get $10350 \times 0.00096618 = 9.999963$. This must be rounded off to the integer 10 to compensate the effect of the previous rounding. The difference $P_{10351}^{20130301 \; 13:59:57}-P_1^{20130228 \; 17:00:00} = P_{10351}^{20130301,1}-P_1^{20130301,1} = 686.50 - 684.00 = 2.5$ divided by $\delta_{ZCN13} = 0.25$ is equal to 10. If a session consists of more than one range, then $P_{N_{s}}^{s}-P_1^{s}$ get contributions from b- and ci-increments and cannot be computed from Table \ref{b-increments} only.

\subsection{Wald, Wolfowitz, Kolmogorov, Prokhorov}

Important theorems related to sums of random numbers of random variables are proved by Abraham Wald \cite{wald1944}, Jacob Wolfowitz \cite{wolfowitz1947}, Kolmogorov and Yuri Prokhorov \cite{kolmogorov1949}. If $\xi_1, \xi_2, \; \dots, \; \xi_n, \; \dots$ is an infinite sequence of I.I.D. random variables, $\zeta_{\nu}=\xi_1+\xi_2+ \; \dots \; + \xi_{\nu}$, the $\nu$, taking only non-negative integer values $\{0, 1, 2, 3, \; \dots\}$, is a random number of the first members of the sequence, the mathematical expectations $\textrm{E}\nu$, $\textrm{E}\xi_i=a$, and $\textrm{E}|\xi_i|=c$ are finite, and for $n>m$ the random variable $\xi_n$ and event $S_m = \{\nu=m\}$ are independent, then the \textit{Wald's identity} takes place: $\textrm{E}\zeta_{\nu}=\textrm{E}\nu\textrm{E}\xi$. Kolmogorov and Prokhorov prove a general result, where $\textrm{E}\xi_n = a_n, \; \textrm{E}|\xi_n|=c_n, \; \textrm{E}(\xi_n-a_n)^2=b_n$ and the Wald's identity is a particular case. Given the probabilities $p_n = P(S_n)=P(\{\nu=n\})$ and denoting $P_n=P(\{\nu \ge n\}) = \sum_{m=n}^{\infty}p_m$: $\textrm{E}\zeta_{\nu}=\sum_{n=1}^{\infty}p_nA_n, \; A_n=\textrm{E}\zeta_n = a_1 + a_2 + \dots + a_n$. The proof, applying the Abel transformation of series \cite[pp. 305 - 306]{fihtengoltz1970}, requires convergence of $\sum_{n=1}^{\infty}P_nc_n$. The absolute convergence is granted by this requirement because $P_n, \; c_n$ are non-negative. If $B_n=b_1+b_2+\dots+b_n$, then they prove that $\textrm{E}(\zeta_{\nu}-A_{\nu})^2=\sum_{n=1}^{\infty}p_nB_n$. In the article \cite{korshunov2009}, an analog of the Wald's identity is derived for a case, where one summand has the infinite mathematical expectation.

\subsection{Illustration of the Wald's identity}

The estimates
\begin{displaymath}
\textrm{E}\zeta_{\nu_{s,r}}=P_{N_{s,r}}^{s,r}-P_1^{s,r}, \; \textrm{E}\nu_{s,r} = N_{s,r} = \frac{t_{N_{s,r}}^{s,r}-t_1^{s,r}}{\overline{\textrm{a-increment}^{s,r}}} + 1 \approx \frac{T_{s,r}^c-T_{s,r}^o}{\overline{\textrm{a-increment}^{s,r}}}, \;
\end{displaymath}
\begin{displaymath}
\textrm{E}\xi_{s,r}=\overline{\textrm{b-increment}^{s,r}},
\end{displaymath}
being substituted into the Wald's identity, give
\begin{equation}
\begin{split}
P_{N_{s,r}}^{s,r}-P_1^{s,r} = \left(\frac{t_{N_{s,r}}^{s,r}-t_1^{s,r}}{\overline{\textrm{a-increment}^{s,r}}} + 1 \right ) \overline{\textrm{b-increment}^{s,r}} \approx \\
\approx \frac{(T_{s,r}^c-T_{s,r}^o) \overline{\textrm{b-increment}^{s,r}}}{\overline{\textrm{a-increment}^{s,r}}} = (T_{s,r}^c-T_{s,r}^o)\rho_{ba}^{s,r},
\end{split}
\label{EqABRatio}
\end{equation}
where $\rho_{ba}^{s,r}$ is the ratio of the mean b- to the mean a-increment. The exact relationship, involving the three estimates, follows from Equations \ref{EqTimeSession} and \ref{EqPriceSession}. The approximate version is supported by Figure \ref{a_m_nsr}. Let us illustrate the latter using ZCN13. On April 5, 2013 the mean a- and b-increments from Tables \ref{a-increments} and \ref{b-increments} are equal to $3.4886$ seconds and $-0.00018459\delta$. The price difference is $P_{21671}^{20130405 \; 13:59:59}-P_1^{20130404 \; 17:00:00}=617.50-618.50=-1=-4\delta$. The single range duration is $T_{20130405 \; 14:00:00}^c - T_{20130404 \; 17:00:00}^o = 75600$ seconds and $75600\frac{-0.00018459}{3.4886}=-4.0001731 \approx -4\delta$. On March 28, 2013 $75600\frac{-0.0077511}{3.8548} = -152.014 \approx -152\delta$ and the difference $P_{19611}^{20130328 \; 13:59:57}-P_1^{20130327 \; 17:00:00} = 676.00 - 714.00 = -38 = -152\delta$.

\subsection{Checking whether adjusted sums of b-increments are Gaussian}

The Wald-Wolfowitz-Kolmogorov-Prokhorov theorems uncover the mean and variance of $\zeta_{\nu}$. This is not yet a distribution unless it is one, like Gaussian, fully characterized by the two moments. Can we expect $\zeta_{\nu}$, being a sum of random variables with finite (?) variance, obeying a Gaussian distribution? In order to compensate the effect of random $\nu$ we should check the mean b-increment, which can be viewed as an adjusted last minus first price difference $\textrm{E}\xi=\frac{\textrm{E}\zeta_{\nu}}{\textrm{E}\nu}$. This also excludes c-contributions. Often, researchers test the differences of today and yesterday close or settlement prices. In the light of the a-b-c-classification, they ignore the random number of b- and fixed number of c- contributions. Such an approach is justified, if $\textrm{E}\nu_{s,r}$ remains constant in ranges and sessions. The mean b-increment is a cleaner candidate for such a test.

For each contract, the mean b-increments from ranges in Table \ref{b-increments} are combined in a sample. Then, the sample moments, ECDF, and EPDF are evaluated. In Table \ref{wco-increments} each Min and Max value has occurred one time and the corresponding columns are replaced with $U^-=\frac{Min - Mean}{StdDev}$ and $U^+=\frac{Max-Mean}{StdDev}$.
\begin{table}[!h]
  \centering
  \topcaption{Sample statistics of mean b-increments extracted from ranges.}
  \begin{tabular}{cccccccccc}
  Ticker & Size & Mean & Min & $U^-$ & Max & $U^+$ & StdDev & Skew & E-K\\
ZCN13 & 157 & 0.0053 & -0.026 & -1.2 & 0.22 & 8.2 & 0.027 & 5.6 & 35\\
ZSN13 & 157 & 0.0049 & -0.31 & -6.1 & 0.37 & 7.1 & 0.051 & 2.1 & 29\\
ZWN13 & 156 & 0.061 & -2 & -3.5 & 6 & 10 & 0.59 & 7.8 & 74\\
ZBM13 & 75 & -0.0019 & -0.29 & -6.1 & 0.24 & 5.2 & 0.047 & -1.34 & 28\\
ESM13 & 144 & 0.0015 & -0.091 & -6.5 & 0.089 & 6.2 & 0.014 & 1.68 & 31\\
GCM13 & 71 & -0.078 & -1.9 & -3.7 & 1.7 & 3.6 & 0.50 & -1.1 & 7.4\\
HGN13 & 91 & 0.022 & -1.7 & -4.1 & 2.3 & 5.8 & 0.39 & 2.0 & 16\\
SIN13 & 92 & 0.097 & -2.4 & -2.2 & 9.5 & 8.2 & 1.2 & 6.1 & 49\\
CLN13 & 67 & 0.0013 & -0.082 & -3.1 & 0.072 & 2.7 & 0.027 & -0.42 & 2.3\\
NGN13 & 70 & 0.00083 & -0.039 & -2.8 & 0.046 & 3.2 & 0.014 & 0.47 & 1.6\\
6AM13 & 61 & -2.2e-005 & -0.0011 & -6.1 & 0.00027 & 1.8 & 0.00017 & -3.8 & 21\\
6BM13 & 62 & 1.1e-005 & -0.00021 & -2.2 & 0.00017 & 1.6 & 0.00010 & -0.45 & -0.93\\
6CM13 & 63 & -1.6e-005 & -0.0016 & -7.0 & 0.00018 & 0.9 & 0.00023 & -5.8 & 39\\
6EM13 & 61 & 2.1e-006 & -0.00023 & -4.4 & 7.7e-005 & 1.4 & 5.4e-005 & -2.2 & 6.8\\
6JM13 & 60 & 1.6e-005 & -0.00021 & -1.6 & 0.00087 & 6.2 & 0.00014 & 4.0 & 24\\
GEM13 & 63 & 9.5e-006 & -0.00014 & -2.5 & 0.00016 & 2.6 & 5.7e-005 & 0.28 & 0.91\\
  \end{tabular}
  \label{wco-increments}
\end{table}
The $U=\max(|U^-|,|U^+|)$ is the greatest deviation from the mean expressed in standard deviations. For the standard normal distribution \cite[p. 972]{abramowitz1972} $P(\{U \ge 5\}) + P(\{U \le -5\}) = 2(1-0.9999997133) = 0.0000005734$. We need 2,000,000 sessions in order to observe in one such or greater deviation. With $\approx 247$ sessions in a year this is 8097 years of trading. However, deviations greater than five standard deviations have been observed for ZCN13, ZSN13, ZWN13, ZBM13, ESM13, HGN13, SIN13, 6AM13, 6CM13, and 6JM13 in less than 160 ranges. GCM13 and 6EM13 have excess kurtosis 7.4 and 6.8 exceeding the Gaussian zero. The remaining CLN13, NGN13, 6BM13, and GEM13 with low excess kurtosis, skewness, and $U$ are candidates for application of the Pearson's $\chi^2$ test. This effort, even, being successful, would not principally change the picture: \textit{not only b-increments but their means obtained with thousands of ticks in ranges are not Gaussian}. Statistical price and time properties can vary within sessions significantly, Figures \ref{explosion}, \ref{ZCN13_2_20130401_price_two}, \ref{noniid}. They change for the same contract experiencing a maximum of liquidity during its life, see column Size in Tables \ref{a-increments} and \ref{b-increments}. These can be the reasons.

\section{The c-Increments}

\begin{figure}[h!]
  \centering
  \includegraphics[width=130mm]{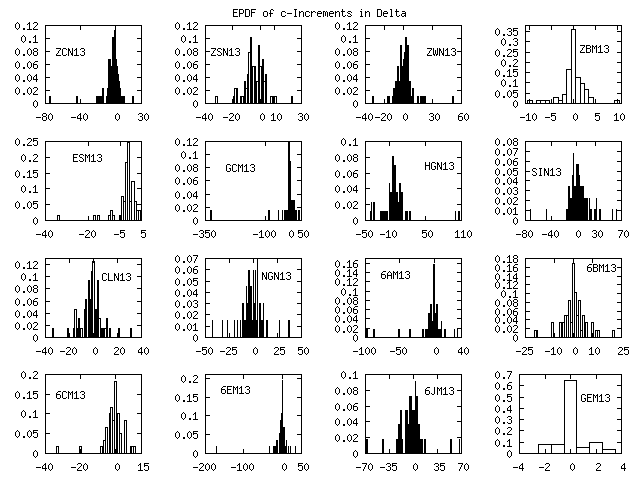}
  \caption[wc_freq_4_4]
   {EPDF of the c-increments expressed in $\delta$ for futures traded in March - July, 2013 on Globex. The c-increments for NGN13 are divided by 10 prior plotting.}
  \label{wc_freq_4_4}
\end{figure}

The c-increments are nonidentical links binding ranges and sessions by price, Table \ref{c-increments}. All of them are indecomposable and associate with larger durations than b-increments. Figure \ref{wc_freq_4_4} depicts their EPDFs.

The absolute mean c-increments are significantly greater than the absolute mean b-increments within sessions, Table \ref{c-increments} vs. \ref{b-increments}. The same cannot be said about the absolute extreme b-increments. They are greater than the absolute mean c-increments and, in many, cases comparable (ZWN13, GEM13) or greater (ZSN13, ZBM13, HGN13, SIN13, CLN13, NGN13, 6BM13, 6CM13) than the absolute extreme c-increments, Figure \ref{extreme_wb_freq_4_4} vs. \ref{wc_freq_4_4}. This means that \textit{the risk of holding a position before a next tick can be greater than for holding it between electronic sessions}. We need to recollect that pauses between electronic sessions are shorter than between pit sessions.

It also makes sense to compare the mean absolute c-increments with $|\textrm{Size} \times \textrm{Mean}|$ from Table \ref{b-increments}. The latter is the absolute difference of the last and first prices. This gives an idea whether the price changes more within or between sessions. For instance, for the ZCN13 sessions on 2013-04-02, 2013-04-03, 2013-04-04, and 2013-04-05 the three products are $35930 \times -0.00030615 = -11\delta$, $9\delta$, $-53\delta$, and $-4\delta$. The four preceding c-increments are $1\delta, \; 3\delta, \; 4\delta,$ and $1\delta$. In these sessions the price made greater moves within sessions than between them. The means of the absolute values are $19.25\delta \approx 19\delta$ and $2.025\delta \approx 2\delta$. The ratio
\begin{equation}
\rho_{bc}^s=\frac{P_{N_s}^s-P_1^s}{P_1^s-P_{N_{s-1}}^{s-1}}=\frac{P_{N_s}^s-P_1^s}{\textrm{c-increment}^s}
\label{EqBCRatio}
\end{equation}
is positive, if both moves are made in one direction, and negative otherwise. The ratio is undefined for zero c-increments. The greater $|\rho_{bc}^s|$ indicates the greater contribution of the session comparing with the pause. In the example $\rho_{bc}^{20130402}=-11, \; \rho_{bc}^{20130403}=3, \; \rho_{bc}^{20130404} = -13.25, \; \rho_{bc}^{20130405} = -4$.

Table \ref{c-increments} contains sample statistics for the entire family of c-increments. The information about the holidays, ch-increments, is the most complete for grains. This includes the Good Friday, Memorial Day, and Independence Day. The latter is not applicable for contracts expired prior July. In addition, the information related to the Memorial Day has been missed by technical reasons for several futures. The ch-increments have an illustrative meaning and cannot support statistical conclusions. A more definite conclusion for this period and contracts is that the mean cw-increments, over weekend, are greater by absolute value than the mean regular business day cr-increments. The exception is HGN13 and 6JM13. The ci-increments, between two ranges in a session, are applicable only to ZCN13, ZSN13, ZWN13, and ESM13. The absolute mean ci-increments are the smallest from cr-, cw-, and ch-increments treated separately.

\section{Price and Time: b- vs. a-Increments}

The mean displacement of a small Einstein's particle suspended in a liquid is proportional to the square root of the time \cite{einstein1905}. It is this property for the mathematical expectation of the price $x$ is obtained by Bachellier in two different routes \cite{bachelier1900}, \cite[pp. 29 - 33 and 33 - 36]{davis2006}. Einstein requires that \textit{"... a time-interval $\tau$ ... is to be very small compared with the observed interval of time, but, ... of such a magnitude that the movements executed by a particle in two consecutive intervals ... are to be mutually independent ...}. A modern definition of the \textit{mathematical Brownian motion} can be found in \cite[p. 1]{rogers2000_1}. In a less rigorous form, $B_t$ is a Brownian motion, if 1) $B_0 = 0$, 2) $B_t$ is a continuous function of $t \ge 0$, 3) for every $t, h \ge 0$ the increments $B_{t + h} - B_t$ are independent of $B_u: \; 0 \le u \le t$ and have a Gaussian distribution with mean 0 and variance $h$. A \textit{Wiener process} $W_t$ relative to a family of information sets $\{I_t\}$ is defined without mentioning a Gaussian distribution: 1) the pair $I_t, W_t$ is a \textit{square integrable martingale} with $W_0=0$ and $\textrm{E}[(W_t-W_s)^2]= t - s, s \le t$, and 2) the trajectories of $W_t$ are continuous over time $t$ \cite[p. 148]{neftci1996}. Neftci references the \textit{L\'{e}vy theorem} stating that any Wiener process is a Brownian motion. Being assumed for prices, these results imply a statistical relationship between b- and a-increments.

\begin{figure}[h!]
  \centering
  \includegraphics[width=130mm]{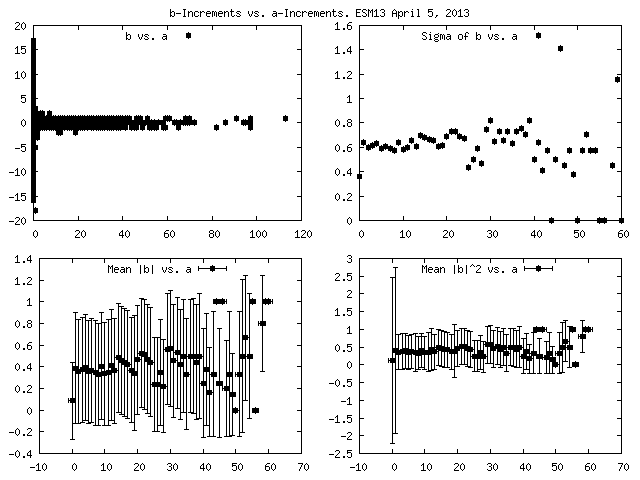}
  \caption[b_vs_a]
   {ESM13 April 5, 2013. Top left: 539882 points (a-increment in seconds, b-increment in $\delta$), due to discreteness many points overlap; top right: the sample standard deviation of b-increments (Table \ref{mwb_vs_a} StdDev) vs. associated a-increment (Table \ref{mwb_vs_a} a-Incr.); bottom left: the sample mean absolute b-increment $\pm$ StdDev vs. associated a-increment $\pm 1$ s, Table \ref{mawb_vs_a}; bottom right: the sample mean square b-increment $\pm$ StdDev (of square) vs. a-increment $\pm 1$ s, Table \ref{mawb2_vs_a}.}
  \label{b_vs_a}
\end{figure}

Each b-increment associates with an a-increment. Plotting the former vs. the latter does not reveal a curve, Figure \ref{b_vs_a} top left.

Sampling b-increments separately for 0, 1, 2 $\dots$ a-increments creates sample \textit{conditional} distributions. If the process is a Brownian motion, then not only each sample should obey a Gaussian distribution but $\textrm{StdDev(b-increment)} \propto \sqrt{\textrm{a-increment}}$. Figure \ref{b_vs_a} top right does not confirm such a proportionality. We cannot rely on Gaussian properties of individual samples either. For instance, Table \ref{mwb_vs_a} associates extreme b-increments -16 and 17 and StdDev 0.36379 with zero a-increment. These are -44 and 47 standard deviations. The two samples of the greatest size 508004 and 20357 are hardly Gaussian.

Einstein did not suggest that the mean particle displacement proportional to the square root of the time has a relationship to prices. If the mean price displacement, the mean b-increment, would be proportional to the square root of time, a-increment, then Figure \ref{b_vs_a} bottom left should show $y \propto \sqrt{x}$. Instead, the points are around a horizontal level.

Working with a Wiener process, one can expect the mean square b-increment to be proportional to the a-increment. In contrast, Figure \ref{b_vs_a} left right presents almost a horizontal line.

\textit{The b- and a-increments escape the relationship between the price changes and the square root of the time known for a Brownian motion. If such properties are observed for price and time increments, which are the sums of b- and a-increments, then there is a limit in scalability and self-similarity in the "micro world of the economic price and time atoms"}.

From Equation \ref{EqABRatio}
\begin{equation}
|\overline{\textrm{b-increment}^{s,r}}| \approx  \frac{|\sum_{i=1}^{N_{s,r}-1} \textrm{b-increment}_i^{s,r}|}{T_{s,r}^c - T_{s,r}^o}\overline{\textrm{a-increment}^{s,r}}.
\label{EqMb_vs_ma}
\end{equation}
Thus, for selected ranges with one $|P_{N_{s,r}}^{s,r} - P_1^{s,r}|$ and small a1- and a2-increments a linear dependence between the mean b- and a-increments is expected. This has no intrinsic dependence between price and time.

\section{A Comment on Diffusion}

\begin{figure}[h!]
  \centering
  \includegraphics[width=130mm]{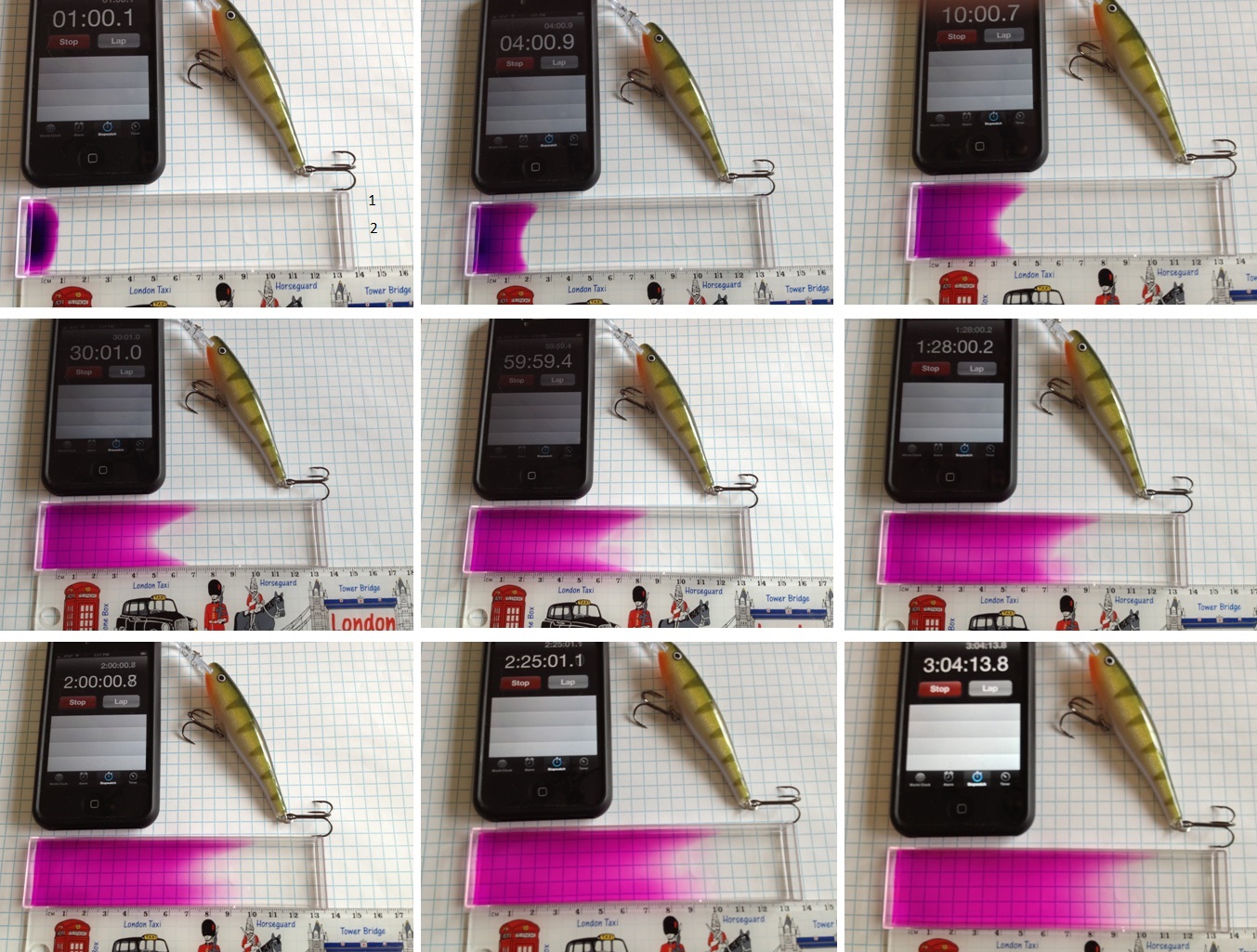}
  \caption[FgDiffusion]
   {The diffusion experiment. Markers 1 and 2 (top, left) indicate two lines along which F1 and F2 are measured.}
  \label{FgDiffusion}
\end{figure}

\begin{figure}[h!]
  \centering
  \includegraphics[width=130mm]{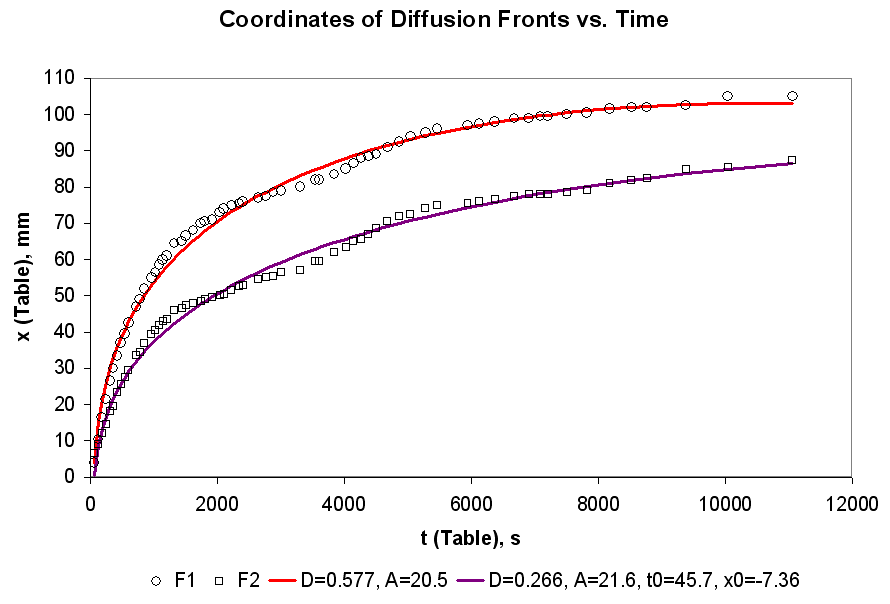}
  \caption[FgDiffusion_plots]
   {Experiment and theory: diffusion front distances vs. time. The optimal regression curves are obtained using Equation \ref{EqDiffusionApproximation}.}
  \label{FgDiffusion_plots}
\end{figure}

The author has conducted a diffusion experiment, Figure \ref{FgDiffusion}, \ref{FgDiffusion_plots}, Table \ref{TblDiffusion}.
The time was measured using Stopwatch of i-Phone 4S with 0.1 second accuracy. A transparent open plastic cover $13.7 \times 3.1 \times 1.3$ cm$^3$ from Papala VMC Corp. for Yellow Perch bait was filled with a 5 mm layer from the home water supply. Crystals of potassium permanganate, KMnO$_4$, were taken on a top of a horizontal toothpick and quickly dropped after rotation on one side of the vessel creating a stripe. The second i-Phone was used to take photos. The distance of the magenta front was measured by a souvenir ruler with 0.5 mm error. Front erosion introduced a larger error and subjectivity into the distance measurements (pictures are available). A care was taken in order family members and a cat would not shake the table. By the same reason ringing and vibration modes of the i-Phone located closely to the cover were switched off. The room temperature was $22.5\pm0.5^o$C $=295.65^o$K.
\begin{table}[!h]
  \centering
  \topcaption{Records of the diffusion experiment: $t$ seconds; $F1, F2$ mm.}
  \begin{tabular}{ccccccccc}
  $t$ & $F1$ & $F2$ & $t$ & $F1$ & $F2$ & $t$ & $F1$ & $F2$ \\
60.1 & 4 & 6.5 & 1620.7 & 68 & 48 & 4500.9 & 89 & 68.5 \\
120.9 & 10.5 & 9 & 1740.2 & 70 & 48.5 & 4680.9 & 91 & 70.5 \\
181 & 16.5 & 12 & 1800.9 & 70.5 & 49 & 4861 & 92.5 & 72 \\
240.8 & 21.5 & 14.5 & 1920.3 & 71 & 49.5 & 5041 & 94 & 72.5 \\
311.5 & 26.5 & 18 & 2040.7 & 73 & 50 & 5280.1 & 95 & 74 \\
361.1 & 30 & 19.5 & 2101.2 & 74 & 50.5 & 5462.3 & 96 & 75 \\
421.1 & 33.5 & 23.5 & 2220.9 & 75 & 51.5 & 5941.1 & 97 & 75.5 \\
480.8 & 37 & 25.5 & 2340.7 & 75.5 & 52.5 & 6122.2 & 97.5 & 76 \\
540.9 & 39.5 & 27.5 & 2401.1 & 76 & 53 & 6368.2 & 98 & 76.5 \\
600.7 & 42.5 & 29.5 & 2642.6 & 77 & 54.5 & 6673.6 & 99 & 77.5 \\
721 & 47 & 33.5 & 2761.1 & 77.5 & 55 & 6901 & 99 & 78 \\
780.8 & 49 & 34.5 & 2879.9 & 78.5 & 55.5 & 7080.9 & 99.5 & 78 \\
849.7 & 52 & 37 & 3001.1 & 79 & 56.5 & 7200.9 & 99.5 & 78 \\
960.9 & 55 & 39.5 & 3301 & 80 & 57 & 7500.8 & 100 & 78.5 \\
1020.8 & 56.5 & 40.5 & 3540.9 & 82 & 59.5 & 7817.4 & 100.5 & 79 \\
1080.6 & 58.5 & 42 & 3599.4 & 82 & 59.5 & 8177.5 & 101.5 & 81 \\
1141.4 & 60 & 43 & 3840.8 & 83.5 & 62 & 8520.6 & 102 & 82 \\
1201.2 & 61 & 43.5 & 4020.8 & 85 & 63.5 & 8761.2 & 102 & 82.5 \\
1320.6 & 64.5 & 46 & 4141.3 & 86.5 & 65 & 9376.6 & 102.5 & 85 \\
1440.8 & 65 & 46.5 & 4260.7 & 88 & 65.5 & 10037.9 & 105 & 85.5 \\
1501 & 66.5 & 47.5 & 4381.5 & 88.5 & 67 & 11053.8 & 105 & 87.5 \\
  \end{tabular}
  \label{TblDiffusion}
\end{table}
The results are plotted, Figure \ref{FgDiffusion_plots}.

\subsection{"Alchemy"}

Intently peering at the eroding magenta front, the author thought whether the two i-Phones, ruler, water, fishing tackle cover, and pinch of permanganate could explain the "madness of men". These artless tools, except the i-Phones, resemble the means of an alchemist. Newton knowing the laws of cooling and mechanics, named today after him, had to lose (or in other versions not to gain) \pounds 20,000 in order to formulate his thesis about the madness. Some convert this to modern US \$5,000,000. Maybe \textit{alchemy} is the right term \cite{soros1994}.

\subsection{Einstein's suspended particle}

Staying on the principals of the molecular-kinetic theory of heat and starting from the Van't Hoff equation (mathematically equivalent to the Mendeleev-Clapeyron equation) written for non-electrolytes, Einstein deduces that the osmotic pressure $p$ remains intact after replacing dissolved molecules with suspended particles at great dilution. This conclusion, unusual for that time, permits him to apply the same equation for the pressure, where the molarity is replaced with the concentration expressed in particles per volume $\nu$: $p=\frac{RT}{N}\nu$. This brings the Avogadro number $N$ to the denominator. $T$ is the absolute temperature. $R$ is the universal gas constant. Reviewing thermodynamical equilibrium, where the free energy vanishes for an arbitrary virtual displacement $\delta x$, he concludes that the osmotic pressure must have an effect of applying the force $K$ to a suspended particle: $K=\frac{1}{\nu}\frac{\partial p}{\partial x}$. Then, he reviews the diffusion flux of the first Fick's law expressed for particles (instead of mass) as $-D\frac{\partial \nu}{\partial x}$, where $D$ is the diffusion coefficient, and equalizes it with the number of particles passing a unit area per unit of time with a certain velocity. He cites Kirchhoff for the velocity $\frac{K}{6\pi kP}$ of a sphere with the radius $P$ moving in the liquid with viscosity $k$ under the force $K$. This is like if one, who sees the magenta front movement on Figure \ref{FgDiffusion}, associates it with a force of certain magnitude. But Einstein sees the source of the force and derives the notorious $D = \frac{RT}{N}\frac{1}{6\pi kP}$. His next step made from the molecular-kinetic consideration leads to a) the second Fick's diffusion law $\frac{\partial f(x, t)}{\partial t} = D \frac{\partial^2 f(x,t)}{\partial x^2}$ with b) $D = \frac{1}{\tau}\int_{-\infty}^{+\infty}\frac{\Delta^2}{2}\phi(\Delta)d\Delta$, where $\Delta$ is a displacement of a particle within the time-interval $\tau$, $\nu=f(x,t)$ is the number of particles per unit volume, $\phi(\Delta)$ is the probability density of the distribution of particles held for $\Delta$. For $x \ne 0$ and $t = 0$: $f(x, t) = 0$, $\int_{-\infty}^{+\infty}f(x,t)dx = n$. He knows the solution of this \textit{initial-value problem} $f(x,t)=\frac{n}{\sqrt{4\pi D}}\frac{e^{-\frac{x^2}{4Dt}}}{\sqrt{t}}$ and treats $2Dt$ as the variance linearly growing in time. Recollect that $\mu_2=\alpha_2 - \alpha_1^2 = \alpha_2$ for $\alpha_1=0$. For him this variance is the mean of the squares of displacements along the X-axis. Thus, $\lambda_x=\sqrt{2Dt}=\sqrt{t}\sqrt{\frac{RT}{N}\frac{1}{3\pi kP}}$. The author presents these details to remind: there exists a solid (less in Einstein's and greater in our time) foundation behind - the molecular-kinetic theory of heat. Einstein states: \textit{"... had the prediction of this movement proved to be incorrect, a weighty argument would be provided against the molecular-kinetic conception of heat"}. The ratio of the deepness of the physical contents to the mathematical complexity in the Einstein's article is the ratio of the price of the Large Hadron Collider to the price of equipment on Figure \ref{FgDiffusion}.

The undertaken experiment deviates from the Einstein's assumptions. The potassium permanganate is electrolyte and dissociates in accordance with the mentioned Svante Arrhenius theory. This increases the so-called \textit{isotonic coefficient} in the Van't Goff equation up to two. The thin water layer still does not eliminate the 3D diffusion. The stripe of the crystals does not prevent the 2D diffusion either. A few crystals remain undissolved awhile forming a complicated concentration profile. The concentration is far from the "great dilution" needed for Einstein to apply the laws of \textit{ideal solutions}. The surface effects caused by the plastic wall touching a solution with the varying permanganate concentration can provoke convection increasing "observable" $D$. Nevertheless, if we can evaluate $D$ from this data, then the molecule size estimate is $\frac{RT}{N}\frac{1}{6\pi kD}$.

\begin{figure}[h!]
  \centering
  \includegraphics[width=130mm]{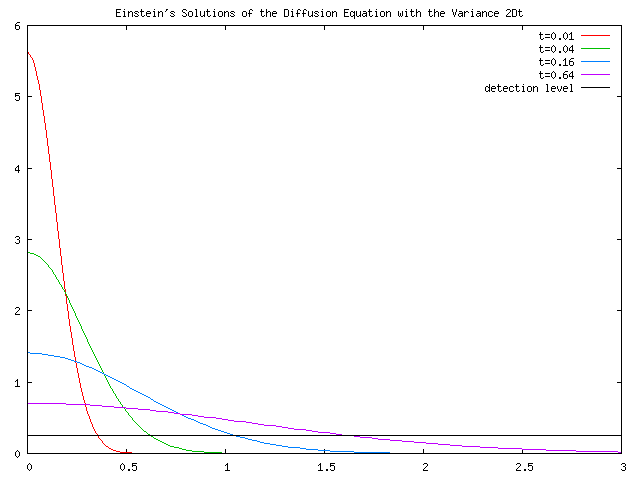}
  \caption[FgDiffusion_theory]
   {Concentration profiles.}
  \label{FgDiffusion_theory}
\end{figure}

The Einstein's solution of the diffusion equation is the \textit{Green's} or the \textit{impulse-response function} \cite[pp. 93 - 95]{farlow1993}. Halves of these profiles times two $f(x,t)=\frac{2ne^{-\frac{x^2}{4Dt}}}{\sqrt{4\pi Dt}}$, $D=1, n = 1$ are on Figure \ref{FgDiffusion_theory}.
An eye detects the concentration $f_d$ at the right side of the profile. This is the intersection of the concentration curve and the horizontal detection line. This solution is not suitable for $t \rightarrow \infty$, where the concentration approaches a constant $> 0$: a) the cover has the length $L$ fixing the volume, b) the amount of permanganate is fixed too. However, when the diffusion front is far from the right wall, the solution is reasonable: the exponent drops quickly. We "measure" not $f$ but color intensity proportional to it. A coefficient of proportionality $s$ is needed. The approximating function $x=F(t)$ is obtained by equalizing densities to $f_d$ and taking $x\ge0$
\begin{displaymath}
f(x,t)=f_d=\frac{se^{-\frac{x^2}{4Dt}}}{\sqrt{4\pi Dt}}, \; x \ll L,
\end{displaymath}
\begin{displaymath}
x = \sqrt{t} \times \sqrt{ D\left(-4\ln\frac{f_d}{s} -2\ln(4\pi Dt)\right)}=\sqrt{Dt} \times \sqrt{A-2\ln(t)},
\end{displaymath}
where $A$ depends on $D$ but has a degree of freedom so that we can ignore it and find $D$ from the multiplier. To compensate the asynchronicity of dropping crystals and starting the stopwatch, and a shift of the ruler two parameters are added $t=t_{table}-t_0$ and $x=x_{table}-x_0$
\begin{equation}
x_{table} = x_0 + \sqrt{D(t_{table}-t_0)} \times \sqrt{A-2\ln(t_{table}-t_0)}.
\label{EqDiffusionApproximation}
\end{equation}
Equation \ref{EqDiffusionApproximation} is the approximating function of the four parameters $D, A, t_0, x_0$. A co-optimization is completed, where $t_0, x_0$ are common for the two curves but $D, A$ vary independently. The total (for two curves) sum of the squares of deviations between $x$ from Table \ref{TblDiffusion} and Equation \ref{EqDiffusionApproximation} dependent on the six parameters is minimized using the Microsoft Excel's Solver. The solver with constrained optimization applies the Lasdon's algorithm \cite{lasdon1978}. The sets $\{D=0.577 \frac{mm^2}{s}, A=20.5\}$, $\{D=0.266 \frac{mm^2}{s}, A=21.6\}$, and common $t_0=45.7s$, $x_0=-7.36mm$ are found for the two fronts. The curves approximate well the experimental points for $x < 0.8L$, Figure \ref{FgDiffusion}. However, the least coefficient $0.266 \frac{mm^2}{s} \approx 2.7 \times 10^{-7} \frac{m^2}{s}$ is 163 times greater than the literature one $1.632 \times 10^{-9} \frac{m^2}{s}$ \cite{lide2005}. The surface tension and convection in the open cell could be responsible for the faster purple color spreading. $P=\frac{8.31\frac{J}{mol \times K}295.65K}{6.02 \times 10^{23} \frac{1}{mol}}\frac{1}{6 \times 3.14 \times 10^{-3}\frac{kg}{m \times s}\times 2.7 \times 10^{-7} \frac{m^2}{s}}=8\times 10^{-13}m$, where one expects $\approx 10^{-10}m$. Well, the Collider worth its money.

\subsection{Solution which Einstein did and Black did not know}

In contrast, Fischer Black was not sure that the equation built by him is the diffusion one and how to solve it \cite{black1987}, \cite[p. 5]{black1989}: \textit{"I spent many, many days trying to find the solution to that equation. I have a Ph.D. in applied mathematics, but had never spent much time on differential equations, so I didn't know the standard methods used to solve problems like that. I have an A.B. in physics, but I didn't recognize the equation as a version of the "heat equation," which has well known solutions "} (VS: like the one known to Einstein). Prior 1973 \cite{black1973} everything was completed (or started). Using a) stock prices following a random walk in continuous time and lognormal at the end of any finite time interval \cite[p. 640, assumption b)]{black1973}, b) stochastic calculus \cite[p. 642, Eq. 4]{black1973}, c) the non-arbitrage postulate, d) certainty of return on the riskless hedged position (pointed out by Robert Merton), and e) other less important for us assumptions, Black and Mayron Scholes come to the partial differential equation, PDE, problem \cite[p. 643, Eq. 7, 8]{black1973}: $w_2=rw-rxw_1-\frac{1}{2}v^2x^2w_{11}, \; w(x,t^*)=x-c, \; x \ge c \; \textrm{or} \; =0, \; x < c$. An omitted sophisticated substitution of variables \cite[p. 643, Eq. 9]{black1973} transforms it to $y_2=y_{11}$ or after switching from the Nobel to ordinary notation $\frac{\partial y}{\partial t} = \frac{\partial^2y}{\partial u^2}, \; y(u,0)=0, u < 0 \; \textrm{or} \; c\big[e^{u(\frac{1}{2}v^2)/(r-\frac{1}{2}v^2)} - 1 \big], u \ge 0$ with the coefficient of diffusion $D=1$. From here, the celebrated option value formula with the \textit{lucky number 13} \cite[p. 644, Eq. 13]{black1973} is derived using the Fourier/Green/Einstein method/solution and back substitution returning to the original variables.

Much earlier, Bachelier solves a very, very similar PDE problem \cite{bachelier1900}. Introducing the \textit{radiation of the probability} ("rayonnement de la probabilit\'{e}") with the logic: \textit{"During the time element $\Delta t$, each price $x$ radiates a quantity of probability proportional to the difference in their probabilities towards the neighboring price"} \cite[p. 40]{davis2006}, he comes to a Fourier equation $c^2\frac{\partial P}{\partial t} - \frac{\partial^2P}{\partial x^2}=0$ involving probability $P$ and a constant $c$. He viewed price increments distributed normally with the variance proportional to time. The latter property, as it is recognized today, gives him a priority in the mathematical treatment of a Brownian motion. The author did not find in the Bachelier's thesis the word "Brownian" as well as a definition of "Speculation".

The geometric or economic Brownian motion introduced by Samuelson was independently suggested by M.F.M. Osborne \cite{osborne1959}. The author wants to mention the independent works of Andre Laurent \cite{laurent1959}, \cite{laurent1957}, \cite{osborne1959_2} and R. Remery \cite{remery1946} (the author could not find the first name and get the Remery's thesis and cites it by \cite{laurent1959}), who had an interesting idea. Laurent writes \textit{"... in Remery's approach departures from the model $\sigma_Y^2=\sigma^2 t$ are interpreted as measures of economic disequilibrium."}.

The stochastic calculus applied by Black and Scholes is based on the result of Kiyosi It\^o \cite[p. 523 - 524]{ito1944}: \textit{"Let $F(x)$ be a function of $x$ such that $F''(x)$ may be continuous. ... The author has proved the equality: $\int_0^tF'(g(\tau,w))d_{\tau}g(\tau,w)=F(g(t,w))-F(g(0,w))-\frac{1}{2}\int_0^tF''(g(\tau,w))d\tau$. In the last term we may see a characteristic property by which we distinguish "stochastic intergal" from "ordinary intergal"."} Here, $g(t,w)$ is \textit{"... any brownian motion ... a (real) stochastic differential process with no moving discontinuity such that $E(g(s,w) - g(t,w))=0$ and $E(g(s,w)-g(t,w))^2=|s-t|$"} \cite[p. 519]{ito1944}. They need to expand the difference in the option values $w(x+\Delta x, t + \Delta t) - w(x,t)$ in order to trace the change in the hedged portfolio value long in one stock share and short in $\frac{1}{w_1}$ options. This can be applied as long as It\^{o}'s conditions exist. The Brownian motions are in scope. In general, the It\^{o} integral $\int HdX$ is defined for quite wide conditions, where the integrand $H$ is \textit{previsible} and the integrator $X$ must be a \textit{semimartingale} \cite[p. 2]{rogers2000}. The Bachelier's, Black-Scholes's, Merton's (with the continuous dividend yield) processes as well as many new variations are in scope. The steps are common a) find a process obeying It\^o's requirements, b) apply his stochastic calculus to the change in the value of instrument, a \textit{derivative}, borrowing risk from the underlying process, c) use the non-arbitrage assumption for building a hedged portfolio, d) assume that the latter, being riskless, gains with a riskless return rate, e) come to the PDE problem, where time and price conditions (initial, final, boundary) depend on the financial instrument to be priced, f) solve it, h) test against the market values.

\subsection{Kreps, Harrison, Pliska}

Currently, there is a well known alternative to the a) - h) plan \cite{harrison1979}, \cite{harrison1981}. Instead of discussing it, the author shares his personal opinion: these results are fundamental for economical sciences and David Kreps, Michael Harrison, and Stanley Pliska, authors of these two articles, deserve full attention of the \textit{Prize Committee for the Sveriges Riksbank Prize in Economic Sciences in Memory of Alfred Nobel}.

\subsection{Testing a-b-c-process via pricing derivatives is too rough}

In all cases we start from a price process. This gives prices of derivative instruments. They can be compared with the market premiums. The goal is to reproduce the latter. Within acceptable tolerance, quite different underlying processes can be approved. When time deterministic curves of volatility and interest rates replace constant parameters in the stochastic differential equation underlying the Black-Scholes model, the latter fits better the dependence of the implied volatility vs. the option expiration time - \textit{volatility term structure}. However, it still greatly underestimates risk because of the lognormal prices. In other words, a protective Stop Loss Order has much more chances to be "touched" or "gapped over" creating shocking slippage, than it follows from the Gaussian assumption. This means that traders of underlying should decide about goodness of a price model not after sophisticated transforming the process into option prices and matching the latter with the market values, but directly requiring from the process to match properties affecting trading results. \textit{Thorough studying the a-b-c-process becomes a must. The maximum trading profit strategy framework and optimal trading elements described in this article is a new method to study it}.

\section{A Comment on Dependence}

Traders want to find dependence between the past and future prices or events determining them. Prices look random. We shall speak about what "random" is in the next section. Fama \cite{fama1965} applies serial correlation model, runs tests, runs by length tests, and the Alexander's technique \cite{alexander1961} with the filter varying from 0.5 to 50 percent in order to conclude that the sample correlation coefficients are small, number of runs and their lengths do not differ significantly from the expected values, and Alexander's strategy, after accounting costs and inability to trade at filter prices, converts profits into losses. This indicates absence of significant dependencies. During the next 48 years his research has been followed by the growth of speculation. An intriguing question is whether the growth was \textit{independent} on the research.  \textit{Determination of dependence between random variables is a fundamental task not only for economics}.

In technical analysis searching for dependencies between prices gets implicit forms. Runs are defined as a sequence of price changes of the same sign. A trader does not care, when a run is interrupted. It is important that the interruptions are "insignificant" and the next runs continue increasing the mark to market profit. When a \textit{head and shoulders pattern} \cite[pp. 74 - 76]{murphy1999} is identified, a trader knowns that it is not always followed by a corresponding price move. There are two events here 1) the pattern and 2) the direction and size of a next price move. Proving or disproving dependence between them is another form of concluding about dependence between the past and future prices. Patterns and \textit{signals} are functions of prices and other information. Many inventive \textit{trading systems} can be found in \cite{kaufman2005}, \cite{williams1999}, \cite{connors1995}, \cite{babcock1989}, \cite{williams1979}, \cite{wilder1978}. Is it possible that one establishes dependence between patterns and independence between prices or vice versa? \textit{Such measures of dependence are welcomed, where our conclusions are in agreement}.

\subsection{R\'{e}nyi's example}

Linear dependence between two random variables $\xi, \eta$ is measured by the coefficient of linear correlation $R(\xi,\eta)$. For Gaussian variables, $R(\xi,\eta) = 0$ ensures their independence. In general, this is not true. Let $\xi$ is uniformly distributed on $[-1, 1]$. $\eta=\xi^2$ is completely determined given $\xi$ but $R(\xi,\eta) = 0$. Alfred R\'{e}nyi's example is the same $\xi$ and $\eta=5\xi^3-3\xi$ \cite[p. 443]{renyi1959}. \textit{Such universal measures are interesting, which cover linear and nonlinear dependencies}.

\subsection{Kolmogorov's advice}

Kolmogorov \cite[p. 256]{kolmogorov1956}: \textit{Let in a large number of repeated trials $n$ event $A$ has occurred $m$ times, and event $B$ has occurred $l$ times, moreover, $k$ times together with event $A$. It is natural to refer to the ratio $\frac{k}{m}$ as conditional frequency of event $B$ under the condition of occurrence of event $A$. ... If there is no relationship between events $A$ and $B$, then it is natural to assume, that event $B$ should appear under the condition of $A$ neither substantially more nor less frequently than in all trials, what approximately means $\frac{k}{m} \approx \frac{l}{n}$ or $\frac{k}{n} = \frac{k}{m} \frac{m}{n} \approx \frac{l}{n} \frac{m}{n}$}. Let event $A$ is an exclusive tick with the a-increment 0 or 1 or $i$ seconds and event $B$ is such a tick with the absolute b-increment 0 or 1 or $j$ $\delta$ with the frequencies $\nu_{A_i}=\frac{m_i}{n}, \; \nu_{B_j}=\frac{l_j}{n}, \; \nu_{A_iB_j}=\frac{k_{ij}}{n}$. Following to Kolmogorov, the difference $\nu_{A_iB_j}-\nu_{A_i} \nu_{B_j}$ for ESM13 is computed in Table \ref{kolmogorov_advice}. This is the difference of an empirical joint distribution frequency and the product of two empirical marginal distribution frequencies. For independent events related probabilities should result in zero. However, the frequencies are only estimates of the probabilities. In theory, one can repeat this consideration for the mean frequencies and get new estimates endlessly. Kolmogorov comments \cite[p. 262]{kolmogorov1956}: \textit{... this will not free us from a necessity on the last stage to turn to the probabilities in their primitive and rough sense}. The difference $\nu_{AB} - \nu_A\nu_B$ can accumulate large error, which will result in large relative error for small $\nu_{AB}$. The values in Table \ref{kolmogorov_advice} vary from 1 to 85 percent. We never get exact zero. The treatment is done separately for sessions on March 4 and May 22 with the least and largest number of increments and for a combined sample of all sessions. Does under such conditions the error 1 - 20 percent mean independence? The quantity $\nu_{AB} - \nu_A\nu_B$ is in the heart of the Hoeffding \cite{hoeffding1948} and Blum-Kiefer-Rosenblat \cite{blum1961}, \cite{csorgo1979} tests of independence but their statistics are for continuous distributions. \textit{The market demands tests of dependence between discrete variables}. The Kolmogorov's quantity is easy to compute for empirical discrete and, after discretization, continuous distributions. How does it relate to the R\'{e}nyi's postulates \cite[p. 443]{renyi1959}?

\subsection{R\'{e}nyi's axioms of dependence}

The R\'{e}nyi's list of properties of an appropriate measure $\delta(\xi,\eta)$ of dependence between $\xi$ and $\eta$ is: A) $\delta(\xi,\eta)$ is defined for any pair of random variables $\xi$ and $\eta$, neither of them being constant with probability 1; B) $\delta(\xi,\eta)=\delta(\eta,\xi)$; C) $0 \le \delta(\xi,\eta) \le 1$; D) $\delta(\xi,\eta)=0$ if and only if $\xi$ and $\eta$ are independent; E) $\delta(\xi,\eta)=1$ if there is a strict dependence between $\xi$ and $\eta$, i.e. either $\xi=g(\eta)$ or $\eta=f(\xi)$ where $g(x)$ and $f(x)$ are Borel-measurable functions (see \cite[p. 38]{kolmogorov1999}); F) if the Borel-measurable functions $f(x)$ and $g(x)$ map the real axis in a one-to-one way into itself, $\delta(f(\xi),g(\eta)) = \delta(\xi,\eta)$; G) if the joint distribution of $\xi$ and $\eta$ is normal, then $\delta(\xi,\eta)=|R(\xi,\eta)|$. We see that $\nu_{AB} - \nu_A\nu_B$ satisfies A, B, and D. R\'{e}nyi determines that the Gebelein's \textit{maximal correlation} $S(\xi,\eta)=\sup_{f,g}(R(f(\xi),g(\eta)))$, where supremum is taken for $f(x)$ and $g(x)$ running all Borel-measurable functions, satisfies the seven postulates. He introduces the notion of \textit{attainable functions}, such that $S(\xi,\eta) = R(f_0(\xi),g_0(\eta))$ holds, and proves two theorems assisting to find them. R\'{e}nyi notices a parallel between conditional expectations and theory of operators reducing the task to finding eigen values and functions of the \textit{completely continuous transformation} $Af=M(M(f(\xi)|\eta)|\xi)$, where $M$ is the mathematical expectation and vertical lines indicate conditional expectations. Specifically, $f_0$ is the eigenfunction belonging to the greatest eigenvalue $S^2=S^2(\xi,\eta)$ of $A$ and $g_0(\eta)=\frac{M(f_0(\xi)|\eta)}{S}$. He establishes the condition, under which the transformation $A$ is completely continuous and the maximal correlation is attained. His method of solving the problem resembles a way, where "mapping" between the notions of two, at the first glance disjointed, branches of mathematics or mathematics and physics, quickly gives a solution in one branch, if it is known or easy in another \cite{galperin2004}. A similar principal is discussed in works of Kara-Murza.

R\'{e}nye confirms that the Linfoot's \textit{informational coefficient of correlation} $L(\xi,\eta)=\sqrt{1-e^{-2I(\xi,\eta)}}$, where $I(\xi,\eta)$ is the \textit{mutual information} between $\xi$ and $\eta$, also has the seven properties. This criterion growths from the foundation built by Claude Shannon \cite{shannon1948} and Khinchine \cite{khinchine1953}, where the Khinchine's exposition by his own words \cite[p. 3]{khinchine1953} is \textit{"more complete and mathematically correct"}. To illustrate the recent developments on the informational correlation and mutual information, the author cites \cite{lu2011} containing more references.

\subsection{Three tests applied to a- and b-increments}

Nonparametric tests based on space partitioning and kernel approaches are compared in \cite{gretton2008}. For a sample of real valued random vectors $X$ and $Y$ with dimensions $d$ and $d'$ and i.i.d pairs $(X_1,Y_1), \dots , (X_n,Y_n)$, the null hypothesis that $X$ and $Y$ are independent, $H_0 : \nu_{AB}=\nu_A\nu_B$, is tested making minimal assumptions regarding the distributions. In terms of a- and b-increments the two partitions are $A_n = \{A_{n,1}, \dots, A_{n,m_n^A}\}$, $B_n = \{B_{n,1}, \dots, B_{n,m_n^B}\}$. Here, $n$ is the number of a- and b-increment pairs in a sample, $m_n^A$ and $m_n^B$ are the numbers of events of a- and b-increments, $d=d'=1$. The events are the sizes of the a- and b-increments is seconds and $\delta$. Only the actual sizes found in a sample form events. For instance, 100 seconds can be followed by 102 seconds with the value 101 not represented. The absent 101 does not increase $m_n^A$. The b-events are treated similarly. The $L_n$, $I_n$, and $\chi_n^2$ statistics of interest are
\begin{displaymath}
L_n(\nu_{AB}, \nu_A\nu_B)=\sum_{A \in A_n}\sum_{B \in B_n}|\nu_{AB} - \nu_A\nu_B|,
\end{displaymath}
\begin{displaymath}
I_n(\nu_{AB}, \nu_A\nu_B)=2\sum_{A \in A_n}\sum_{B \in B_n}\nu_{AB}\log\frac{\nu_{AB}}{\nu_A \nu_B},
\end{displaymath}
\begin{displaymath}
\chi_n^2(\nu_{AB}, \nu_A\nu_B)=\sum_{A \in A_n}\sum_{B \in B_n}\frac{(\nu_{AB}-\nu_A \nu_B)^2}{\nu_A \nu_B}.
\end{displaymath}
Arthur Gretton and L\'{a}szl\'{o} Gy\"{o}rfi prove that \textit{almost surely} $L_n(\nu_{AB}, \nu_A\nu_B) > \sqrt{2\ln2}\sqrt{\frac{m_n^A m_n^B}{n}}=\epsilon_{L_n}$, if $\lim_{n \rightarrow \infty}\frac{m_n^A m_n^B}{n} = 0, \; \lim_{n \rightarrow \infty}\frac{m_n^A}{\ln(n)} = \lim_{n \rightarrow \infty}\frac{m_n^B}{\ln(n)} = \infty$, rejects the $H_0$ hypothesis of independence. They also suggest to reject independence, if $I_n(\nu_{AB}, \nu_A\nu_B) > \frac{m_n^A m_n^B (2\ln(n + m_n^A m_n^B) + 1)}{n}=\epsilon_{I_n}$. Finally, they derive $\xi_{\chi_n^2}=\frac{n \chi_n^2(\nu_{AB}, \nu_A\nu_B) - m_n^A m_n^B}{\sqrt{2m_n^A m_n^B}} \rightarrow \textrm{Gaussian}(\alpha_1 = 0, \mu_2 = 1)$ meaning convergence on distribution. The author has computed these quantities separately for classes with absolute and signed b-increments, Tables \ref{TblDependenceAbs} and \ref{TblDependence}.

\begin{table}[!h]
  \centering
  \topcaption{$L_1$, log-likelihood, and Pearson $\chi_n^2$ quantities for testing independence between a- and absolute b-increments.}
  \begin{tabular}{ccccccccccc}
  Ticker & $n$ & $m_n^A$ & $m_n^B$ & $m_n^{AB}$ & $L_n$ & $I_n$ & $\chi_n^2$ & $\epsilon_{L_n}$ & $\epsilon_{I_n}$ & $\xi_{\chi_n^2}$ \\
ZCN13 & 2799609 & 1098 & 44 & 2884 & 0.16 & 0.083 & 1.7 & 0.15 & 0.53 & 1.5e+4 \\
ZSN13 & 2693356 & 973 & 61 & 3272 & 0.18 & 0.081 & 3.4 & 0.17 & 0.68 & 2.6e+4 \\
ZWN13 & 1537493 & 1255 & 27 & 3153 & 0.18 & 0.09 & 4.2 & 0.17 & 0.65 & 2.5e+4 \\
ZBM13 & 5803196 & 907 & 25 & 1888 & 0.077 & 0.067 & 1 & 0.074 & 0.13 & 2.8e+4 \\
ESM13 & 27437768 & 444 & 31 & 1023 & 0.076 & 0.044 & 0.082 & 0.026 & 0.018 & 1.3e+4 \\
GCM13 & 5439767 & 1048 & 140 & 3620 & 0.14 & 0.037 & 4.6 & 0.19 & 0.87 & 4.6e+4 \\
HGN13 & 1804971 & 1341 & 55 & 4318 & 0.25 & 0.1 & 13 & 0.24 & 1.2 & 6.2e+4 \\
SIN13 & 1445265 & 1389 & 75 & 4696 & 0.26 & 0.1 & 10 & 0.32 & 2.1 & 3.1e+4 \\
CLN13 & 3459101 & 1176 & 46 & 4366 & 0.18 & 0.079 & 2.9 & 0.15 & 0.49 & 3e+4 \\
NGN13 & 1417453 & 1371 & 71 & 4079 & 0.22 & 0.11 & 0.87 & 0.31 & 2 & 2.6e+3 \\
6AM13 & 3698272 & 367 & 47 & 1092 & 0.18 & 0.085 & 0.26 & 0.08 & 0.15 & 5.1e+3 \\
6BM13 & 4151484 & 455 & 37 & 1289 & 0.18 & 0.081 & 0.16 & 0.075 & 0.13 & 3.6e+3 \\
6CM13 & 2432275 & 584 & 45 & 1344 & 0.17 & 0.094 & 0.37 & 0.12 & 0.33 & 3.8e+3 \\
6EM13 & 8936861 & 313 & 58 & 840 & 0.11 & 0.051 & 0.088 & 0.053 & 0.067 & 4e+3 \\
6JM13 & 6428595 & 306 & 34 & 991 & 0.15 & 0.057 & 0.38 & 0.047 & 0.052 & 1.7e+4 \\
GEM13 & 293054 & 1640 & 4 & 2885 & 0.073 & 0.082 & 0.27 & 0.18 & 0.59 & 6.3e+2 \\
  \end{tabular}
  \label{TblDependenceAbs}
\end{table}

\begin{table}[!h]
  \centering
  \topcaption{$L_1$, log-likelihood, and Pearson $\chi_n^2$ quantities for testing independence between a- and b-increments.}
  \begin{tabular}{ccccccccccc}
  Ticker & $n$ & $m_n^A$ & $m_n^B$ & $m_n^{AB}$ & $L_n$ & $I_n$ & $\chi_n^2$ & $\epsilon_{L_n}$ & $\epsilon_{I_n}$ & $\xi_{\chi_n^2}$ \\
ZCN13 & 2799609 & 1098 & 79 & 3797 & 0.16 & 0.084 & 2.5 & 0.21 & 0.95 & 1.7e+4 \\
ZSN13 & 2693356 & 973 & 108 & 4454 & 0.18 & 0.082 & 4.8 & 0.23 & 1.2 & 2.8e+4 \\
ZWN13 & 1537493 & 1255 & 50 & 4151 & 0.18 & 0.092 & 6.5 & 0.24 & 1.2 & 2.8e+4 \\
ZBM13 & 5803196 & 907 & 46 & 2372 & 0.077 & 0.068 & 2 & 0.1 & 0.23 & 3.9e+4 \\
ESM13 & 27437768 & 444 & 57 & 1351 & 0.076 & 0.044 & 0.094 & 0.036 & 0.033 & 1.1e+4 \\
GCM13 & 5439767 & 1048 & 240 & 4672 & 0.14 & 0.037 & 8.7 & 0.25 & 1.5 & 6.6e+4 \\
HGN13 & 1804971 & 1341 & 98 & 5641 & 0.25 & 0.11 & 21 & 0.32 & 2.2 & 7.4e+4 \\
SIN13 & 1445265 & 1389 & 131 & 6039 & 0.26 & 0.1 & 16 & 0.42 & 3.7 & 3.7e+4 \\
CLN13 & 3459101 & 1176 & 80 & 5848 & 0.18 & 0.08 & 5.4 & 0.19 & 0.85 & 4.3e+4 \\
NGN13 & 1417453 & 1371 & 120 & 5393 & 0.22 & 0.11 & 1.6 & 0.4 & 3.4 & 3.7e+3 \\
6AM13 & 3698272 & 367 & 85 & 1533 & 0.18 & 0.085 & 0.42 & 0.11 & 0.26 & 6.1e+3 \\
6BM13 & 4151484 & 455 & 68 & 1818 & 0.18 & 0.081 & 0.24 & 0.1 & 0.23 & 3.8e+3 \\
6CM13 & 2432275 & 584 & 81 & 1792 & 0.17 & 0.095 & 0.6 & 0.16 & 0.59 & 4.6e+3 \\
6EM13 & 8936861 & 313 & 103 & 1171 & 0.11 & 0.051 & 0.11 & 0.071 & 0.12 & 3.8e+3 \\
6JM13 & 6428595 & 306 & 63 & 1394 & 0.15 & 0.057 & 0.65 & 0.064 & 0.097 & 2.1e+4 \\
GEM13 & 293054 & 1640 & 6 & 3541 & 0.073 & 0.089 & 0.4 & 0.22 & 0.88 & 7.7e+2 \\
  \end{tabular}
  \label{TblDependence}
\end{table}

To verify the preconditions for the $L_n$ test application, review that, for example, for absolute b-increments of ESM13 $\frac{m_n^A m_n^B}{n} = \frac{444 \times 31}{27437768} \approx 0.0005$ vs. zero, and $\frac{m_n^A}{\ln(n)} \approx 26, \frac{m_n^B}{\ln(n)} \approx 2$ vs. $\infty$. We have to conclude cautiously. Based on these results, independence should be rejected for ESM13, 6AM13, 6BM13, 6CM13, 6EM13, and 6JM13. It cannot be rejected for GCM13, NGN13, and GEM13. The remaining ZCN13, ZSN13, ZWN13, ZBM13, and CLN13,  are at the decision boundary. Comparison of $I_n$ with $\epsilon_{I_n}$ suggests that the independence hypothesis cannot be rejected for ESM13, 6BM13, and 6JM13. 6AM13, and 6EM13 are at the boundary and the remaining contracts do not show dependence. The $\xi_{\chi_n^2}$ are based on an asymptotic behavior. The author could not interpret these values. Instead, he has maximally verified  the written C++ program and has not found bugs. All the quantities are computed for discrete random variables. It seems that, at least, for the $L_n$ test we have all rights for doing it \cite{gretton2008}.

The author skips the perspective kernel based tests \cite{gretton2008}. \textit{While the data does not confirm the Brownian type of dependence between the indecomposable a- and b-increments for ESM13, the hypothesis of their independence cannot be accepted either}.

\section{A Comment on Randomness}

Years ago, being a student, the author wanted to find an ontological explanation of randomness. If quantum mechanics pretends that, Einstein is wrong in his search for a causal explanation of the observed randomness, and postulates the latter as an objective reality, then we can try an opposite program and find an objective basis of randommeness. A system of classical particles giving the birth to the Laplace determinism is a simple candidate to begin with. If we can find something creating a chance in it, then the goal is reached. This cannot be a "random" component clarifying nothing. It should be known to science. The student found nothing but ... infinity. If the system contains infinite number of particles interacting deterministically or with infinite number of other systems, then there is always a possibility of an upcoming signal/interaction, which can affect the behavior of the system in a random manner. This does not deal with our poor knowledge of the coordinates, impulses, and potential energy functions of the particles. This does not deal with our computers truncating real numbers and causing large discrepancies of trajectories after tiny changes of initial conditions like in the unstable chaotic systems. The future in such a system is \textit{objectively random}. He has derived the formula: \textit{if the universe is infinite, then randomness is objective}. Years later the student was proud to know that the most productive modern mathematical definitions of random sequences could not avoid using infinity \cite{kolmogorov1987}, \cite{uspenskii1990}. A philosophical and historical tour into the topic of $\infty$ is prepared by David Foster Wallace \cite{wallace2010}.

\subsection{To "shy criticism"}

A "random process" is a "random function" $x(t)$ of an independent variable $t$ \cite{korn1968}. This clarifies nothing. Each trial supplies $X(t)$ referred to as the process realization. Accordingly, each process can be considered as an aggregate of either realizations or "random variables" dependent on the parameter $t$. The joint probability distribution of these variables must be given. This does not clarify "randomness" but points to two complementary ways to study "random processes". However, both considerations fail, when we think about the market. Each a-b-c-process is a unique realization of a "random process". We cannot collect realizations under the \textit{same} conditions. Judging about the process by a single realization is not expected in this method. Is the second way, random variables dependent on $t$, suitable? Ironically, here we judge about each "random variable" by a single observed value. A student will confirm that this is incorrect application of the probability theory. This paragraph is a supplement to \textit{"Shy Criticism into the Address of the Probability Theory"} in \cite[pp. 155 - 158]{uspenskii1990} prompted by the market. It would be useful to estimate the randomness or complexity of an individual object such as a unique realization of the a-b-c-process without comparison the latter with an ensemble of non-existing realizations required by the probability theory, for instance, by observing initial growing segments of the a-b-c-realization.

\subsection{The role of distribution}

If the heads to tails ratio in a growing sequence of coin trials is $\approx \frac{1}{9}$, then many say that the sequence is not random and indicates an unfair coin. If the coin is manufactured with the $\approx \frac{1}{9}$ property, then the randomness is accepted. A pseudo random generator can be programmed to supply 0s and 1s with the frequencies $p = 0.1$ and $1-p = 0.9$ \cite[p. 120]{knuth1998}. These are Bernoulli trials, where \textit{our decision to assign the label "random" to a sequence depends on a probability distribution}. In a \textit{Markov situation}, probabilities to get 0 and 1 on the $n$th place depend on previous outcomes. On Figure \ref{ZCN13_20130328_price_limit_down} prior 11:00:00 prices are random. Between 11:00:00 and 11:15:00 prices are "crazy" random. After 11:15:00 the price does not change until closing. A constant price is not a random variable but the situation is random: trading is not over and sessions, where the price leaves the limit, occur. Would the limit price 676 hold was unknown at 12:00:00 on March 28, 2013. In each case there is "own" randomness \cite[p. 109]{uspenskii1990}. Without losing generality, Bernoulli trials with $p = \frac{1}{2}$ are good to clarify \textit{whether an individual sequence of zeros and units can be random}. Kolmogorov and Uspenskii \cite[p. 391]{kolmogorov1987} introduce the set of all finite words of 0 and 1, \textit{chains}, $\Xi$, the set of all infinite \textit{sequences} $\Omega$, and their union $\Sigma = \Xi \cup \Omega$ \cite[p. 107]{uspenskii1990}. They classify sequences in $\Omega$ as: a) \textit{stochastic S}, b) \textit{typical T}, and c) \textit{chaotic C} and discuss how well the $S$, $T$, $C$ represent a set of random sequences $R$ for which the known probability laws are valid.

\subsection{Stochasticity}

A sequence $(a_n)$ is stochastic, $(a_n) \in S$, based on the \textit{stability of frequencies}. The frequency of zeros $f_0$ in the beginning segments of $(a_n)$ must tend to $\frac{1}{2}$: $(0)$ is not stochastic; (01), to a common dissatisfaction, is. Thus, the restriction is added: $f_0=\frac{1}{2}$ must be observed in subsequences selected using \textit{admissible rules}. The keyword is "admissible". The rule "pick up all 0s" creates a subsequence with $f_0=1$ or an empty one. It is not admissible similar to all rules requiring the knowledge of the value of the member to be selected. The rule "pick up members from even positions" is admissible. It rejects both $(0)$ and $(01)$. Therefore, $S$ is defined with respect to a family of admissible selection rules. The trunk of this direction is \cite{mises1919}, \cite{church1940}, \cite{kolmogorov1963}, \cite{loveland1966}, \cite{kolmogorov1987}, \cite{uspenskii1990}. von Mises sees the selection problem but does not suggest a rules family. Alonzo Church proposes the algorithm with the domain $\Xi$ and range $\{Yes, No\}$. Given the chain  $a_0, a_1, \dots , a_{n-1}$, it selects the member $a_n$ based on the specified function $\phi:\Xi \rightarrow \{Yes, No\}$. A decidable set $D$ must be designated so that $\phi(x)=Yes$, if $x \in D$, and $\phi(x)=No$, if $x \in \Xi \setminus D$. In the Church's stochastic class $CS$, named \textit{Mises-Church random sequences} \cite[p. 399]{kolmogorov1987}, the order of members in a subsequence is the same as in the sequence. A computable permutation of a $CS$-sequence may create a non $CS$-one. Some $CS$-sequences do not satisfy the \textit{law of the iterated logarithm} \cite[pp. 432 - 434]{gnedenko1988}. The $CS$ is too wide for $R$ \cite[p. 397]{kolmogorov1987}: $R \ne CS$, and $R \subset CS$. Kolmogorov and Loveland independently extend the Church's definition. Now, the order of reviewing the elements of a sequence can be chosen. This class $KS$ is named \textit{"Mises-Kolmogorov-Loveland random sequences"} \cite[p. 399]{kolmogorov1987}. A computable permutation of a $KS$-sequence is again a $KS$-one. Thus, $KS \ne CS$. It is narrower. The question whether a sequence obtained from a $KS$-sequence after application of the admissible rules is always a $KS$-one remains open. $CS$-sequences have this property. Examples of $KS$-sequences violating the laws of probability theory are unknown: $R \subset KS \subset CS$. The question \textit{"Is $KS=R$ true or false?"} remains open \cite[p.398]{kolmogorov1987}.

\subsection{Typicalness}

A sequence is typical, $(a_n) \in T$, if it does not belong to the \textit{maximal set} having \textit{effectively measure zero} or being \textit{effectively negligible}. We are indebted by this direction to Per Martin-L\"{o}f \cite{martin-lof1966}, \cite{kolmogorov1987}, \cite{uspenskii1990}. It borrows notions of the measure theory \cite{kolmogorov1999} and the theory of algorithms \cite{uspenskii1987}, particularly, computable functions \cite{uspenskii1960}. A set is "zero", if its measure is zero. The term "effectively" implies existence of a computable function generating set elements. The term "maximal" associates with a union of individual sets of effective measure zero. Let $x \in \Xi$ is a chain of the length $l(x)$ and $\Omega_x$ is a subset of all infinite continuations of $x$. The empty sequence in $\Omega$ is denoted $\Lambda, \; l(\Lambda)=0$. The $\Omega_x$ form the class of Borel subsets of the space $\Omega$. A unique measure $\mu$ can be defined by the values $\mu(\Omega)=1$ and $\mu(\Omega_x) = m(x)$, where $m(x)$ is a function such that for any $x$ $m(\Lambda)=1$, $m(x) = m(x0)+m(x1)$, $m(x) \ge 0$ \cite[p. 188]{uspenskii1990}. The set $A \subset \Omega$ is zero, if for any $\epsilon > 0$, there exists the sequence $x_0, x_1, \dots$ of chains for which $A \subset \Omega_{x_0} \cup \Omega_{x_1} \cup \dots$, and $\sum m(x_i) < \epsilon$. A countable union of zero sets is a zero set. Let $\epsilon$ is \textit{rational} and there exists a computable function $X: (\epsilon, i) \rightarrow X(\epsilon, i)$ with the range in $\Xi$. The $\mu$ is a \textit{computable distribution} \cite[p. 119]{uspenskii1990} on $\Omega$, $A \subset \Omega$. The $A$ is effectively measure zero, if for any $\epsilon > 0$ $A \subset \Omega_{X(\epsilon, 0)} \cup \Omega_{X(\epsilon, 1)} \cup \dots$, and $\sum_i \mu(\Omega_{X(\epsilon,i)} < \epsilon$. The Martin-L\"{o}f's theorem (one of the forms) states: \textit{if the distribution $\mu$ is computable, then there exists the maximal by inclusion effectively zero set containing any other effectively zero set. In other words, the union of all effectively zero sets is an effectively zero set}. This implies that the complement in $\Omega$ to the maximal effectively zero set is effectively measure one set. The sequences from the latter are suggested by Martin-L\"{o}f to be named the random sequences and by others \textit{Martin-L\"{o}f random sequences} \cite[p. 393]{kolmogorov1987}. This class is narrower than $KS$ and $CS$. Computable \textit{pseudo} and \textit{quasi} random sequences are not random by Martin-L\"{o}f \cite[pp. 199 - 200 (Russian), p. 178 (English)]{martin-lof1966}. Examples are uniform \cite{tausworthe1965}, \cite{lecuyer1988}, \cite{park1988}, \cite{lecuyer1996}, \cite{matsumoto1998}, normal, exponential, and some other periodic sequences \cite{box1958}, \cite{marsaglia1964}, \cite{marsaglia2000}, \cite{doornik2005}, and low discrepancy sequences \cite{sobol1979}, \cite{niederreiter1992}, see also \cite{lidl2008}. This corresponds to a common sense. \textit{The C++ Standard Committee}, \textit{JTC1/SC22/WG21}, has made a present to C++, \cite{stroustrup2000}, programmers. It has included a comprehensive \textit{random number generation} framework, section 26.5, into the \textit{ISO/IEC 14882:2011 Programming Language C++ draft}. The C++ \textit{typedef mt19937\_64} aliasing the 64 bits uniform generator template class \textit{std::mersenne\_twister\_engine}, see \cite{matsumoto1998}, has required behavior: \textit{the 10000th consecutive invocation of a default-constructed object of type mt19937\_64 shall produce the value 9981545732273789042}. There is no unpredictability here and the sequence, which has the tremendously successful but still period $2^{19937}-1$, belongs to an effectively null set. If an a-b-c-process would be a computable sequence (chain for futures), then it would be non-random, atypical, in the Martin-L\"{o}f's system. Typicalness by Schnorr and Solovay is discussed in \cite{uspenskii1990}.

Reviewing a uniform pseudo random chain passed many of George Marsaglia's \textit{Diehard} \cite{marsaglia1995} and Pierre L'Ecuyer's and Richard Simard's \textit{TestU01} tests of randomness, one may formulate the task of recovering an algorithm responsible for the chain. Such algorithms assume a \textit{seed}, \textit{set of states}, and \textit{transition rules}. All these items have to be discovered. Traders dream about an algorithm, mind state lit up with foresight, or trained intuition breaking the rules of at least short a-b-c-chains. An axiomatic postulation that this is impossible will never satisfy traders thinking about such a delicate topic. How can one prove that this is impossible? Absence of a definite proof is one of the reasons why new generations of traders continue the journey for the Holy Grail. The author's observation is that the reason is the market itself: its behavior frequently creates big profit opportunities. The author names it the main law of the speculative market and suggests to measure it using the maximum profit strategy.

\subsection{Chaoticity}

A sequence is chaotic, $(a_n) \in C$, if it is "complex". The Kolmogorov's idea is that randomness is absence of regularities. Regularities shorten a program computing a sequence. Traders dream about a program, rule, system, pattern, or signal, which anticipates future price moves. Kolmogorov associates complexity of a sequence with the length of the shortest description. He switches from the Shannon's and Khinchine's interpretation of entropy as a an informational measure based on the probability theory to a new foundation, where the theory of algorithms underlines both the information and probability theory. From this position, he talks about \textit{complexity} \cite{kolmogorov1965} and latter \textit{entropy} \cite{kolmogorov1968} of an \textit{individual object}, which cannot be treated by the classical probability theory. The creator of the probability axioms proposes a new logical basis. The author believes that the a-b-c-chains are such individual objects.

Let $K$ is a natural number and measure of the complexity of finite binary chains like $y_n=a_0a_1\dots a_{n-1} \in \Xi, \; K:\Xi \rightarrow N$. At most $n$ bits are sufficient to specify $y_n$. A complexity measure of such a chain should not be greater than the length $n$. This is because in the absence of short descriptions, the chain $y_n$ can be fully described by writing down all its $n$ digits. Logically, for a chaotic sequence the complexity of its beginning segments $K(a_0), \; K(a_0a_1, \; \dots)$ should grow fast. Such sequences, lacking regularities, are candidates for naming them random. It is in this sense the theory of probability gets the theory of algorithms as a new logical basis.

In contrast with the entropy, playing a role of the measure of information and defined for a collection of objects obeying probability laws, in a discrete case $H = -\sum_{i}p_i\log_2(p_i)$, the new entropy is a property of an individual object and represents its complexity. It becomes a synonym of complexity of an object relative to an optimum description process. It is in this sense the theory of information gets the theory of algorithms as a new foundation.

Let $Y$ be a fixed set of objects with descriptions $x$ belonging to $\Xi$ . An object $y^*$ may have multiple descriptions $x_1^*,x_2^*,\dots$. Each $x^*$ may describe several objects $y_1^*,y_2^*,\dots$. The total number of pairs $(x,y)$ is the cartesian product $\Xi \times Y$. The pairs, where $x$ is a description of $y$, form a subset $E \subset \Xi \times Y$. The set $E$ is referred to as the descriptive process \cite[p. 394]{kolmogorov1987}.  Then, $x$ is a description of the object $y$ relative to the descriptive process $E$. The complexity of an object relative to a descriptive process is defined as the length of the shortest description $K_E(y)=\min(l(x) | (x,y)\in E)$. The complexity is set to infinity, if there is no a description of the object in the descriptive process. If $U$ is a family of descriptive processes, then the Kolmogorov's theorem \cite{kolmogorov1965} states that there exists an $A \in U$ such that, for every $E \in U$, $K_A(y) \le K_E(y) + c_E$, where the constant $c_E$ does not depend on $y$. Kolmogorov \cite[p. 664]{kolmogorov1968}: \textit"Credit for noting this relatively simple condition evidently belongs to Solomonov and me". The article of Ray Solomonoff \cite{solomonoff1964} is in scope. $\frac{1}{2}$-Bernullian sequences, a starting Martin-L\"{o}f's point, are independently developed by George Chaitin \cite{chaitin1966}. The \textit{entropy} of $y$ is its complexity relative to an optimum descriptive process. It is determined to within an additive constant \cite[p. 394]{kolmogorov1987}. Now, a binary sequence $a_0a_1\dots$ is \textit{chaotic} if and only if $K(a_0a_1\dots a_{n-1}) \ge n - c$ for some constant $c$ and all $n$. It depends on the family $U$. The latter can be selected differently and one proposal, \textit{monotone complexity} or \textit{monotone entropy}, is from Leonid Levin \cite{levin1973} (see also \cite{levin1974}). It is denoted by $L$ and must have two properties: 1) \textit{preservation of comparability} and 2) \textit{countability}. A descriptive process $E$ preserves comparability, if it has the property: $(x_1,y_1)\in E \wedge (x_2,y_2) \in E \wedge (x_2$ is a continuation of $x_1) \Rightarrow (y_2$ is a continuation of $y_1) \vee (y_1$ is a continuation of $y_2)$. A set is countable if it is in the range of values of a computable sequence. Finally, the class $C$ contains all sequences, where, for all $n$, $L(a_0a_1\dots a_{n-1}) \ge n - c$ and $c$ does not depend on $n$ \cite[p. 395]{kolmogorov1987}. The \textit{Levin-Schnorr theorem} establishes coincidence of $T$ and $C$. These sequences obey the known laws of probability theory \cite{vovk1987} and $R=C=T$ gets a rigorous definition. 

\subsection{A finite case}

Appealing to infinity cannot satisfy a practitioner. Kolmogorov and Uspenskii review a hypothetical finite binary chain, which one could intuitively name "random". Gradually replacing 1s with 0s, they get the chain of 0s, which nobody names "random". They did not find any intermediate state with a boundary between "random" and "not random" and conclude that there is more sense to speak about reduction of complexity within such a process. This leads to the Kolmogorov's \textit{defect of randomness} or \textit{defect of being chaotic}: $d(y|M)=\log_2|M| - H(y|M)$, where $M$ is an arbitrary set of binary chains, $y \in M$, $|M|$ is cardinality of $M$, and $H(y|M)$ is a computable conditional entropy of $y$. The latter is given by Formula (2) in \cite{kolmogorov1968} (VS: $x$ and $y$ are swapped): $H(y|x) = \min_{A(P,x)=y} l(P)$, where $P$ is a program building the object $y$, given object $x$, $A$ is an optimum method of programming, and $l(P)$ is the length of the program. The method is optimum if, for any other method $F$, $H_A(y|x) \le H_F(y|x) + c_F$, where $c_F$ is a constant independent on either $x$ or $y$. The index $A$ can be dropped since entropies corresponding to two optimum methods differ by a constant. The \textit{$\Delta$-randomness} is introduced for a small number $\Delta$ so that $d(y|M) \le \Delta$. A chain of ZCN13 prices on March 28, 2013 after 11:13:24 is represented by 1150 ticks with the total volume 1943 and constant price 676. This does not account complexity of non-constant a-increments. The 1150 constant prices can be "programmed" by a short description comparing with possible chains of prices obtained by adding chains of b-increments, obeying Hurwitz distribution with a binomial simulation of the sign of b-increments, to the initial price 676 at 11:13:24. This would make cardinality of the set large (finite due to the limit price required by the CME contract specifications), but $H(Price|M)$ small. We can expect big defect of randomness.

820 references can be found in \cite{li2008}. A recent book on these topics is \cite{downey2010}. The notions described in this section are waiting their application to the market time-series, a-b-c-chains. 

\subsection{A Kolmogorov's typo?}

Reading the high caliber paper of Kolmogorov \cite{kolmogorov1968}, the author could not overcome a feeling that there is the same typo on pages 664 (English) and 6 (Russian) in the sentence: \textit{Our basic formula (1) implies a "universal programming method" A}. Formula (1) relates to a traditional definition of entropy in a probabilistic sense. Formula (2) $H(x|y) = \min_{A(P,y)=x} l(P)$ is the innovation. Kolmogorov promises on pages 662 (English) and 4 (Russian): \textit{We will return again to the interpretation of the notation $A(P, y) = x$.} The discussed sentence is the place, where the exposition returns to $A$. Logically, the sentence should reference Formula (2) but not (1). Otherwise, understanding is made difficult.

\subsection{The message}

A remarkable and unexpected circumstance is that in all cases the randomness of a sequence is defined in terms of the theory of algorithms \cite[p. 389]{kolmogorov1987}. Once an individual random object is defined, the constant down to the limit price on March 28, 2013 can be classified as a low complexity chain. This does not eliminate the task of searching an ontological basis of randomness by philosophizing followers of Alfred R\'{e}nyi \cite[Letters on Probability]{renyi1980}, and Donald Knuth \cite[p. 18, 25, pp. 183 - 184]{knuth2001}.
 
Writing this section, the author imagined the effort of Jimmie Savage sending out of Yale a dozen postcards about the  Louis Bachelier's 1900 Sorbonne thesis until Paul Samuelson was \textit{the only fish to respond to Savage's cast} \cite[p. vii, Samuelson's Foreword]{davis2006}. This section is to attract attention of traders and economists to individual random objects successfully introduced by the theories of algorithms and probability during the last 50 (coincidence) years. The market is a candidate for utilization of these achievements, which has already started \cite{vyugin2006}. Internet and \url{http://arxiv.org/} are better means than postcards.

\section{The Maximum Profit Strategy}

A \textit{trading strategy} is a chain of bought and sold units. Analysts mean by a trading strategy a non-anticipative process reflecting account positions and/or their changes. Replicating self-financing admissible strategies is a key for pricing derivative instruments \cite[p. 16]{hunt2000}. Traders and developers of systems making automatic decisions mean by strategy rules for \textit{buy}, \textit{sell}, and \textit{do nothing} actions.

\subsection{Profit and loss}

Trading actions can be expressed by whole positive and negative numbers of bought and sold contracts or shares. These actions are buy and sell transactions. Zeros mark do nothing actions, which are not transactions. Each transaction assumes a cost. A transaction triplet of an a-b-c-process can be extended to $(t_i, P_i, V_i, C_i, U_i)$, where $C_i$ and $U_i$ is a transaction cost per unit and strategy action. For futures, $C_i$ is a constant or function of $U_i$. For stocks, $C_i$ can be a fixed percent of $P_i$ and function of $U_i$. Given $P_i, C_i, U_i$, the profit and loss of the strategy, $PL$, can be computed as \cite[p. 35]{salov2007}
\begin{equation}
PL = k \left(P_n \sum_{i=1}^n U_i - \sum_{i=1}^n P_i U_i \right ) - \sum_{i=1}^n C_i |U_i| - C_n |\sum_{i=1}^n U_i |,
\label{EqPL}
\end{equation}
where $k$ is the dollar value of one point, Table \ref{table_futures_properties}. For ES it is $\frac{\$12.50}{\delta=0.25}=\$50.00$. A net action is the algebraic sum of actions. If $\sum_{i=1}^n U_i = 0$, then a position expressed in the number of contracts remains intact after application of the strategy. Formula \ref{EqPL} handles the total non zero net action by adding the mark-to-market offsetting action at the last price $P_n$. Being out of the market, buying or selling short a futures contract requires the same \textit{initial margin}, commissions, and fees. A position must hold the mark-to-market equity above the \textit{maintenance margin} - a fraction of the initial one. If price moves against the position and the equity drops below the maintenance margin, which must be aggregated for all open positions, then a broker demands to add money to the account or some positions will be liquidated by the broker. This is the \textit{margin call}. It is unwise to open a large position leaving no equity on the account.
\begin{table}[!h]
  \centering
  \topcaption{Three weeks of ESZ13 last prices occurred at 16:14:59 CT each and five hypothetical strategies. The cost per contract per transaction is \$4.66. Initial and maintenance margins are \$1127.50 and \$1025.00.}
  \begin{tabular}{ccrrrrr}
   Date & Price & BuyHold & Emotional & MPS0 & MPS1 & MPS2\\
2013-11-04 & 1763.00 & 1 & 1 & -1 & -1 & -1 \\
2013-11-05 & 1756.75 & 0 & 0 & 2 & 2 & 2 \\
2013-11-06 & 1764.00 & 0 & 0 & -2 & -2 & -2 \\
2013-11-07 & 1744.75 & 0 & 0 & 2 & 3 & 3 \\
2013-11-08 & 1766.75 & 0 & -1 & 0 & 0 & 2 \\
2013-11-11 & 1767.75 & 0 & 1 & -2 & -6 & -8 \\
2013-11-12 & 1764.50 & 0 & -1 & 2 & 8 & 9 \\
2013-11-13 & 1779.50 & 0 & 1 & 0 & 0 & 3 \\
2013-11-14 & 1787.75 & 0 & 0 & 0 & 0 & 3 \\
2013-11-15 & 1794.25 & 0 & 0 & -2 & -14 & -25 \\
2013-11-18 & 1789.25 & 0 & -1 & 0 & 0 & -3 \\
2013-11-19 & 1785.00 & 0 & 0 & 0 & 0 & -3 \\
2013-11-20 & 1779.75 & 0 & 0 & 2 & 26 & 45 \\
2013-11-21 & 1793.75 & 0 & 0 & 0 & 0 & 15 \\
2013-11-22 & 1801.00 & -1 & 0 & -1 & -16 & -40 \\
 & PL & 1890.68 & 484.54 & 6150.44 & 34424.00 & 58910.80 \\
 & DD & -962.50 & -962.50 & -4.66 & -4.66 & -4.66 \\
 & Trades & 1 & 3 & 8 & 8 & 14 \\
 & Wins & 1 & 2 & 8 & 8 & 14 \\
 & Losses & 0 & 1 & 0 & 0 & 0 \\
  \end{tabular}
  \label{TblPL}
\end{table}
Table \ref{TblPL} illustrates Formula \ref{EqPL}.

\subsection{Buy and hold}

The smallest account for trading one contract is equal to the initial margin. With \$1127.50 the \textit{buy and hold strategy} faces margin calls on November 5, 6, and 7, 2013. At least \$1942.16 was a must to sustain the largest drawdown, DD, on November 7 with the long position entered at 1763 on November 4. Inability to transfer additional money to the account would result in the position liquidation and actual loss. A well credited account would accumulate a decent profit.

\subsection{Emotional trading}

It also requires \$1942.16. After buying at 1763 and experiencing \$312.50 equity drop on the next day, a hope returned with the price 1764 on Wednesday November 6 indicating the mark-to-market profit: $\$50 \times (-1763 + 1764) - 2 \times \$4.66 = \$40.68$. Thursday ruins the \textit{plan to become rich quickly}: a) the equity drops from the previous maximum 1764 by \$962.50, and b) potential loss on the position including \$3.50 commissions and \$1.16 \textit{Exchange and Clearing Fees} (total \$4.66) for the offsetting transaction is $\$50 \times (-1763 + 1744.75) - 2 \times \$4.66 = -\$912.50 - \$9.32 = -\$921.82$. One did not like to take the profit \$40.68 only in order to see the loss \$921.82. Being exhausted, the trader would be glad to take any profit. The opportunity arrives on Friday November 8: it is $\$50 \times (-1763 + 1766.75) - \$9.32=\$178.18$. The position is closed and profit is pocketed with \textit{all worries left in the past}. Ironically, after the weekend the price continues moving up one point, \$40.68 (after costs). Regretting about the premature exit, the trader \textit{jumps on the train} but only in order to see the next day loss $\$50 \times (-1767.75 + 1764.50) - \$9.32 = -\$171.82$ - depression and taking the loss. But on Wednesday November 13 the price is unbelievably high \$1779.50. "Oh, how much money would be made, if there would be just a little bit more patience and bravery. I have to be on the long position. Is the date 13 unlucky for a buy? Enter at 1779.50!" Finally, the price moves up to 1794.25 and begins a new retrace. At 1789.25 on November 18, 2013 the trader takes the profit $\$50 \times (-1779.50 + 1789.25) - \$9.32 = \$478.18$. The total PL is $\$178.18 - \$171.82 + \$478.18 = \$484.54$.

\subsection{MPS0}

What are the most profitable transactions for the 15 prices? The maximum number of traded contracts is $\lfloor \frac{\textrm{Account}}{\textrm{IMargin}} \rfloor=\textrm{int}(\frac{\textrm{Account}}{\textrm{IMargin}})= \textrm{floor}(\frac{\textrm{Account}}{\textrm{IMargin}})$. Investing more with the right transactions increases the profit. The answer is determinate, if the number of contracts on long and short positions is restricted to one. Then, Column MPS0 is a chain of transactions giving the maximum profit, if the transaction cost per contract is \$4.66. This is essentially a \textit{reversal strategy} switching from long to short and vice versa positions after entering the market. The last optimal mark-to-market transaction is added, if the position is not closed prior the end of the chain. A MPS always "exits" the market after "entering" it. It may not have one transaction. It has no losing trades. If the a-b-c-chain has no price fluctuations exceeding transaction costs, then MPS degenerates to the \textit{do nothing strategy}, which is always available. The drawdown still can be non-zero, if the transaction cost is non-zero. In Table \ref{TblPL} MPSs start with a single contract and initial margin \$1127.50.

\subsection{MPS1}

A MPS0 gradually accumulates profits. If they are reinvested at the \textit{price swing points}, then a greater profit potential is exploited. The size of positions grows as long as the account reaches new multiples of the initial margin. This leads to MPS1, \textit{the first P\&L reserve strategy}, which has the same number of transactions as MPS0. 

\subsection{MPS2}

Often, it is beneficial to increase a position between swing points, if the grown equity and initial margin requirements permit it. This results in MPS2, \textit{the second P\&L reserve strategy}. It may contains more transactions than MPS0. \textit{There is no way to make a greater profit trading futures using a self-financing strategy other than MPS2}. 

The proofs, l- and r-algorithms generating the optimal transactions, and basic properties of MPS0, MSP1, and MPS2 are in \cite{salov2007}, see Appendix F. MPS0 is a fundamental market characteristic and basis for MPS1, and MPS2. One cannot apply these strategies because the last optimal transaction depends on future prices. Inability of knowing in advance future prices does not make the latter less objective market properties. \textit{Similarly, MPS is an objective market property dependent on prices, transaction costs, and trading regulations such as initial and maintenance margin requirements}.

\section{The MPS0 Properties}

The author does not repeat in this section the MPS0 properties outlined in \cite{salov2007}. Only new properties and data are discussed.

\subsection{Transactions Spectra}

\begin{figure}[h!]
  \centering
  \includegraphics[width=130mm]{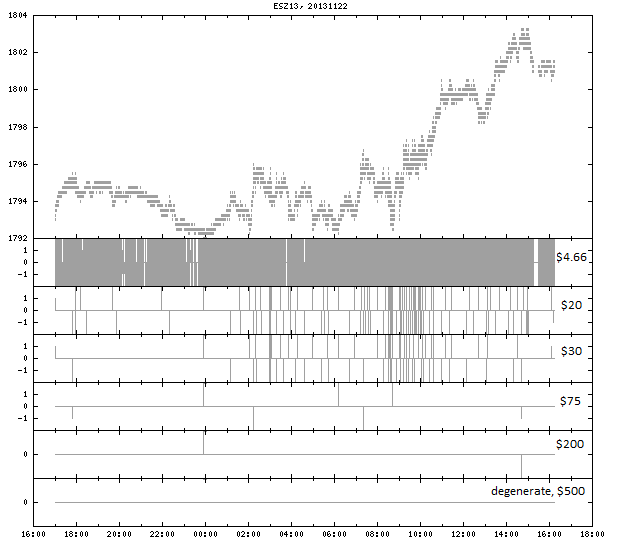}
  \caption[fg_ESZ13_mps]
   {ESZ13 traded on Friday November 22, 2013. Spectra of MPS0 transactions depending on the transaction cost per contract: \$4.66, \$20, \$30, \$75, \$200, and \$500 (do nothing strategy). Setting line color red and violet at the left and right side and making a smooth color gradient proportional to the seconds elapsed along the trading session would strength the spectral analogy.}
  \label{fg_ESZ13_mps}
\end{figure}

Zero transaction costs give Pardo's potential profit (a number), \cite[pp. 125 - 126]{pardo1992}, \cite{salov2007}. We still can get a list of related optimal transactions (a vector) \cite{salov2007}. Pardo did not formulate the latter task. Increasing cost linearly decreases profit of optimal trades and non-linearly affects the number of optimal transactions. Strategy actions can be depicted as vertical lines directed up and down for buy and sell transactions with the height proportional to the number of traded contracts. Such plots resemble atomic linear spectra and cause associations with quantum properties of electrons in atoms. They are insensitive to decreasing or increasing transaction cost within certain limits and then abruptly several lines appear or disappear indicating a new \textit{"quantum state"}. An analogy is the \textit{photoelectric effect}. The main contribution into its experimental study was done by Alexander Stoletov, 1888 - 1891. Einstein has explained the phenomenon theoretically \cite{einstein1905a} (\textit{"The Nobel Prize in Physics 1921 was awarded to Albert Einstein "for his services to Theoretical Physics, and especially for his discovery of the law of the photoelectric effect"."}, \url{http://www.nobelprize.org/nobel_prizes/physics/laureates/1921/}). In our case, the explanation is the discrete properties of prices and discrete nature of trading: inability to trade fractional number of contracts. With cost approaching infinity, we get a do nothing strategy, zero maximum profit, and no lines in the spectrum of transactions for any given a-b-c-chain, Figure \ref{fg_ESZ13_mps}.

\subsection{Maximum profit vs. transaction cost}

In a private conversation, Vadim Zharnitsy has suggested that Pardo's potential profit corresponds to the total variation of a function. The latter is defined on the interval $[a, b]$ as $V_a^b(f)=\sup_{partition} \sum_{i=0}^{n_{partition} - 1}|f(x_{i+1} - f(x_i)|$. For a discrete a-b-c-chain a natural, finest, and single partition is given by all ticks. The transaction cost is zero. In order to convert the variation obtained from prices $P_i$ into dollars, one needs to multiply it by $k$. In terms of transactions, this can be interpreted so that at each tick a previous position is offset and a new one in entered. Even, for this artificial zero cost case, the l- and r-algorithms return the same profit but a different spectrum of transactions. These algorithms assign monotonically increasing or decreasing local prices to one trade reducing the number of trades. This note clarifies that removing $k|P_{i+1} - P_i| \le C_{i+1} + C_i$ from the sum can give a wrong result, if the cost $C_i > 0$. The later condition makes the total variation of a function interpretation inapplicable and the l- and r-algorithms more complex. Their algorithmic complexity remains linear $O(n)$. This creates a computational advantage over the "less certain" \textit{genetic algorithms} \cite{holland1969}, \cite{holland1973}, \cite{goldberg1989} accommodated to this task. The author, originally building MPS manually, often was able to find a better strategy after one was claimed the best. Only after proving a few theorems and establishing the essential reversal property of MPS0 \cite{salov2007} the results got desirable stability. Reversing a position, offsetting existing and entering opposite one at one price, is a known aggressive and risky trading technique. It can cause a financial disaster, if the market does not confirm the swing point. The last MPS0 transaction has the mark-to-market accounting sense. Whether it is the right swing price depends on the future. However, all previous swing points are fixed. New prices cannot change them. This property allows to use MPS0 transactions as trading signals suitable for building real strategies. In contrast with MPS0, the latter can lose money. A detailed example is described in \cite{salov2008}.

The total variation of a function is a good analogy. It should be used with care because the a-b-c-process is a stochastic but not deterministic function. We have seen that functions of a simpler Brownian motion cannot be integrated in time properly using only the limits of the Riemann or Stieltjes sums. The Ito's approach justifies additional terms and one possible meaning of a stochastic integral. This should be remembered to treat MPS0 profits stochastically. While these words denote a task of theoretical stochastic treatment of the a-b-c-process and MPS, the author believes that more time is required to study the experimental a-b-c-properties. Otherwise, there is a danger of hasty assumptions "important" for simplification of building a theory but useless for trading. Insufficiently justified assumptions accompanied by scientific terms is a form of pseudoscience. They can be forgiven being delusions but not intentions. 

\begin{figure}[h!]
  \centering
  \includegraphics[width=130mm]{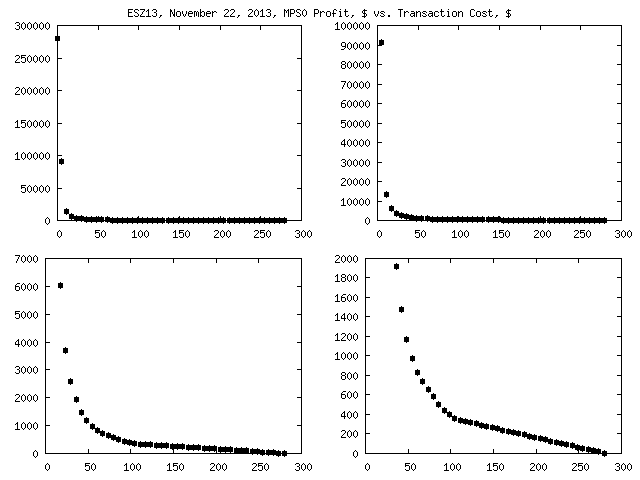}
  \caption[mps0_vs_cost_ESZ13_20131122_2_2]
   {ESZ13 traded on Friday November 22, 2013. The same dependence MPS0 profit vs. transaction cost is zoomed in four parts.}
  \label{mps0_vs_cost_ESZ13_20131122_2_2}
\end{figure}

A plot of a typical dependence of the MPS0 profit vs. transaction cost is on Figure \ref{mps0_vs_cost_ESZ13_20131122_2_2}.
The left most point is the zero cost and Pardo's potential profit \$279875.00. This value cannot be achieved because at least Exchange fees are applied as a cost of each transaction. The second point is the transaction cost \$4.66 or \$9.32 per trade. These are realistic numbers for retail traders. Sometimes the author hears that a tremendous profit could be made with a perfect anticipation of price moves. Usually, this is not followed by any concrete number. With the retail futures cost, trading a single ESZ13 contract on each open position on November 22, 2013 would profit \$91,452.56.

The largest profit is generated by MPS2. It quickly increases the trading positions above the limits implied. Table "Position Limits, Positions Accountability and Reportable Level" from CME Rulebook \url{http://www.cmegroup.com/rulebook/CME/I/5/5.pdf} is a reference guide. On September 23, 2013, CME Group announced about increasing the position limits for S\&P 500 Stock Price index futures and E-mini S\&P 500 Stock Price Index futures from 20,000 to 28,000 and from 100,000 to 140,000 contracts. This should be counted in net futures equivalents. It is difficult to imaging a 100,000 contracts order. It would strike the price. In accordance with the CFTC and SEC report \cite{us2010}, \textit{"The 2010 Flash Crash"} also known as \textit{"The Crash of 2:45"} was caused by a total sell of 75,000 E-mini S\&P 500 contracts by a mutual fund complex. This was a hedge to an existing equity position. The E-mini price plunged 3\% down in less than four minutes. 75,000 is 1.3\% of the total volume $\approx 5,700,000$ of E-mini S\&P 500 contracts traded during the dramatic Thursday May 6, 2010.

Reinvesting is a powerful mechanism to increase profits during the same period of time, if the transactions are right. There are 38064 ESZ13 transactional ticks with the total volume 129950 on November 22, 2013 between 08:30:00 and 08:52:23. MPS2 starting with one contract produces the summary:
\begin{verbatim}
ACCOUNT
Ticker              = ES
Initial account     = 1127.5
Initial margin      = 1127.5
Maintenance margin  = 1025
Point value         = 50
STATISTICS
Total P&L                        = 1.55152e+008
Total P&L/unit                   = 14721.1
Gross profit                     = 1.55152e+008
Gross profit/unit                = 14721.1
Gross loss                       = 0
Gross loss/unit                  = 0
Total number of trades           = 2884
Number of winning trades         = 2884
Number of losing trades          = 0
Average profit                   = 53797.5
Average profit/unit              = 5.10441
Average loss                     = 0
Average loss/unit                = 0
Largest winning trade            = 2.00583e+006
Largest winning trade/unit       = 40.68
Largest losing trade             = 0
Largest losing trade/unit        = 0
Max number of consecutive wins   = 2884
Max number of consecutive losses = 0
Maximum consecutive profit       = 1.55152e+008
Maximum consecutive profit/unit  = 14721.1
Maximum consecutive loss         = 0
Maximum consecutive loss/unit    = 0
Total elapsed seconds            = 1343
Total trade seconds on positions = 1675
Total trade profit seconds       = 1675
Total trade loss seconds         = 0
Maximum account value            = 1.55153e+008
Minimum account value            = 1122.84
Largest drawdown                 = -4.66
Average drawdown                 = -0.000122432
\end{verbatim}
Given arbitrary valid prices, costs, margins, account, and strategy the program summaries 31 statistics. The detailed report is too long. For MPS0, MPS1, and MPS2 the list may not get losing or break even trades. Thus, on November 22, 2013 in 20 minutes and 23 seconds after the S\&P 500 futures pit session opening the Globex's ESZ13 market had offered to a retailed trader with \$1127.50 the \$155,152,000 opportunity. This was an ordinary session. The program has been terminated because the last position reached 137,221 contracts. This is about the new CME position limit. This exceeds the total volume 129950 for this interval. The C++ program building MPS2 and using 32 bit registers with the largest unsigned value $2^{32}-1 = 4294967295$ cannot make a reliable calculation already for two, three hours of ESZ13 ticks due to the \textit{overfilling}. The size of transactions and profit growth exponentially in time. Accordingly, MPS1 and MPS2 are the tools complimentary to MPS0. They help to study intermediate ticks, where an established position can be increased prior a new swing price in the long or short direction of the position and never against it.

\subsection{"It is not a bug but a feature!"}

Programmers sometimes use this idiom. The total elapsed seconds in the list are 1343 = 08:52:23 - 08:30:00 = $22 \times 60 + 23$ but the seconds on the positions are 1675. How is this possible? If a strategy adds to a position at a different time and possibly price, then the position is complex. The strategy $(1_0, 1_1,-2_2)$ buys two contracts at times 0 and 1 and sells two contracts at time 2. For accounting purposes the three transactions are subdivided into two trades $(1_0,-1_2)$ and $(1_1,-1_2)$ with holding times $2 - 0 = 2$ and $2 - 1 = 1$. The sum is 3. However, the elapsed time is $2 - 0 = 2$. For a while, the two trades exist simultaneously, \textit{parallel in time}. The program computes position holding times in such a manner. The author believes that the true power of reinvesting is hidden in this parallelization. Reinvesting creates positions, which can be considered as parallel in terms of the smaller original positions. Like a multi core system, supercomputer, breaking the task in smaller parts and solving each on a dedicated CPU parallel in time, the reinvesting triggers several positions, threads, coexisting in time. A system with positive mathematical expectation of trades trading a fixed number of contracts gets power after reinvesting the gained profits. The optimal capital allocation for a next bet was considered by John Kelly \cite{kelly1956} and has the roots in the Shannon's work \cite{shannon1948}. The Kelly's formula ignores discreteness and permits trading fractional number of contracts. Because of that, it cannot wipe out an account \cite[Chapter 4]{salov2007}. The latter chapter considers optimal allocation of capital under a condition, where $\textrm{The Next Number of Contracts} = \lfloor \frac{\textrm{Allocation Fraction} \times \textrm{Account}}{\textrm{IMargin}}\rfloor$ can become zero, ruining account, and using the ECDF of wins and losses estimated from a list of previous trades. Other money management considerations are presented by Ralph Vince \cite{vince1992}, \cite{vince1995} and Ryan Jones \cite{jones1999}.

Returning to the ratio $\frac{\textrm{Total trade profit seconds}}{\textrm{Total elapsed seconds}} = \frac{1675}{1343} \approx 1.25$, the author suggests to name it the \textit{time coefficient of trading parallelization}.

\subsection{Do nothing threshold}

For a constant cost, the right boundary on Figure \ref{mps0_vs_cost_ESZ13_20131122_2_2} can be computed without building MPS0. Given an a-b-c-chain, the do nothing threshold is equal to $\frac{k(P^{max} - P^{min})}{2}$. From Figure \ref{fg_ESZ13_mps}, $P_{20131122 \; 14:42:13}^{max} = 1803.25$ and $P_{20131121 \; 23:54:21}^{min} = 1803.25$ give $\frac{\$50 \times (1803.25 -1792)}{2} = \$281.25$. If the cost would be \$281.24 per transaction per contract, then the strategy $(1_{20131121 \; 23:54:21}, -1_{20131122 \; 14:42:13})$ with the initial margin \$1127.50 would profit 2 cents. But the cost \$281.25 leads to a break even, exact zero after the cost, trade. Break even trades are excluded from the MPS0, MPS1, and MPS2.

\subsection{Number of optimal trades vs. transaction cost}

\begin{figure}[h!]
  \centering
  \includegraphics[width=130mm]{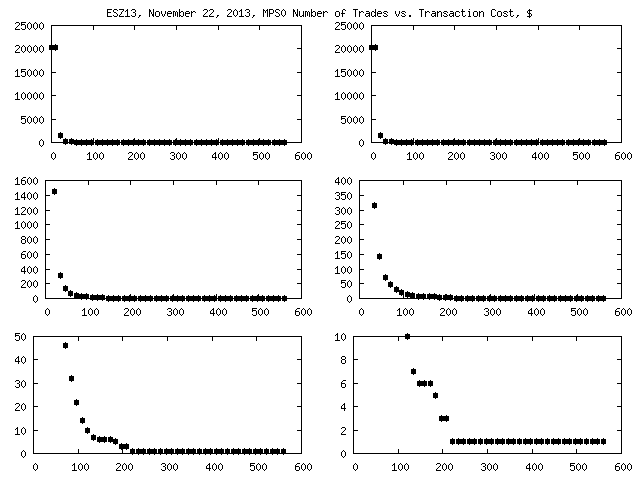}
  \caption[n_vs_cost_ESZ13_20131122_3_2]
   {ESZ13 traded on Friday November 22, 2013. The same dependence MPS0 number of trades vs. transaction cost is zoomed in six parts.}
  \label{n_vs_cost_ESZ13_20131122_3_2}
\end{figure}

The number of trades is the same for the zero and \$4.66 costs, Figure \ref{n_vs_cost_ESZ13_20131122_3_2}. This is because the break even trades are not included into MPS0 and already a one $\delta$ b-increment is enough for making the profit $\$50 \times 0.25 - 2 \times \$4.66 = \$3.18$. Due to the discreteness and inability to trade fractional number of contracts, recollect unrealistic feature of the Kelly's formula, we get a dependence, where several costs may generate the same number of trades. Naturally, that the profit on such intervals decreases linearly with the increasing cost, while the number of trades remains intact. Only after a critical cost the spectrum of transactions changes again. While the same trades are profitable for both costs zero and \$4.66 and their number 20217 is common, 
the difference of profits is $\$279875.00 - \$91452.56 = \$188422.44$. Division of this difference by two and 20217 yields exactly the cost per transaction per contract \$4.66.

\section{The Optimal Trading Elements}

In own publications, the author distinguishes \textit{transactions} and \textit{trades}. Transactions are individual buy and sell actions. Trades are pairs of offsetting transactions with the zero net action. An MPS exits a market, if it enters it. Its net action is zero. A trading position is \textit{simple}, if it contains contracts traded at one time and price. Otherwise, it is \textit{complex}. A position can be offset in a few transactions. There is a task to combine transactions into trades. For a zero net action, this is always possible. Several methods are known. Since MPS0 reverses simple positions between entry and exit, an aggregation into a sequence of trades is possible. The strategy $(1_0, -2_1, 2_2, -1_3)$ is divided into three sequential trades $(1_0, -1_1),(-1_1, 1_2),(1_2,-1_3)$. These trades divide an a-b-c-chain into a set of adjoint chains. In general, several MPSs may generate the same profit using the same a-b-c-chain and costs. This is possible, if sequential ticks have one price or the ticks with one price are separated by ticks where the price does not fluctuate enough to compensate the cost. The ESZ13 cost \$4.66 and prices $(1799.00_0, 1803.25_1, 1803.25_2, 1799.75_3)$ imply three MPS0s $(1_0,-2_1,1_3)$, $(1_0,-1_1,-1_2,1_3)$, $(1_0,-2_2,1_3)$ and one profit \$368.86. In each case there are two sequential trades returning \$203.18 and \$165.68. Sequential neighbors constructed by the l- or r-algorithms have a common tick.

Given the a-b-c- and cost chains (for futures the latter is a constant), MPS generates a chain of \textit{optimal trades}. \textit{An optimal trading element is a set of all properties associated with an optimal trade returned by an MPS} \cite[p. 39]{salov2011b}, \cite[p. 26]{salov2012}. The key feature is partitioning the a-b-c-process into fragments, where the maximum profit could be made. The MPS0 partition points are illustrated by the transactions spectra, Figure \ref{fg_ESZ13_mps}. The word "all" leaves the list of properties open. These can be a-b-c-properties, news, .... Due to the MPS0 reversal property, optimal trading elements, OTE, are bound to trades with alternating entering buy and sell transactions. Accordingly, each OTE is either a buy, BOTE, or sell, SOTE, one. BOTE and SOTE from MPS0 alternate. It is impossible to get neighboring OTEs of one sign. \textit{In terms of this approach, a trader buys and sells not gold, crude oil, or corn but OTEs}.

From now, the transaction cost will be used in two senses: \textit{filtering cost, f-cost}, and \textit{trading cost, t-cost}. The filtering cost is a an analytical tool specified for an MPS0 in order to get OTEs. OTE chains differ with respect to the f-cost. Evaluation of OTE profits is done using the t-cost. For futures we consider constant t- and f-costs in dollars per transaction per contract, $C^t \le C^f$.

\section{Properties of the MPS0 Optimal Trades}

\subsection{OTE trade direction}

This is the sign of the first transaction of an optimal trade. It divides OTEs on BOTEs and SOTEs. A new BOTE is born, when the price deviates from the last MPS0 low swing price more than it is defined by the f-cost. If the f-cost is \$75, then after reaching 1792 on 11/21/2013 at 23:54:21 the ESZ13 price should move up to 1795, when one detects the new BOTE. Only at this time, the previous SOTE is fixed and cannot change. If the price after 23:54:21 would go lower, then the previous SOTE would grow in profit and new reversal point would be established. Then, the three points up should be counted from that price. Similarly, after reaching 1803.25 on 11/22/2013 at 14:42:13 the price should move down to 1800.25 in order one would detect a new SOTE. At that time and price the previous BOTE would be fixed. This resembles the Alexander's filter \cite{alexander1961}. Alexander understood the potential of reversing from long to short and vice versa positions. Highly estimating the Alexander's contribution, the author wants to emphasize what deviates it from the current proposals: a) the Alexander's filter is not an equivalent of MPS0 and can lose money; b) it shares with MPS0 the reversal property but no proofs were given that the Alexander's filter could be a part of MPS0; in fact, a rigorous proof that MPS0 is a \textit{true reversal system} is given by the author \cite[pp. 25 - 27, Property 4]{salov2007}; c) given the trading cost, there is no way to distinguish optimal swing prices using the Alexander's filter; at best, one can find that some filtered price moves started losing money; the l- and r-algorithms solve this task; d) Alexander applies percent filter; the l- and r-algorithms accommodate constant, percent, and arbitrary cost chains in order to search for optimal trades on an interval; e) MPS0's OTEs are considered as objective market offers; there are no such notions in the work of Alexander; f) the MPS0's profits and frequencies of OTEs are suggested as measures of market activity attracting to trading and speculation; there is no such angle of view in \cite{alexander1961}; g) the frequent large profit opportunities are considered essential for the market existence; Alexander does not discuss such a condition; h) there is no attempt in Alexander's work to summarize these properties in the formulation of the main law of the speculative market.

Exclusion of break even and losing trades from the list of optimal trades of an MPS0 implies that \textit{the sum of b-increments and the mean b-increment of any BOTE is strictly positive. For any SOTE the same quantities are negative}.

\subsection{OTE profits}

\begin{figure}[h!]
  \centering
  \includegraphics[width=130mm]{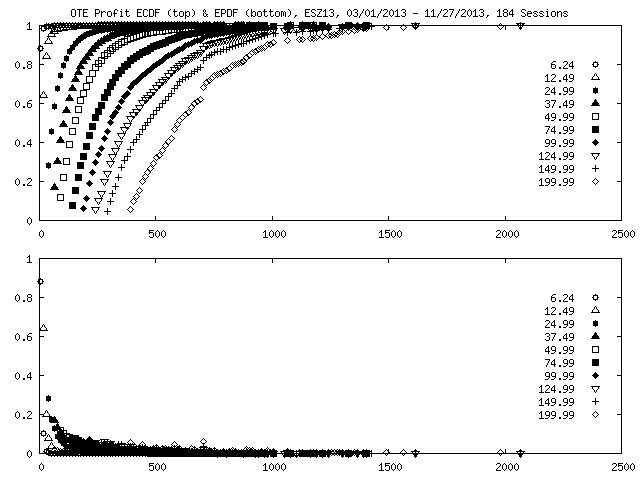}
  \caption[ote_ecdf_epdf_ESZ13]
   {ESZ13 traded in 184 sessions between March 1 and November 27, 2013. ECDF and EPDF of the OTE profits for a variety of f-costs. The t-cost is \$4.66.}
  \label{ote_ecdf_epdf_ESZ13}
\end{figure}

The OTE profit is computed using the entry and exit prices of the optimal trade, the $k$ dollar price point value, and the trading cost, t-cost. Once a new OTE is detected, born, within a session is becomes the current one and the previous OTE is fixed and cannot change. Three scenarios are possible: 1) the profit of the current OTE will grow, 2) the current OTE will be replaced by a new one of the opposite type, 3) the session will terminate. OTEs are obtained using $C^f$ and evaluated using $C^t$. The following is true for OTE trades: $k|P_{entry} - P_{exit}| > 2C^f$. The profit of such an OTE is $PL^{OTE}=k|P_{entry} - P_{exit}| - 2C^t > 2(C^f - C^t)$. Since $C^t \le C^f$, $PL^{OTE} > 0$. The absolute difference of prices is a whole multiple of $\delta$, $k|P_{entry} - P_{exit}|_{min} = kn_{min}\delta > 2C^f$, and $n_{min} > \frac{2C^f}{k\delta}$. If the ratio is a natural number or zero for $C^f=0$, then $n_{min}=\frac{2C^f}{k\delta} + 1$. For a fractional number, $n_{min}=\lfloor \frac{2C^f}{k\delta} \rfloor + 1$. Without losing generality, we can apply the last formula in both cases. Thus,
\begin{equation}
\centering
PL_{min}^{OTE} = k\delta(\lfloor \frac{2C^f}{k\delta}\rfloor + 1) - 2C^t; \; PL_i^{OTE} = PL_{min}^{OTE} + k\delta i, \; i = 0, 1, 2, \dots .
\label{EqPLOTE}
\end{equation}
Similar to prices, b-increments, and \textit{PL} the OTE profit is discrete. For ESZ13 $k = \$50, \delta = 0.25, C^t = \$4.66, C^f=\$49.99$, we get $n_{min} = \lfloor \frac{2 \times \$49.99}{\$50 \times 0.25} \rfloor + 1 = 8$, $PL_{min}^{OTE}=\$90.68$, $PL_i^{OTE} = \$90.68 + \$12.50i$. For given f- and t-costs, we are interested in the frequency of the OTE profit for each $i$. This helps judging, which of the three scenarios is likely, and is needed for building sets of trading rules - new systems based on OTE profits, Table \ref{TblOTEvsFCostESZ13}.
\begin{table}[!h]
  \centering
  \topcaption{ESZ13 traded between March 1 and November 27, 2013. Basic statistics of intra-session OTE profits are computed for 184 sessions using tabulated f-costs and t-cost = \$4.66.}
  \begin{tabular}{rrcrrrcccc}
   F-Cost & $N_{OTE}$ & Mean & Min & $n_{min}$ & Max & $n_{max}$ & StdDev & Skew. & E-Kurt.\\
6.24 & 1602254 & 5.1779 & 3.18 & 1411961 & 865.68 & 1 & 9.1922 & 24.8 & -1.506 \\
12.49 & 140804 & 25.915 & 15.68 & 90255 & 865.68 & 1 & 27.817 & 9.76 & 18.75 \\
24.99 & 17126 & 85.807 & 40.68 & 4859 & 940.68 & 1 & 70.034 & 4.02 & 24.99 \\
37.49 & 7170 & 143 & 65.68 & 1214 & 1178.18 & 1 & 99.989 & 2.97 & 13.62 \\
49.99 & 4147 & 195.98 & 90.68 & 479 & 1178.18 & 1 & 123.71 & 2.45 & 8.35 \\
62.49 & 2780 & 245.13 & 115.68 & 257 & 1328.18 & 1 & 144.3 & 2.08 & 5.658 \\
74.99 & 2020 & 291.7 & 140.68 & 159 & 1403.18 & 1 & 162.18 & 2.01 & 5.474 \\
87.49 & 1557 & 334.92 & 165.68 & 107 & 1403.18 & 1 & 177.89 & 1.81 & 4.113 \\
99.99 & 1244 & 376.19 & 190.68 & 75 & 1403.18 & 1 & 193.55 & 1.85 & 4.423 \\
112.49 & 1023 & 415.11 & 215.68 & 76 & 1403.18 & 1 & 209.16 & 1.72 & 3.419 \\
124.99 & 826 & 461.47 & 240.68 & 48 & 2065.68 & 1 & 221.73 & 1.89 & 5.683 \\
137.49 & 696 & 501.73 & 265.68 & 26 & 2065.68 & 1 & 232.58 & 1.77 & 4.761 \\
149.99 & 604 & 536.56 & 290.68 & 27 & 2065.68 & 1 & 239.89 & 1.73 & 4.37 \\
162.49 & 524 & 573.1 & 315.68 & 22 & 2065.68 & 1 & 248.87 & 1.8 & 5.01 \\
174.99 & 467 & 603.85 & 340.68 & 29 & 2065.68 & 1 & 257.65 & 1.76 & 4.559 \\
187.49 & 396 & 650.09 & 365.68 & 18 & 2065.68 & 1 & 266.45 & 1.71 & 3.98 \\
199.99 & 367 & 672.29 & 390.68 & 21 & 2065.68 & 1 & 265.61 & 1.7 & 3.939 \\
  \end{tabular}
  \label{TblOTEvsFCostESZ13}
\end{table}
The ECDF and EPDF of the OTE profits are plotted on Figure \ref{ote_ecdf_epdf_ESZ13} for selected f-costs.
Let us interpret one of the ECDFs corresponding to the f-cost \$49.99. As long as the price moves two ESZ13 points against the last local extreme, a corresponding BOTE or SOTE is born and comes to the list of OTEs considered for this f-cost. We see that in 184 sessions from 4147 registered OTEs, 11.6\% have the least profit \$90.68. 88.4\% OTEs have the profit exceeding this one at least by one $\delta$. This means that once the OTE is born, in 88.4\% it grows by one tick or more. The mean OTE profit is \$195.98. One cannot take the profit \$90.68 because this move is used for detection of a new OTE. However, even the difference between the mean and this threshold is $\$195.98 - \$90.68 = \$105.30$. At first glance, this looks like a \textit{money machine}. Place and trail an \textit{entry stop limit order} counting order prices from the last price extreme. With high liquidity contracts such as ESZ** traded in a November, set equal stop and limit order prices. If the order is placed in advance (a minute or so) and the price touches the stop price, then the chances of filling the order at the limit price are high. This fixes the previous OTE and "announces" the birth of the new one. In 88.4\% the price moves higher for BOTE and lower for SOTE and one knows the new OTE type because the previous is fixed and the types alternate. There is a problem with this trivial trading system alternating positions at the OTE birth price. If the 11.4\% scenario is realized and an opposite position is taken at a new "birth level", then the loss will be the two points plus the double t-cost \$109.32. If the price moves one $\delta$ further from the birth price, then the loss supposed to be one $\delta = \$12.50$ less, \$96.82. On average, this is achieved using a trailing stop order. If one takes an opposite position at a new birth price, then the chain of outcomes for f-cost \$49.99 in dollars is $-109.32, -96.82, -84.39, ..., -109.32 + 12.5i$, where $i=0,1,\dots$. It becomes positive for $i \ge 9$. The outcomes being multiplied by corresponding EPDF mass values and summed together give the mathematical expectation of the strategy after accounting transaction costs: -\$4.024 per trade per contract. This negative number can be obtained easier after noticing that $\$90.68 + \$12.50i - (-\$109.32 +\$12.5i) = \$200$ for all $i$. Thus, we can subtract \$200 from the rounded off mean \$195.98 in Table \ref{TblOTEvsFCostESZ13} and get \$-4.02. \textit{Buying long BOTE and selling short SOTE at the "birth price" results in a negative mathematical expectation of PL for f-cost \$49.99 and t-cost \$4.66}. The same strategy based on other means from Table \ref{TblOTEvsFCostESZ13} loses too.

One needs to decrease the cost more than by \$4.02 per contract per trade. Since commissions and fees \$4.66 are given, a trader would have to either buy BOTE at lower than the birthday price or, being on a position, not wait losing \$109.32. One should buy and sell only \textit{some} BOTEs and SOTEs. After touching the BOTE birth level, prices often retrace without changing the OTE type. This provides lower entry buy prices. Having a few ticks above the birth level one can take the profit. Strategies do not have to buy BOTE but can sell them and vice versa. What happens between the birth (once it is detected) and previous extreme prices, after the birth, on the boundary of neighboring OTEs, a-b-c-chains during BOTE and SOTE phases, variability and persistence of properties are in scope. Details of such an analysis will become trading secrets. 

\subsection{OTE frequencies}

Buying and/or selling BOTEs and/or SOTEs, a trader should be interested how frequent and sizable these offers are. We got an answer about the size. The mean number of OTE per session for f-cost \$49.99 is equal to $\frac{4147}{184} = 22.5$. E-mini S\&P 500 December contract becomes liquid in September, when the September contract expires. From now and until December this contract is a \textit{nearby}. Combining ESZ13 OTEs from March 2013 and after September 20013 reduces the mean number of OTE elements for nearby contracts.
\begin{figure}[h!]
  \centering
  \includegraphics[width=130mm]{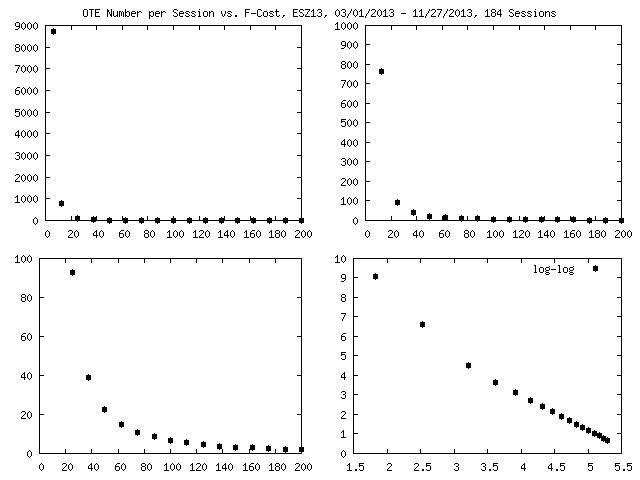}
  \caption[ote_n_vs_fct_ESZ13_2_2]
   {ESZ13 traded in 184 sessions between March 1 and November 27, 2013. Mean number of OTE per session vs.  f-cost. Three zoomed charts and lol-log (bottom right) dependence. Two linear fragments on log-log chart have slopes 3.1 and 1.8.}
  \label{ote_n_vs_fct_ESZ13_2_2}
\end{figure}
Looking at the close to linear log-log dependence on Figure \ref{ote_n_vs_fct_ESZ13_2_2} we need to remember that no OTEs is generated at the threshold f-cost. On the right of the threshold the number is constant zero, and logarithm cannot be evaluated. The f-cost must be multiplied by two in order to get the dollar equivalent of the price difference. Thus, on average two \$400 per contract price moves take place every session. This is eight ES points or $32\delta$. More than six \$200 moves, 4 points or $16\delta$, is expected. This can keep an intra-day trader busy. The number of smaller profit OTEs is considerable and this is where large positional ultra short term traders can operate. Instead of waiting large moves, they agree to extract profits from smaller moves but more frequently.

\subsection{OTE durations}

Durations of OTEs are the sums of a-increments associated with the optimal MPS0 trades.
\begin{table}[!h]
  \centering
  \topcaption{ESZ13 traded between March 1 and November 27, 2013. Basic statistics of intra-session OTE durations in seconds are computed for 184 sessions using tabulated f-costs.}
  \begin{tabular}{rrcrrrcccc}
   F-Cost & $N_{OTE}$ & Mean & Min & $n_{min}$ & Max & $n_{max}$ & StdDev & Skew. & E-Kurt.\\
6.24 & 1602254 & 7.9497 & 0 & 977135 & 59441 & 1 & 204.54 & 111 & 19.51 \\
12.49 & 140804 & 90.296 & 0 & 63592 & 76261 & 1 & 734.21 & 33.3 & 243.5 \\
24.99 & 17126 & 730.6 & 0 & 4088 & 76261 & 1 & 2357.8 & 10.5 & 175.6 \\
37.49 & 7170 & 1719.9 & 0 & 1133 & 99065 & 1 & 3946.8 & 7 & 88.45 \\
49.99 & 4147 & 2920.9 & 0 & 541 & 99065 & 1 & 5546.4 & 4.79 & 39.1 \\
62.49 & 2780 & 4214 & 0 & 331 & 102225 & 1 & 7080.6 & 3.71 & 22.72 \\
74.99 & 2020 & 5682.4 & 0 & 211 & 102225 & 1 & 8445.5 & 3.02 & 14.7 \\
87.49 & 1557 & 7182.7 & 0 & 121 & 102225 & 1 & 9918.1 & 2.62 & 10.18 \\
99.99 & 1244 & 8802.5 & 0 & 86 & 102225 & 1 & 11184 & 2.29 & 7.618 \\
112.49 & 1023 & 10513 & 0 & 66 & 102225 & 1 & 12804 & 2.02 & 5.351 \\
124.99 & 826 & 12672 & 0 & 45 & 102225 & 1 & 14384 & 1.82 & 3.966 \\
137.49 & 696 & 14674 & 0 & 31 & 102225 & 1 & 15587 & 1.68 & 3.126 \\
149.99 & 604 & 16221 & 0 & 31 & 102225 & 1 & 16532 & 1.56 & 2.567 \\
162.49 & 524 & 17683 & 0 & 30 & 102225 & 1 & 17446 & 1.46 & 2.129 \\
174.99 & 467 & 19200 & 0 & 30 & 102225 & 1 & 18772 & 1.34 & 1.469 \\
187.49 & 396 & 21500 & 0 & 30 & 102225 & 1 & 19895 & 1.17 & 0.9223 \\
199.99 & 367 & 22003 & 0 & 26 & 102225 & 1 & 20112 & 1.15 & 0.852 \\
  \end{tabular}
  \label{TblOTEDurationsvsFCostESZ13}
\end{table}
\begin{figure}[h!]
  \centering
  \includegraphics[width=130mm]{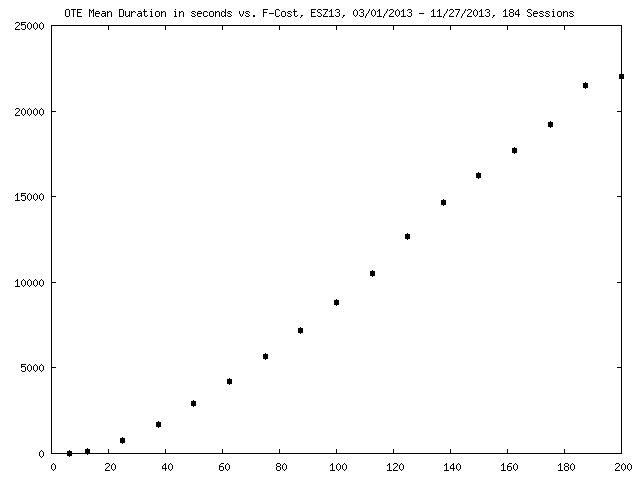}
  \caption[ote_duration_vs_fct_ESZ13]
   {ESZ13 traded in 184 sessions between March 1 and November 27, 2013. Mean OTE duration in seconds vs.  f-cost in dollars.}
  \label{ote_duration_vs_fct_ESZ13}
\end{figure}
It is interesting that Figure \ref{ote_duration_vs_fct_ESZ13} and one in \cite[p. 31]{salov2012} built for nearby Live Cattle futures traded in different months coincide on the f-cost interval [10, 130].

\subsection{OTE parametric dependencies}

The coincidence of dependencies of durations vs. f-costs for ESZ13 and a bunch of Live Cattle futures traded in 2011 and 2012 rises a question whether other OTE properties are similar for the same f-costs across different futures. Trading BOTEs and SOTEs, one concerns about their profits, frequencies, durations. Concrete commodities step out on the second plan. The f-cost becomes a common parameter, which can be set equal for a-b-c-chains of different commodities. This promote parametric mappings between OTE properties like $ECDF^{ESZ13} \Leftrightarrow \textrm{f-cost} \Leftrightarrow  ECDF^{ZCZ13}$ leading to $ECDF^{ESZ13} \Leftrightarrow ECDF^{ZCZ13}$. Equation \ref{EqPLOTE} can be reversed
\begin{equation}
i = \frac{PL_i^{OTE} + 2C^t}{k\delta} - 1 - \lfloor \frac{2C^f}{k\delta} \rfloor.
\label{EqIfromPLOTE}
\end{equation}
This supports linear transformations like $ECDF^{OTE}(PL_i^{OTE}) \rightarrow ECDF^{OTE}(i)$. Equalizing $i$ for the futures 1 and 2 and selecting the same $C^f$ gives
\begin{equation}
\frac{PL_i^{OTE1} + 2C_1^t}{k_1\delta_1} = \frac{PL_i^{OTE2} + 2C_2^t}{k_2\delta_2} + \lfloor \frac{2C^f}{k_1\delta_1} \rfloor  - \lfloor \frac{2C^f}{k_2\delta_2} \rfloor.
\label{EqPLOTE2Contracts}
\end{equation}
For a group of futures like ES, C, S, W, $k_1 = k_2 = \$50, \; \delta_1 = \delta_2 = 0.25$ and $PL_i^{OTE1} = PL_i^{OTE2} + 2(C_2^t - C_1^t)$. For other futures the slope of the line is $\frac{k_1\delta_1}{k_2\delta_2}$ and the intercept depends on $C^f$. Given $C^f$ and $C^t$ the $PL_i^{OTE}$ play the role of \textit{characteristic values}, where the probability mass is concentrated: MPS0 will not return OTEs with other than the profits given by Equation \ref{EqPLOTE}. Smooth dependences like on Figures \ref{ote_ecdf_epdf_ESZ13}, \ref{ote_n_vs_fct_ESZ13_2_2}, \ref{ote_duration_vs_fct_ESZ13} prompt for smooth parametric dependencies built for two futures like ESZ13 and ZCZ13. They would make BOTEs and SOTEs \textit{pure money elements} for a speculator. To equalize futures in terms of risk, Richard Dennis has invented \textit{trading units} described for the \textit{turtle system} in \cite[pp. 117 - 120]{faith2007}, \cite[p. 86]{covel2007}. Here, we check whether two futures, where MPS0 returns OTEs of almost one size, have the same empirical OTE frequencies. We were dealing with a parametric curve discussing the Weibull distribution, Equations \ref{EqWeibullSkewness}, \ref{EqWeibullEKurtosis}.

\begin{figure}[h!]
  \centering
  \includegraphics[width=130mm]{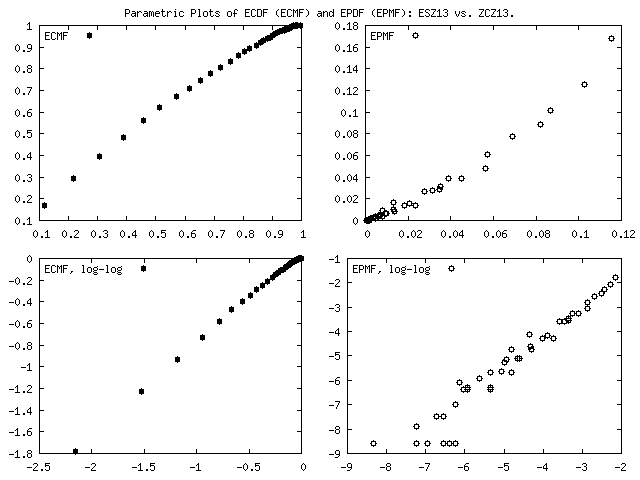}
  \caption[ote_param_ESZ13_ZCZ13_2_2]
   {ESZ13 vs. ZCZ13. Parametric curves. ECMF, EPMF, and their log-log variations, $C^f=\$49.99, \; C_{ESZ13}^t=\$4.66, \; C_{ZCZ13}^t=\$5.33$.}
  \label{ote_param_ESZ13_ZCZ13_2_2}
\end{figure}

It is possible due to statistical nature of OTE samples that certain profits are not represented in a list returned after application of an MPS0. In order to build a parametric dependence properly, it is needed to evaluate using Equation \ref{EqIfromPLOTE} the index $i$ given the profit of OTE. Then, for two futures the values corresponding to the same $i$ should be mapped, Figure \ref{ote_param_ESZ13_ZCZ13_2_2}.
Such comparative dependencies help to see better where the similarity between ESZ13 and ZCZ13 having identical $k$ and $\delta$ ends. Of course, for a \textit{commercial} and a \textit{fundamental analyst} trading futures is not only a money game. Coffee, corn, crude oil, gold, and cocoa will remain for them different products. If the dependence is getting close to a line with the slope one, then this would indicate equivalence of two futures with respect to a considered property. We see that the dependencies are strong but the curves are not quite straight and the slope is not exactly one.

\subsection{OTE Profits vs. Durations}

\begin{figure}[h!]
  \centering
  \includegraphics[width=130mm]{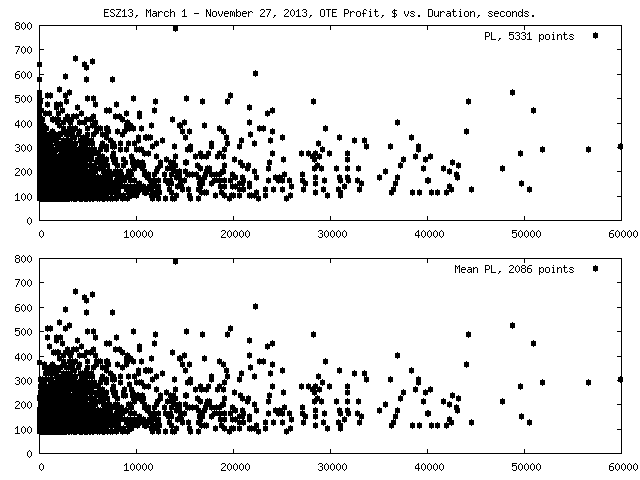}
  \caption[ote_pl_vs_duration_ESZ13]
   {ESZ13. Top: OTE profits vs. durations. Bottom: Mean OTE profits vs. durations, $C^f=\$49.99, \; C^t=\$4.66$.}
  \label{ote_pl_vs_duration_ESZ13}
\end{figure}

Plots of ESZ13 OTE profits and mean profits vs. durations do not indicate a dependence, Figure \ref{ote_pl_vs_duration_ESZ13}.
The total number of ESZ13 transactional ticks used for OTE computations in 184 sessions between March 1 and October 27, 2013 is equal to 20,045,845. This is extracted from the total number of ticks 20,142,102.

\section{The Maximum Trading Profit Framework}

The \textit{maximum trading profit framework}, MPSF, is a complement to the market analysis. Its purpose is to apply MPS in order to search for new trading signals for building trading systems. To some extent, it opposes to the analysis of trend and volatility in situations, where no consistent definitions of the latter two notions are presented and their computation becomes impossible leaving us with intuitive qualitative meaning only. The MPSF is a quantitative approach.

The MPSF solves three tasks. The first task is to build an MPS for the a-b-c- and cost-chains. This is achieved using the l- or r-algorithms. The second task is to combine the optimal transactions into optimal trades dividing by their entry and exit transactions the original a-b-c-fragment. This creates a chain of alternating BOTE and SOTE. The third task is to study the BOTE and SOTE properties.

The term \textit{framework} is borrowed from the \textit{Unified Modeling Language}: \cite[p. 284]{rumbaugh1999} \textit{"A generic architecture that provides an extensible template for applications within a domain."}. The majority of the constituting programs are written by the author in C++ \cite{stroustrup2000} and some in \textit{AWK} \cite{aho1988}, \textit{sed} \cite{dougherty1997}, and \textit{Python} \cite{martelli2006}.

\section{A Comment on Disequilibrium}

The author has mentioned the method discussed by Galperin \cite{galperin2004}, where mapping between the notions of two "unrelated" branches of mathematics and  physics "transfers" a known solution from one to the other branch, where it is unknown. Such a mapping is not a mechanical process. It is based on illuminating moments of (let us call it) the human being intuition. Seven charming examples as well the main idea of the "billiard formula for measuring distances in Lobachevsky spaces" are found in \cite[2. Lyric deviation: on usefulness of changing the point of view]{galperin2004}. In the author's opinion, the R\'{e}nyi's proof of Theorem 1 \cite[pp. 447 - 449]{renyi1959} belongs to this category. Figure \ref{explosion}, as it was mentioned, resembled to the author chemical and nuclear chain reactions responsible for explosions. What exactly should one "map" in order to support this analogy?

\subsection{Chain reactions}

 Let us give the word to authority \cite[p. 491]{semenov1956}: \textit{"Where branched chain reactions are concerned, there are two possibilities: (1) the rate of branching exceeds that of termination, which results in very rapid development of the chain avalanche; (2) the rate of termination is greater than that of branching, so that the avalanche cannot develop and the reaction cannot even take place ..."}. "Branching" maps to the "number of market participants and their orders". News ''ignites" the stream of orders. The latter affects the a-b-c-process in a manner, where it "expresses a confirmation". This confirmation, the price move, increases "branching" - triggers more filled orders after "exploring" the depth of the book and new coming. The a-b-outcome becomes even more irregular. "Termination" associates with the number of already filled orders so that the number of remaining orders to be sent reaches a normal but much smaller number, which should correspond to the large termination rate. This disturbs a diffusion by an explosion and after short time everything returns to a diffusion. This qualitative picture implies that we deal not with a single jump of price but an \textit{envelope} seen on Figure \ref{explosion} with a dimension also in time but partly hidden by the one second inaccuracy of reports.

\begin{figure}[h!]
  \centering
  \includegraphics[width=100mm]{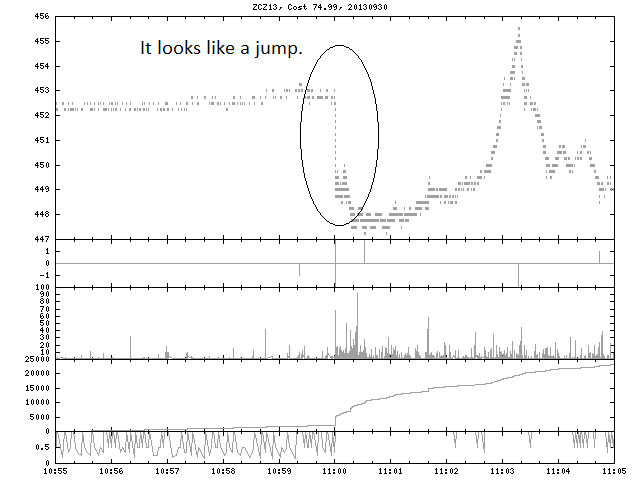}
  \caption[ZCZ13_1_20130930_74_99_price]
   {ZCZ13 traded on September 30, 2013. Price vs. time. Five minutes before the news at 11:00:00 CT and five minutes after.}
  \label{ZCZ13_1_20130930_74_99_price}
\end{figure}
\begin{figure}[h!]
  \centering
  \includegraphics[width=100mm]{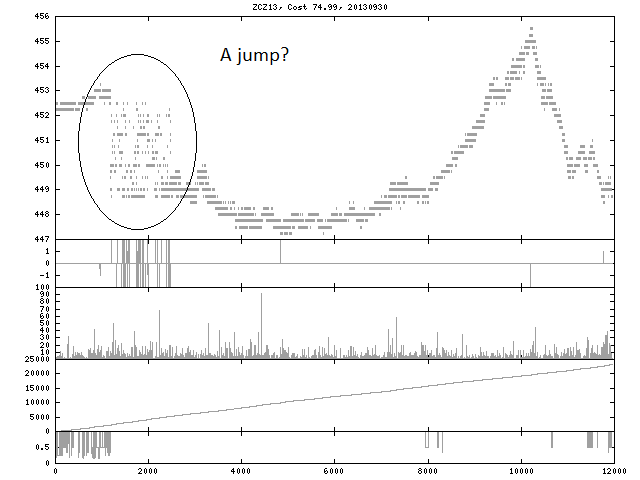}
  \caption[ZCZ13_1_20130930_74_99_price2]
   {ZCZ13 traded on September 30, 2013. Price vs. tick index. This format helps to see the complex price structure right after 11:00:00 CT.}
  \label{ZCZ13_1_20130930_74_99_price2}
\end{figure}
\begin{figure}[h!]
  \centering
  \includegraphics[width=100mm]{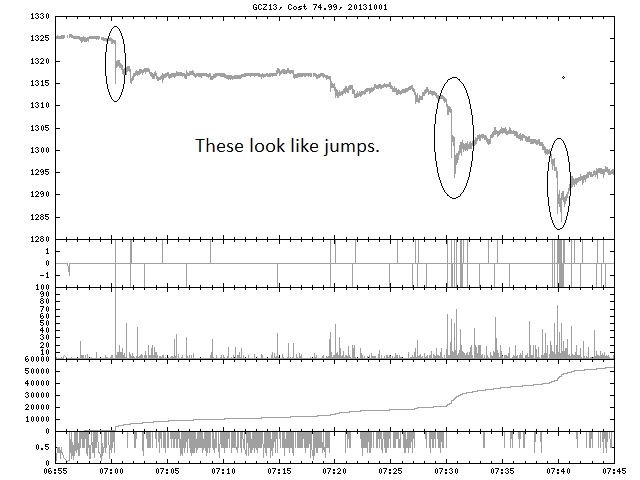}
  \caption[GCZ13_1_20131001_74_99_price]
   {GCZ13 traded on October 1, 2013. Price vs. time. Three consecutive price drops within fifty minutes.}
  \label{GCZ13_1_20131001_74_99_price}
\end{figure}
\begin{figure}[h!]
  \centering
  \includegraphics[width=100mm]{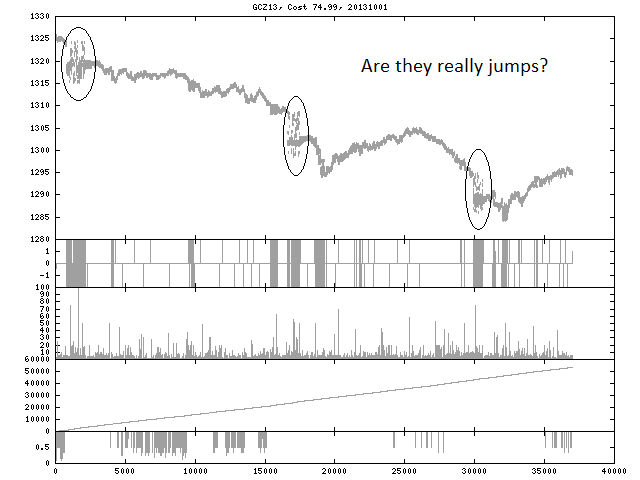}
  \caption[GCZ13_1_20131001_74_99_price2]
   {GCZ13 traded on October 1, 2013. Price vs. tick index. The format reveals a complex structure of three price drops occurred within fifty minutes.}
  \label{GCZ13_1_20131001_74_99_price2}
\end{figure}
\begin{figure}[h!]
  \centering
  \includegraphics[width=100mm]{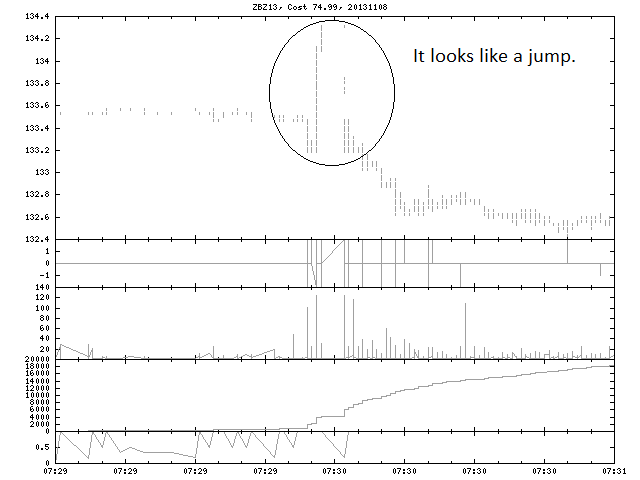}
  \caption[ZBZ13_1_20131108_74_99_price_price]
   {ZBZ13 traded on November 8, 2013. Price vs. time. One minute before the new at 07:30:00 CT and one minute after.}
  \label{ZBZ13_1_20131108_74_99_price}
\end{figure}
\begin{figure}[h!]
  \centering
  \includegraphics[width=100mm]{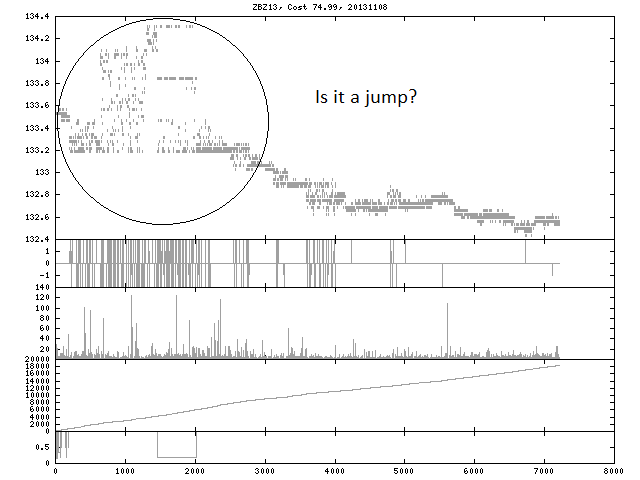}
  \caption[ZBZ13_1_20131108_74_99_price2]
   {ZBZ13 traded on November 8, 2013. Price vs. tick index. The format reveals a complex price structure within two minutes.}
  \label{ZBZ13_1_20131108_74_99_price2}
\end{figure}

\subsection{Price jumps}

Merton, understanding the geometric Brownian motion limitations, has added \textit{jumps} to diffusion \cite{merton1976}. A jump is an attempt to extend a model descriptive power for cases, where price increments deviate from a Gaussian distribution, using Poisson-driven process. For a friendly exposition of mathematics of rare events see \cite[pp. 144 - 167]{neftci1996}. There are no physical equivalents of such jumps for the Einstein's suspended particle. Already in this, one should perceive market specifics. It would be like one mixes permanganate solution on Figure \ref{FgDiffusion} by a spoon. Examination of Figures \ref{explosion}, \ref{ci_b_ZCN13_20130328}, \ref{noniid}, \ref{ZCZ13_1_20130930_74_99_price}, \ref{ZCZ13_1_20130930_74_99_price2}, \ref{GCZ13_1_20131001_74_99_price}, \ref{GCZ13_1_20131001_74_99_price2}, \ref{ZBZ13_1_20131108_74_99_price}, \ref{ZBZ13_1_20131108_74_99_price2} leads to conclusion that microstructure of the rare event is not a single jump separating two diffusions (before and after) but either a chain of jumps characterized by their high rate or another diffusion characterized by much greater variance and driving a wedge into the main process. Also, it is not necessarily a jump from one to a very different level, where the diffusion flow continues. In situations like "The 2010 Flash Crash" mentioned, the price plunging quickly takes back a considerable part of the move \cite[p. 19, Chart: E-Mimi Volume and Price]{us2010}, see also Figures \ref{ZBZ13_1_20131108_74_99_price} and \ref{ZBZ13_1_20131108_74_99_price2}.

\subsection{How, why, what: the answers}

If the author correctly understood Laurent's comment, Remery wanted to measure deviation from equilibrium by deviation of the price process from a Brownian motion. If variance of the latter is positive, then, even, for zero drift we expect to get not a line. This will supply to traders a basis for their activity. They will trade. Literally understood Sharpe's definition of equilibrium would indicate non-equilibrium conditions. Only a horizontal line of prices with non-zero cost can prevent trading. Following to Sharpe's logic, the latter is a characteristic of an equilibrium. Following to the proposals of this article such a state would correspond to the do nothing strategy with zero profit. Thus, the maximum profit after non-zero transaction costs is a measure of deviation from equilibrium. This is the answer on the first question of Introduction: How to express the non-equilibrium state in terms of trading.

The second question, Why the speculators continue trading, is answered by the MPS measuring the frequency and size of the market offers in seconds and dollars. Both characteristics remain too high and attractive. As long as it is so, the speculators cannot lose their interest. The MPS is also an answer on the third question, What is the objective market property explaining such an aspiration. This is an objective and fundamental market property formulated in terms of trading operations, and profits - the primary notions understood and appreciated by traders.

The author is observing that Economics is ready to pass the road of Thermodynamics, Chemistry, and Biology, where the "quiet" evolutionary notions of equilibrium and isolated systems have been disturbed by the revolutionary conceptions of non-equilibrium systems and irreversible processes, see \cite{prigogine1977}. This development does not erase the picture drawn by predecessors but adds to it necessary details.

\section{A Comment on Speculation}

Characterizing Bachellier's thesis, Mandelbrot and Richard Hudson write \cite[p. 44]{mandelbrot2004}: \textit{"It was about the money-grubbing form of speculation, the trading of government bonds on the Paris exchange, or Bourse, a thriving den of capitalism modeled after a Greek temple and located on the opposite river bank, geographically and intellectually, from the famed Sorbonne. Then and now in France, unbridled speculation had an unsavory reputation. While investment was socially desirable, pure gambling, or agiotage, was not"}. Modern computer and technological development erases geographical differences between the opposite riverbanks. Massive migration of mathematicians, physicists and chemists in finance should fill the intellectual gap, if it exists (this delicate topic requires other definitions and measures). Hopefully, this raises the exchange's riverbank without lowering the Sorbonne's side. The author says "without lowering", in a good sense: not because zeros are subtracted, they would not raise the opposite side either, but because the feeding scientific tradition is still alive. However, even good teaching traditions can die \cite{arnold1998}.

Livermore \cite[p. 3]{livermore1940}: \textit{"Speculation is the most uniformly fascinating game in the world"}; \cite[p. 7]{livermore1940} \textit{"Speculation is nothing more than anticipating coming movements"}. An independent critical review of the \textit{Livermore system}, exposed in \cite{livermore1940}, can be found in \cite{thomson1983}. In the author's opinion, it would be naive to try to repeat Livermore's success using his system. This system worked for him. To separate his rules from trading wisdom trained during 40 years would be fruitless. One should not be afraid to trust his or her own feelings, if behind there is a long term focus to the market. Larry Livingston, immortal personage skillfully copied from Jesse Livermore by Edwin Lef\`{e}vre, says \cite{lefevre1923}: \textit{"... there is nothing new in Wall Street. There can't be because speculation is as old as the hills. Whatever happens in the stock market to-day has happened before and will happen again."}. This is not a definition but a \textit{constructive} view. Speculation will exist as long as the market will offer frequent and substantial money making opportunities measured by MPS.
 
\paragraph{Acknowledgments.}  I would like to thank Nigel Goldenfeld for critical proposals improving the paper, Austin Gerig for the endorsement of the original manuscript, Grigori (Grisha) Galperin for presenting copies of \cite{vorobec1992}, \cite{galperin2001}, \cite{galperin2004} and periodic supply of mathematical tasks helping to keep the brain active, Alexander Tumanov and Vadim Zharnitsky for the off-print \cite{tumanov2006}. The author has discussed several topics in different years and presented here with Goldenfeld, Tumanov, Zharnitsky, Timur Misirpashaev, Perry Kaufman, Robert Pardo, Igor Toshchakov.

Robert Pardo has introduced the maximum potential profit (a number) \cite[pp. 125 - 126]{pardo1992}. 20 years ago these two pages have inspired the author's interest and research, where the maximum trading profit strategy (a vector) has been created \cite{salov2007}, \cite{salov2008}, the optimal trading elements have been introduced \cite{salov2011b}, \cite{salov2012}, the maximum profit strategy framework has been designed and implemented \cite{salov2007}, \cite{salov2011b}, the a-b-c-classification has been proposed \cite{salov2011}, \cite{salov2011b}, \cite{salov2012}, \cite{salov2012b}, and the main law of the speculative market has been formulated \cite{salov2011b}, \cite{salov2012}.

\section{Appendix A. Sample Statistics of a-Increments}
\begin{center}
\begin{longtable}{|l|r|c|r|r|r|r|c|c|c|}
\caption[Sample Statistics of a-Increments]{Sample Statistics of a-Increments in Seconds} \label{a-increments} \\

 \hline
 \multicolumn{1}{|c|}{Date} &
 \multicolumn{1}{c|}{Size} &
 \multicolumn{1}{c|}{Mean} &
 \multicolumn{1}{c|}{Min} &
 \multicolumn{1}{c|}{$n_{min}$} &
 \multicolumn{1}{c|}{Max} &
 \multicolumn{1}{c|}{$n_{max}$} &
 \multicolumn{1}{c|}{StdDev} &
 \multicolumn{1}{c|}{Skew.} &
 \multicolumn{1}{c|}{E-Kurt.} \\
 \hline 
 \endfirsthead

 \multicolumn{10}{c}%
 {\tablename\ \thetable{} -- continued from previous page} \\
 \hline
 \multicolumn{1}{|c|}{Date} &
 \multicolumn{1}{c|}{Size} &
 \multicolumn{1}{c|}{Mean} &
 \multicolumn{1}{c|}{Min} &
 \multicolumn{1}{c|}{$n_{min}$} &
 \multicolumn{1}{c|}{Max} &
 \multicolumn{1}{c|}{$n_{max}$} &
 \multicolumn{1}{c|}{StdDev} &
 \multicolumn{1}{c|}{Skew.} &
 \multicolumn{1}{c|}{E-Kurt.} \\
 \hline 
 \endhead

 \hline \multicolumn{10}{|r|}{{Continued on next page}} \\ \hline
 \endfoot

 \hline
 \endlastfoot

\multicolumn{10}{|c|}{ZCN13, 17:00:00 - 14:00:00 CT } \\ \hline

2013-03-01 & 10350 & 7.3041 & 0 & 8140 & 1324 & 1 & 45.495 & 13.4 & 234.2\\
2013-03-04 & 13422 & 5.6324 & 0 & 10978 & 4386 & 1 & 58.506 & 41.1 & 2559\\
2013-03-05 & 9779 & 7.7305 & 0 & 7929 & 1833 & 1 & 54.602 & 16.2 & 356.4\\
2013-03-06 & 17750 & 4.2584 & 0 & 14913 & 2597 & 1 & 41.858 & 29.7 & 1356\\
2013-03-07 & 14815 & 5.1029 & 0 & 12339 & 1841 & 1 & 37.201 & 19.1 & 583.7\\
2013-03-08 & 15893 & 4.7567 & 0 & 13186 & 2486 & 1 & 41.664 & 27.3 & 1153\\
2013-03-11 & 16429 & 4.6016 & 0 & 13710 & 1311 & 1 & 34.522 & 16.5 & 366.6\\
\end{longtable}
\end{center}

\section{Appendix B. Sample Statistics of b-Increments}
\begin{center}
\begin{longtable}{|l|r|c|r|r|r|r|c|c|c|}
\caption[Sample Statistics of b-Increments]{Sample Statistics of b-Increments Expressed in $\delta$} \label{b-increments} \\

 \hline
 \multicolumn{1}{|c|}{Date} &
 \multicolumn{1}{c|}{Size} &
 \multicolumn{1}{c|}{Mean} &
 \multicolumn{1}{c|}{Min} &
 \multicolumn{1}{c|}{$n_{min}$} &
 \multicolumn{1}{c|}{Max} &
 \multicolumn{1}{c|}{$n_{max}$} &
 \multicolumn{1}{c|}{StdDev} &
 \multicolumn{1}{c|}{Skew.} &
 \multicolumn{1}{c|}{E-Kurt.} \\
 \hline 
 \endfirsthead

 \multicolumn{10}{c}%
 {\tablename\ \thetable{} -- continued from previous page} \\
 \hline
 \multicolumn{1}{|c|}{Date} &
 \multicolumn{1}{c|}{Size} &
 \multicolumn{1}{c|}{Mean} &
 \multicolumn{1}{c|}{Min} &
 \multicolumn{1}{c|}{$n_{min}$} &
 \multicolumn{1}{c|}{Max} &
 \multicolumn{1}{c|}{$n_{max}$} &
 \multicolumn{1}{c|}{StdDev} &
 \multicolumn{1}{c|}{Skew.} &
 \multicolumn{1}{c|}{E-Kurt.} \\
 \hline 
 \endhead

 \hline \multicolumn{10}{|r|}{{Continued on next page}} \\ \hline
 \endfoot

 \hline
 \endlastfoot

\multicolumn{10}{|c|}{ZCN13, $\delta = 0.25$, 17:00:00 - 14:00:00 CT } \\ \hline

2013-03-01 & 10350 & 0.00096618 & -4 & 1 & 3 & 2 & 0.43674 & -0.0438 & 8.206\\
2013-03-04 & 13422 & -0.0014156 & -12 & 1 & 12 & 1 & 0.6841 & -0.0108 & 79.82\\
2013-03-05 & 9779 & 0.0019429 & -3 & 3 & 4 & 1 & 0.43673 & 0.143 & 8.788\\
2013-03-06 & 17750 & -0.0033239 & -4 & 1 & 4 & 1 & 0.40217 & -0.0939 & 10.29\\
2013-03-07 & 14815 & 0 & -4 & 1 & 3 & 6 & 0.4075 & -0.0898 & 9.869\\
2013-03-08 & 15893 & 0.0022022 & -10 & 1 & 11 & 1 & 0.75575 & -0.0684 & 32.43\\
2013-03-11 & 16429 & 0.002313 & -4 & 2 & 4 & 3 & 0.41607 & 0.0864 & 12.04\\
\end{longtable}
\end{center}

\section{Appendix C. Sample Statistics of c-Increments}
\begin{center}
\begin{longtable}{|l|c|r|c|r|r|r|r|c|c|c|}
\caption[Sample Statistics of c-Increments]{Sample Statistics of c-Increments Expressed in $\delta$} \label{c-increments} \\

 \hline
 \multicolumn{1}{|c|}{Type} &
 \multicolumn{1}{|c|}{Ticker} &
 \multicolumn{1}{c|}{Size} &
 \multicolumn{1}{c|}{Mean} &
 \multicolumn{1}{c|}{Min} &
 \multicolumn{1}{c|}{$n_{min}$} &
 \multicolumn{1}{c|}{Max} &
 \multicolumn{1}{c|}{$n_{max}$} &
 \multicolumn{1}{c|}{StdDev} &
 \multicolumn{1}{c|}{Skew.} &
 \multicolumn{1}{c|}{E-Kurt.} \\
 \hline 
 \endfirsthead

 \multicolumn{11}{c}%
 {\tablename\ \thetable{} -- continued from previous page} \\
 \hline
 \multicolumn{1}{|c|}{Type} &
 \multicolumn{1}{|c|}{Ticker} &
 \multicolumn{1}{c|}{Size} &
 \multicolumn{1}{c|}{Mean} &
 \multicolumn{1}{c|}{Min} &
 \multicolumn{1}{c|}{$n_{min}$} &
 \multicolumn{1}{c|}{Max} &
 \multicolumn{1}{c|}{$n_{max}$} &
 \multicolumn{1}{c|}{StdDev} &
 \multicolumn{1}{c|}{Skew.} &
 \multicolumn{1}{c|}{E-Kurt.} \\
 \hline 
 \endhead

 \hline \multicolumn{11}{|r|}{{Continued on next page}} \\ \hline
 \endfoot

 \hline
 \endlastfoot

C & ZCN13 & 89 & -3.3258 & -75 & 1 & 20 & 1 & 10.642 & -4.11 & 24.22\\
CR & ZCN13 & 69 & -1.5362 & -21 & 1 & 9 & 1 & 4.5972 & -0.921 & 3.424\\
CW & ZCN13 & 17 & -7.0588 & -44 & 1 & 20 & 1 & 13.117 & -0.971 & 2.95\\
CH & ZCN13 & 3 & -23.333 & -75 & 1 & 10 & 1 & 45.369 & -1.52 & NA\\
CI & ZCN13 & 66 & 0.4697 & -8 & 1 & 12 & 1 & 3.202 & 0.726 & 2.39\\ \hline
C & ZSN13 & 89 & -4.1236 & -32 & 1 & 23 & 1 & 7.3032 & -0.175 & 3.131\\
CR & ZSN13 & 69 & -2.913 & -18 & 1 & 23 & 1 & 5.888 & 1.13 & 4.292\\
CW & ZSN13 & 17 & -9.8824 & -32 & 1 & 11 & 1 & 9.9554 & 0.0168 & 0.4474\\
\end{longtable}
\end{center}
The CR, CW, and CH are regular, weekend, and holiday c-increments. The C combines the three previous types. The CI is the c-increments between ranges. It is unavailable, if there is one range in a session.

\section{Appendix D. Conditional Sample Statistics of b-Increments}

\begin{center}
\begin{longtable}{|c|r|c|r|r|r|r|c|c|c|}
\caption[Sample Statistics of b-Increments vs. Time]{Sample Statistics of b-Increments Expressed in $\delta$ Found for Associated a-Increments Expressed in Seconds. ESM13 April 5, 2013.} \label{mwb_vs_a} \\

 \hline
 \multicolumn{1}{|c|}{a-Incr.} &
 \multicolumn{1}{c|}{Size} &
 \multicolumn{1}{c|}{Mean} &
 \multicolumn{1}{c|}{Min} &
 \multicolumn{1}{c|}{$n_{min}$} &
 \multicolumn{1}{c|}{Max} &
 \multicolumn{1}{c|}{$n_{max}$} &
 \multicolumn{1}{c|}{StdDev} &
 \multicolumn{1}{c|}{Skew.} &
 \multicolumn{1}{c|}{E-Kurt.} \\
 \hline 
 \endfirsthead

 \multicolumn{10}{c}%
 {\tablename\ \thetable{} -- continued from previous page} \\
 \hline
 \multicolumn{1}{|c|}{a-Incr.} &
 \multicolumn{1}{c|}{Size} &
 \multicolumn{1}{c|}{Mean} &
 \multicolumn{1}{c|}{Min} &
 \multicolumn{1}{c|}{$n_{min}$} &
 \multicolumn{1}{c|}{Max} &
 \multicolumn{1}{c|}{$n_{max}$} &
 \multicolumn{1}{c|}{StdDev} &
 \multicolumn{1}{c|}{Skew.} &
 \multicolumn{1}{c|}{E-Kurt.} \\
 \hline 
 \endhead

 \hline \multicolumn{10}{|r|}{{Continued on next page}} \\ \hline
 \endfoot

 \hline
 \endlastfoot

0 & 508004 & -5.5118e-005 & -16 & 3 & 17 & 5 & 0.36379 & 1.23 & -2.551 \\
1 & 20357 & -0.0010316 & -18 & 1 & 3 & 3 & 0.64112 & -1.09 & 30.4 \\
2 & 4760 & 0.010714 & -3 & 1 & 2 & 1 & 0.60021 & -0.0273 & -0.06725 \\
3 & 2028 & -0.0039448 & -1 & 381 & 2 & 3 & 0.61231 & 0.0409 & -0.2043 \\
4 & 1122 & -0.011586 & -1 & 228 & 2 & 1 & 0.62995 & 0.0303 & -0.4094 \\
5 & 718 & -0.018106 & -1 & 133 & 1 & 120 & 0.59374 & 0.0049 & -0.1622 \\
6 & 544 & -0.045956 & -1 & 112 & 1 & 87 & 0.60363 & 0.0196 & -0.2643 \\
7 & 406 & 0.064039 & -1 & 57 & 2 & 2 & 0.59283 & 0.126 & 0.2436 \\
\end{longtable}
\end{center}
The sum of Size is equal to 539882 also found in Tables \ref{a-increments} and \ref{b-increments}. The sample standard deviation is set to zero for cases of one point. The sample skewness and excess kurtosis are marked "undefined", if the number of points is insufficient for their estimation.

\begin{center}
\begin{longtable}{|c|r|c|r|r|r|r|c|c|c|}
\caption[Sample Statistics of Absolute b-Increments vs. Time]{Sample Statistics of Absolute b-Increments Expressed in $\delta$ Found for Associated a-Increments Expressed in Seconds. ESM13 April 5, 2013.} \label{mawb_vs_a} \\

 \hline
 \multicolumn{1}{|c|}{a-Incr.} &
 \multicolumn{1}{c|}{Size} &
 \multicolumn{1}{c|}{Mean} &
 \multicolumn{1}{c|}{Min} &
 \multicolumn{1}{c|}{$n_{min}$} &
 \multicolumn{1}{c|}{Max} &
 \multicolumn{1}{c|}{$n_{max}$} &
 \multicolumn{1}{c|}{StdDev} &
 \multicolumn{1}{c|}{Skew.} &
 \multicolumn{1}{c|}{E-Kurt.} \\
 \hline 
 \endfirsthead

 \multicolumn{10}{c}%
 {\tablename\ \thetable{} -- continued from previous page} \\
 \hline
 \multicolumn{1}{|c|}{a-Incr.} &
 \multicolumn{1}{c|}{Size} &
 \multicolumn{1}{c|}{Mean} &
 \multicolumn{1}{c|}{Min} &
 \multicolumn{1}{c|}{$n_{min}$} &
 \multicolumn{1}{c|}{Max} &
 \multicolumn{1}{c|}{$n_{max}$} &
 \multicolumn{1}{c|}{StdDev} &
 \multicolumn{1}{c|}{Skew.} &
 \multicolumn{1}{c|}{E-Kurt.} \\
 \hline 
 \endhead

 \hline \multicolumn{10}{|r|}{{Continued on next page}} \\ \hline
 \endfoot

 \hline
 \endlastfoot

0 & 508004 & 0.084964 & 0 & 468795 & 17 & 5 & 0.35373 & 12 & -2.515 \\
1 & 20357 & 0.38645 & 0 & 12592 & 18 & 1 & 0.51155 & 2.57 & 67.9 \\
2 & 4760 & 0.35819 & 0 & 3059 & 3 & 1 & 0.48171 & 0.631 & -1.433 \\
3 & 2028 & 0.37179 & 0 & 1277 & 2 & 3 & 0.48646 & 0.57 & -1.574 \\
4 & 1122 & 0.39483 & 0 & 680 & 2 & 1 & 0.49085 & 0.454 & -1.737 \\
5 & 718 & 0.35237 & 0 & 465 & 1 & 253 & 0.47804 & 0.619 & -1.622 \\
6 & 544 & 0.36581 & 0 & 345 & 1 & 199 & 0.4821 & 0.559 & -1.696 \\
7 & 406 & 0.34483 & 0 & 268 & 2 & 2 & 0.48616 & 0.784 & -1.062 \\
\end{longtable}
\end{center}
The sample moments of the absolute b-increments are presented individually for each a-increment from the interval [0, 60] seconds. The a-increment 57 seconds did not occur. The values for the remaining [61, 113] a-increments are given as one distribution. This row shows that the maximum absolute b-increment at these waiting times was one. Thus, the remaining mean absolute b-increments being computed for individual a-increments are guaranteed to be $\le 1$. In fact, their mean 0.34483 is not very different from other means.

\begin{center}
\begin{longtable}{|c|r|c|r|r|r|r|c|c|c|}
\caption[Sample Statistics of Squares of b-Increments vs. Time]{Sample Statistics of Squares of b-Increments Expressed in $\delta$ Found for Associated a-Increments Expressed in Seconds. ESM13 April 5, 2013.} \label{mawb2_vs_a} \\

 \hline
 \multicolumn{1}{|c|}{a-Incr.} &
 \multicolumn{1}{c|}{Size} &
 \multicolumn{1}{c|}{Mean} &
 \multicolumn{1}{c|}{Min} &
 \multicolumn{1}{c|}{$n_{min}$} &
 \multicolumn{1}{c|}{Max} &
 \multicolumn{1}{c|}{$n_{max}$} &
 \multicolumn{1}{c|}{StdDev} &
 \multicolumn{1}{c|}{Skew.} &
 \multicolumn{1}{c|}{E-Kurt.} \\
 \hline 
 \endfirsthead

 \multicolumn{10}{c}%
 {\tablename\ \thetable{} -- continued from previous page} \\
 \hline
 \multicolumn{1}{|c|}{a-Incr.} &
 \multicolumn{1}{c|}{Size} &
 \multicolumn{1}{c|}{Mean} &
 \multicolumn{1}{c|}{Min} &
 \multicolumn{1}{c|}{$n_{min}$} &
 \multicolumn{1}{c|}{Max} &
 \multicolumn{1}{c|}{$n_{max}$} &
 \multicolumn{1}{c|}{StdDev} &
 \multicolumn{1}{c|}{Skew.} &
 \multicolumn{1}{c|}{E-Kurt.} \\
 \hline 
 \endhead

 \hline \multicolumn{10}{|r|}{{Continued on next page}} \\ \hline
 \endfoot

 \hline
 \endlastfoot

0 & 508004 & 0.13235 & 0 & 468795 & 289 & 5 & 2.3418 & 73.6 & 6.578 \\
1 & 20357 & 0.41101 & 0 & 12592 & 324 & 1 & 2.3398 & 130 & 1.797e+4 \\
2 & 4760 & 0.36029 & 0 & 3059 & 9 & 1 & 0.5007 & 1.75 & 17.94 \\
3 & 2028 & 0.37475 & 0 & 1277 & 4 & 3 & 0.50218 & 1.01 & 2.106 \\
4 & 1122 & 0.39661 & 0 & 680 & 4 & 1 & 0.50023 & 0.723 & 0.4784 \\
5 & 718 & 0.35237 & 0 & 465 & 1 & 253 & 0.47804 & 0.619 & -1.622 \\
6 & 544 & 0.36581 & 0 & 345 & 1 & 199 & 0.4821 & 0.559 & -1.696 \\
7 & 406 & 0.35468 & 0 & 268 & 4 & 2 & 0.53731 & 1.94 & 8.355 \\
\end{longtable}
\end{center}

\section{Appendix E. Following Kolmogorov's Advice}

\begin{center}
\begin{longtable}{|c|r|r|r|r|r|r|r|r|r|c|}
\caption[Following Kolmogorov's Advice]{Following Kolmogorov's Advice for Conditional Frequencies. ESM13, March - June 2013.} \label{kolmogorov_advice} \\

 \hline
 \multicolumn{1}{|c|}{Date} &
 \multicolumn{1}{c|}{A} &
 \multicolumn{1}{c|}{B} &
 \multicolumn{1}{c|}{$n$} &
 \multicolumn{1}{c|}{$m=n_A$} &
 \multicolumn{1}{c|}{$l=n_B$} &
 \multicolumn{1}{c|}{$k=n_{AB}$} &
 \multicolumn{1}{c|}{$\nu_A$} &
 \multicolumn{1}{c|}{$\nu_B$} &
 \multicolumn{1}{c|}{$\nu_{AB}$} &
 \multicolumn{1}{c|}{\%} \\
 \hline 
 \endfirsthead

 \multicolumn{11}{c}%
 {\tablename\ \thetable{} -- continued from previous page} \\
 \hline
 \multicolumn{1}{|c|}{Date} &
 \multicolumn{1}{c|}{A} &
 \multicolumn{1}{c|}{B} &
 \multicolumn{1}{c|}{$n$} &
 \multicolumn{1}{c|}{$m=n_A$} &
 \multicolumn{1}{c|}{$l=n_B$} &
 \multicolumn{1}{c|}{$k=n_{AB}$} &
 \multicolumn{1}{c|}{$\nu_A$} &
 \multicolumn{1}{c|}{$\nu_B$} &
 \multicolumn{1}{c|}{$\nu_{AB}$} &
 \multicolumn{1}{c|}{\%} \\
 \hline 
 \endhead

 \hline \multicolumn{11}{|r|}{{Continued on next page}} \\ \hline
 \endfoot

 \hline
 \endlastfoot

2013-03-04 & 0 & 0 & 2607 & 1224 & 1540 & 983 & 4.7e-1 & 5.9e-1 & 3.8e-1 & 26 \\
2013-03-04 & 0 & 1 & 2607 & 1224 & 813 & 230 & 4.7e-1 & 3.1e-1 & 8.8e-2 & -66 \\
2013-03-04 & 1 & 0 & 2607 & 88 & 1541 & 53 & 3.4e-2 & 5.9e-1 & 2.0e-2 & 2 \\
2013-05-22 & 0 & 0 & 750068 & 720763 & 664659 & 646644 & 9.6e-1 & 8.9e-1 & 8.6e-1 & 1 \\
2013-05-22 & 0 & 1 & 750068 & 720763 & 80853 & 69939 & 9.6e-1 & 1.1e-1 & 9.3e-2 & -11 \\
2013-05-22 & 0 & 2 & 750068 & 720763 & 3727 & 3409 & 9.6e-1 & 5.0e-3 & 4.5e-3 & -5 \\
2013-05-22 & 0 & 3 & 750068 & 720763 & 632 & 583 & 9.6e-1 & 8.4e-4 & 7.8e-4 & -4 \\
2013-05-22 & 0 & 4 & 750068 & 720763 & 137 & 129 & 9.6e-1 & 1.8e-4 & 1.7e-4 & -2 \\
\end{longtable}
\end{center}

Events A and B are a- and absolute b-increments expressed in seconds and $\delta$. The last session on June 21, 2013 with 7347 increments is added to the combined sample "ALL". Compare $n$ with Size in Tables \ref{a-increments} and \ref{b-increments}. The value in Column \% is $\frac{\nu_{AB}-\nu_A\nu_B}{\nu_{AB}}100\%$, where $\nu_A=\frac{m}{n}, \; \nu_B=\frac{l}{n}, \; \nu_{AB}=\frac{k}{n}.$ In order to reduce the table size, only the combinations with $k=n_{AB} \ge 50$ are reported.

\section{Appendix F. A Minor Correction}

After publication \cite{salov2007} in February 2007, the author detected that older r- and l-algorithms included missed the if-statement needed, if the last two prices were equal. The following modification was sent to the publisher in May 2007. The bold font and comments mark the missing text.

Page 43. c. If $i = $ end, then (set $U_{min} = \textrm{polarity} * U * (\textrm{polarity} - 1) / 2$; set $U_{max} = -\textrm{polarity} \times U \times (\textrm{polarity} + 1) / 2$; \textbf{If min = max, then (set } $\mathbf{U_{max} = -polarity \times U) } )$ go to STEP 2).

Page 44. c. If $i = $ begin, then (set $U_{min} = \textrm{polarity} \times U \times (\textrm{polarity} - 1) / 2$; set $U_{max} = -\textrm{polarity} \times U \times (\textrm{polarity} + 1) / 2$; \textbf{If min = max, then (set } $\mathbf{U_{max} = -polarity \times U) })$ go to STEP 2.)

Page 46.
\begin{verbatim}
if(i == prices.size() - 1) {
    units[minI] = polarity * (int)nContracts *
        (polarity - 1)/2;
    units[maxI] = -polarity * (int)nContracts *
        (polarity + 1)/2;
    // THE FOLLOWING TWO LINES MUST BE INSERTED!
    if(minI == maxI)
        units[maxI] = -polarity * (int)nContracts;
}
\end{verbatim}

Page 47.
\begin{verbatim}
if(i == 0) {
    units[minI] = polarity * (int)nContracts *
        (polarity - 1)/2;
    units[maxI] = -polarity * (int)nContracts *
        (polarity + 1)/2;
    // THE FOLLOWING TWO LINES MUST BE INSERTED!
    if(minI == maxI)
        units[maxI] = -polarity * (int)nContracts;
}
\end{verbatim}

{

\bigskip

\noindent\textbf{Valerii Salov} received his M.S. from the Moscow State University, Department of Chemistry in 1982 and his Ph.D. from the Academy of Sciences of the USSR, Vernadski Institute of Geochemistry and Analytical Chemistry in 1987.  He is the author of the articles on analytical, computational, and physical chemistry, the book Modeling Maximum Trading Profits with C++, \textit{John Wiley and Sons, Inc., Hoboken, New Jersey}, 2007, and papers in \textit{Futures Magazine} and \textit{ArXiv}.

\noindent\textit{v7f5a7@comcast.net}

\end{document}